\documentclass[prc,aps,nofootinbib,showkeys,showpacs]{revtex4}   
\usepackage{amssymb,mathrsfs}
\newcommand{\be}{\begin{equation}}
\newcommand{\ee}{\end{equation}}
\newcommand{\ba}{\begin{array}}
\newcommand{\ea}{\end{array}}
\newcommand{\q}{\quad}
\newcommand{\AIh}{{\cal A}}
\newcommand{\DIh}{{\cal D}}

\newcommand{\lpl}{\{\!\!\{}
\newcommand{\lpr}{\}\!\!\}}
\newcommand{\Id}{1\!\!1}

\begin{document}


\title{Correlation functions from a unified variational principle:\\ trial Lie groups}

\author{R. ~Balian \footnote{email: roger.balian@cea.fr}}
\affiliation{Institut de Physique Th\'eorique, CEA Saclay, 91191 Gif-sur-Yvette cedex, France}

\author{M. ~V\'en\'eroni \footnote{deceased, April 2015}}
\affiliation{Institut de Physique Nucl\'eaire, Universit\'e Paris-Sud and 
IN2P3-CNRS, F-91406 Orsay cedex, France}

\begin{abstract}
Time-dependent expectation values and correlation functions for
many-body quantum systems are
evaluated by means of a unified variational principle. It optimizes
a generating functional depending on sources associated with the
observables of interest. It is built by imposing through Lagrange
multipliers constraints that account for the initial state (at equilibrium or
off equilibrium) and for the backward Heisenberg evolution of the
observables. The trial objects are respectively akin to a density operator and
to an operator involving the observables of interest and the sources. We
work out here the case where trial spaces constitute Lie groups. This
choice reduces the original degrees of freedom to those of the
underlying Lie algebra, consisting of simple observables; the
resulting objects are labeled by the indices of a basis of this
algebra. Explicit results are obtained by expanding in powers of the
sources. Zeroth and first orders provide thermodynamic quantities and
expectation values in the form of mean-field approximations, with dynamical 
equations having a classical Lie-Poisson structure. At second order,
the variational expression for two-time correlation functions separates---as does
its exact counterpart---the approximate dynamics of the observables
from the approximate correlations in the initial state. Two building
blocks are involved: (i) a commutation matrix which stems from the
structure constants of the Lie algebra; and (ii) the second-derivative
matrix of a free-energy function. The diagonalization of both
matrices, required for practical calculations, is worked out, in a way
analogous to the standard RPA. The ensuing structure of the variational
formulae is the same as for a system of non-interacting bosons (or of harmonic
oscillators) plus, at non-zero temperature, classical gaussian
variables. This property is explained by mapping the original Lie algebra
onto a simpler Lie algebra. The results, valid for any trial Lie
group, fulfill consistency properties and encompass several
special cases: linear responses, static and time-dependent
fluctuations, zero- and high-temperature limits, static and dynamic
stability of small deviations.
\end{abstract}


\maketitle

\tableofcontents

\section{Introduction} \label{sec1}

Variational methods have proved their flexibility and efficiency in
many domains of physics, chemistry  and applied mathematics, in
particular when no small parameter allows perturbative approaches.
In physics one may wish, for systems of fermions, bosons or spins,
to evaluate variationally various quantities, such as thermodynamic
properties, expectation values, fluctuations or correlation
functions of some observables of interest. One may deal with the
ground state, with equilibrium at finite temperature, or with
time-dependences in off-equilibrium situations.

We want, moreover, to perform these evaluations {\it consistently}.
Suppose for instance that we have optimized the free energy of a
system by determining variationally its approximate state within
some trial class; nothing tells us that this state is also suited to
a consistent evaluation of other properties. Is it possible,
remaining in the same trial class, to optimize some other quantity
than the free energy, for instance a statistical fluctuation?

The wanted properties can be of different types. For instance, one
may be interested in both the expectation values and the
correlations of some set of basic observables. Consistency then
requires the simultaneous optimization of these quantities. In order
to evaluate them in a unified framework it appears natural, as in
probability theory and statistical mechanics, to rely on
characteristic functions, or more generally for time-dependent
problems, on functionals that generate correlation functions. As in
field theory, time-dependent sources ${\xi}_j(t)$ are associated
with the basic observables ${Q}_j$. Expansion of the generating
functional in powers of these sources will supply expectation values
and correlation functions of the set ${Q}_j$ in a consistent fashion.

Our strategy, therefore, will be {\it the variational optimization
of a generating functional} $\psi\{\xi\}$ for (connected) correlation
functions. To face this problem we will rely on a general method
\cite{GRSq, BV88q} allowing the systematic construction of a
variational principle that optimizes some wanted quantity. In this
procedure, the equations that characterize this quantity are
implemented as constraints by means of Lagrange multipliers.

The desired generating functional is expressed as $\displaystyle \psi\{\xi\}
\equiv \ln {\rm Tr}{\boldmath A}(t_{\rm i})\,D$ in terms of two
entities, the state $D$ in the Heisenberg picture and the
"generating operator" $\displaystyle A(t) \equiv  T \exp [ i\int^\infty_{t} dt'
\sum_j \xi_j(t')\,
 {\boldmath Q}_j^{\rm H}(t',t) ] $ taken at the initial time $t=t_{\rm i}$.
The operators ${\boldmath Q}_j^{\rm H}(t',t)$ entering $A(t)$ are
the observables of interest in the Heisenberg picture, $t'$ being
the running time and $t$ the reference time at which ${\boldmath
Q}_j^{\rm H}(t,t)$ reduces to the observable ${Q}_j$ in the
Schr\"odinger picture. The variational determination of
$\psi\{\xi\}$ together with its expansion at successive
orders in the sources ${\xi}_j(t')$ provides the various desired
outcomes: namely, at zeroth order (for $A(t)=I$) thermodynamic
potentials if $D$ is a (non-normalized) Gibbs state, or ground state
energy in the zero-temperature limit; at first order expectation
values; at second order correlation functions for an initial
off-equilibrium state $D$, etc... Static correlations within the
state $D$ will be obtained from sources located at the origin
$t_{\rm i}$ of times.

Implementing the variational principle requires the use of formally
simple equations which characterize the two ingredients $D$ and
$A(t_{\rm i})$ of the generating functional $\psi\{\xi\}$, and which
will be taken as constraints accounted for by Lagrange multipliers.
The state $D=\exp(-{\beta}K)$ will be characterized by Bloch's
equation for $D({\tau})=\exp(-{\tau}K)$; the associated Lagrange
multiplier is an operator depending on an imaginary time $\tau$. In
order to characterize the second ingredient $A(t_{\rm i})$ of
$\psi\{\xi\}$, we have defined $A(t)$ for an arbitrary initial time
$t$. The observables ${\boldmath Q}_j^{\rm H}(t',t)$ entering the
"generating operator" $A(t)$ then satisfy a {\it backward Heisenberg
equation} [Eq. (\ref{BH5q})] in terms of the reference time $t$ which
will eventually be fixed at $t_{\rm i}$. This backward Heisenberg
equation plays a crucial role as it produces for $A(t)$ the formally
simple differential equation (\ref{A3q}); the associated
Lagrange multiplier is a time-dependent operator. The equations for
the density operator $D({\tau})$ and for the generating operator
$A(t)$ are complemented by the boundary conditions $D({0})=I$ and
$A(+{\infty})=I$, where $I$ is the unit operator. The "time"
${\tau}$ of $D({\tau})$ varies forward, the time $t$ of $A(t)$
backward.

Unrestricted variations of the trial operators ${\DIh}(\tau)$ and
${\AIh}(t)$, and of their associated multipliers, yield the exact
generating functional, the stationarity conditions being the exact
dynamical equations for $D({\tau})$ and $A(t)$. These operators
should be restricted within a trial subspace to make the evaluations
feasible, and then the resulting equations become coupled. Their
solution will be simplified by expansion in powers of the sources
${\xi}_j(t)$, yielding a tractable, unified and consistent
treatment.
\vspace{.35cm}

In this article, we choose as trial subspace a {\it Lie group} of
operators. This will be the sole approximation. The formalism
will be developed for an arbitrary Lie group (Secs.\,\ref{sec3}-\ref{sec5}). 
Explicit calculations are then allowed for a sufficiently simple underlying
Lie algebra.

For arbitrary systems and for any trial Lie group, mean-field like
approximations are recovered at zeroth and first orders in the
sources (Sec.\,\ref{sec4}) for thermodynamic quantities and static or
dynamic expectation values (for instance, selecting for fermions at
finite temperature the Lie algebra of single-particle operators
yields back in this general framework the static and time-dependent
Hartree-Fock theories).

However, at second order in the sources, the formalism generates non
trivial results for fluctuations, for static correlations and for
two-time correlation functions (Sec.\,\ref{sec5}). Remarkably, new
variational approximations come out for these quantities, although
the trial operators belong to the same simple class as the one that
provides standard results for the expectation values (exponentials
of single-particle operators in the example of fermions). Within the
Lie group the trial operators adapt themselves to each question being
asked so as to optimize the answer -- while the use of the generating
functional ensures the consistency of the results thus obtained.

The second half of this article (Secs.\,\ref{sec6}-\ref{sec10}) is 
devoted to the properties of the outcomes of the present variational theory
with trial Lie groups. In particular, the static and dynamic stabilities
are related to each other (Sec.\,\ref{sec9}); a quasi-bosonic structure
(Sec.\,\ref{sec8}) and a classical structure (Sec.\,\ref{sec10}) are exhibited for any system;
various consistency properties are reviewed (Sec.\,\ref{sec6}).

Some past works are related to the present variational approach, which 
generalizes and unifies them within a natural framework. For
fermions at finite temperature, the optimization of expectation values
has been shown to lead to the static HF and dynamic TDHF equations
 \cite{BV85q}, while variational expressions for fluctuations \cite{BV92q}
 and correlation functions \cite{BV93q} have been derived. In particular,
 the large fluctuations observed in heavy ion nuclear reactions have thus been
 correctly reproduced \cite{BV84q,MK85q,BF85q,BS08q,Si11q,Si12q}, as recalled in Sec.\,\ref{sec6.2.5}.
 Applications of the variational principle of Sec.\,\ref{sec2.3} have been made
 to bosonic systems \cite{BF99q}, to Bose condensation \cite{BB10q, BB10q2},
 to field theory in ${\phi}^{4}$, including two-time correlation functions
 \cite{CMq,MB98q}, and to restoration of broken particle-number invariance
 for paired fermions at finite temperature \cite{BFV99q}. Let us also mention an
 application to cosmology \cite{EJPq} and an extension to control theory
\cite{KG95q,SGLq}. 
In the case of small fluctuations around the 
mean-field trajectory for fermions, the variational principle leads to time-dependent RPA 
corrections similarly to other approaches assuming either initial sampling of 
quantum fluctuations  \cite{DLI,DLII} or directly solving time-dependent 
RPA \cite{DLIII} (for a recent review, see \cite{DLIV}).
\vspace{.35cm}

The main results are recapitulated in Sec.\,\ref{sec11}, before the conclusion in 
Sec.\,\ref{sec12}.

\section{A unified variational approach} \label{sec2}

\subsection{Generating functional} \label{sec2.1}

We consider a quantum physical system prepared at the initial time $t_{\rm i}$ in a state
represented by a density operator of the form $\tilde{D} \propto \exp(-{\beta}K)$
in the Hilbert space $\mathscr{H}$. The dynamics is generated by the Hamiltonian $H$, possibly time-dependent.
If the system is in canonical (or grand canonical)
equilibrium, one has $K=H$ (or $K=H-{\mu}N$); for dynamical problems $K$ needs  
not commute with $H$.
Ground state problems are traited by letting ${\beta} \to \infty$.

Partition functions will be evaluated as ${\rm Tr}\,D$ for the unnormalized state
\begin{equation}  \label{IN12q}
 D = e ^{\,-\,{\beta}\,K}\,.
\end{equation}
The normalized state will be denoted as $\tilde{D}=D/{\rm Tr}\,D$. We
are mainly interested in expectation values, fluctuations  and
correlation functions of some set of observables denoted as $Q^{\rm
S}_j$ in the Schr\"odinger picture. We will work in the Heisenberg
picture in which the observables
\begin{equation}\label{IN13q}
Q_j^{\rm H}(t_{\rm f},t_{\rm i}) =U^\dagger(t_{\rm f},t_{\rm
i})\,Q_j^{\rm S}\,U(t_{\rm f},t_{\rm i})
\end{equation}
depend on two times, the initial reference time $t_{\rm i}$ and the
final running time $t_{\rm f}$. In the unitary evolution operator
\begin{equation}\label{Q1q}
U(t_{\rm f},t_{\rm i})=
  T e^{\textstyle -\frac{i}{\hbar}\int_{t_{\rm i}}^{t_{\rm f}} dt H(t)} \,
\end{equation}
$T$ denotes time ordering from right to left.

In order to generate consistently the desired quantities, we associate with each observable $Q_j$ a time-dependent source $\xi_j(t)$ and we introduce the 
{\it generating operator}
\begin{equation}      \label{A1q}
 {\boldmath A}(t)
\equiv  T e^{\textstyle \,i\int^\infty_{t} dt' \sum_j \xi_j(t') \,
 {\boldmath Q}_j^{\rm H}(t',t)}.
\end{equation}
Then, the {\it generating functional}
\begin{equation}    \label{GO2q}
 \psi\{\xi\} \equiv \ln {\rm Tr}{\boldmath A}(t_{\rm i})\, D \, ,
\end{equation}
which depends on the functions $\xi_j(t)$, encompasses the quantities of interest.
In particular the partition function ${\rm Tr}\,D$, the expectation values
\begin{equation}  \label{IN14q}
\langle Q_j \rangle_{t}
              ={\rm Tr}\,Q_j^{\rm H}(t,t_{\rm i})\tilde{D}
\end{equation}
at the time $t$, and the two-time correlation functions
\begin{equation}\label{Q4q}
C_{jk}(t',t'')= {\rm Tr}\,T\,Q_j^{\rm H}(t',t_{\rm i})\, Q_k^{\rm
H}(t'',t_{\rm i})\,\tilde{D} - \langle Q_j \rangle_{t'}  \langle
Q_k\rangle_{t''}\, ,
\end{equation}
are obtained as functional derivatives with respect to the sources
according to the successive terms of the expansion of $\psi\{\xi\}$,
\begin{equation}   \label{GO3q}
\begin{array}{rl}
\textstyle
\psi\{\xi\} =&
\textstyle
\!\!\ln {\rm Tr}\,D
+i\int_{t_{\rm i}}^\infty dt \sum_j \xi_j(t)  \langle Q_j \rangle _{t}
\vspace{.25cm} \\
&
\textstyle \quad
-{1\over 2} \int_{t_{\rm i}}^\infty dt' \int_{t_{\rm i}}^\infty dt''
\sum _{jk} \xi_j(t')\, \xi_k(t'') \,C_{jk}(t',t'') +\ldots \,.
\end{array}
\end{equation}
Variances ${\Delta}Q^{2}_j(t)$, hence fluctuations ${\Delta}Q_j(t)$, are found as $C_{jj}(t,t)$; 
static expectation values and correlations in the state $\tilde{D}$ are found by letting 
both times equal to $t_{\rm i}$. Linear responses are also covered by the formalism.

\subsection{The constraints} \label{sec2.2}

In order to optimize simultaneously all the quantities embedded in the generating functional
$\psi\{\xi\}$, use is made of a general procedure \cite{GRSq,BV88q}
inspired by the Lagrange multiplier method. The purpose is to construct an
expression whose stationary value provides the quantity we are looking for,
namely here  $\psi\{\xi\} = \ln{\rm Tr}\,A(t_{\rm i})D$.
To implement in this expression the quantities $D=e^{-{\beta}K}$ and $A(t)$,
we will characterize them by equations regarded as constraints.

To characterize the state $D$, we introduce a trial "time"-dependent
operator $\DIh(\tau)$ compelled to satisfy the initial condition
${\DIh}(0)=I$ and the {\it Bloch equation}
\begin {equation} \label{BL12q}
\frac {{\rm d}{\DIh}(\tau)}{{\rm d}\tau } + K\,{\DIh}(\tau)=0\,,
\end{equation}
where the imaginary "time" $\tau$ runs from $0$ to $\beta$. One
recovers $D={\exp}(-{\beta}K)$ from the solution of (\ref{BL12q})
for ${\tau}={\beta}$.

Let us turn to the generating operator $A(t)$ defined by (\ref{A1q}) in terms
of the observables $Q_j^{\rm H}(t',t)$. In this form, $A(t)$ looks difficult to 
handle and we wish to characterize it by a formally
simple equation that can be taken as a constraint.
To this aim, the operators $Q_j^{\rm H}(t',t)$ are regarded
as functions of the initial time $t$ rather than of the final running time $t'$
which is kept fixed at $t'=t_{\rm f}$. They thus satisfy
the {\it backward Heisenberg equation}  \cite{BV85q}
\begin{equation}     \label{BH5q}
\frac {dQ_j^{\rm H}(t_{\rm f},t)} {dt} = - \frac {1} {i\,\hbar} \,
[Q_{j}^{\rm H}(t_{\rm f},t), H] \,.
\end{equation}
This differential equation, together with its {\it final boundary
condition} $Q^{\rm H}_j(t_{\rm f},t_{\rm f})=Q^{\rm S}_j(t_{\rm f})$
at $t=t_{\rm f}$, is equivalent to the definition
(\ref{IN13q}) of $Q^{\rm H}_j(t_{\rm f},t)$. Contrary to
the standard forward Heisenberg equation (a differential equation in terms of $t'$), 
the backward equation (\ref{BH5q}) holds even when the observable
$Q_j^{\rm S}$ and/or the Hamiltonian $H$ depend on time in the
Schr\"odinger picture. Note that in the backward Eq.(\ref{BH5q}),
$H$ is written in the Schr\"odinger picture if it is time-dependent.

In the present context, the forward equation for  $Q^{\rm H}_j(t',t)$
would be of no help
in dealing with the definition (\ref{A1q}) of the generating operator $A(t)$
whereas the backward Heisenberg equation
(\ref{BH5q}) readily provides  the differential equation \cite{BV93q}
\begin{equation}    \label{A3q}
\frac {dA(t)}{dt}+\frac {1} {i\,\hbar} \,[A(t),\,H] +iA(t)\sum _j
\xi _j(t)\,Q^{\rm S}_j=0\, ,
\end{equation}
which, together with the boundary condition $A(+\infty)=I$,
is equivalent to (\ref{A1q}).
The generating functional $\psi\{\xi\}$ defined by (\ref{GO2q})
involves $A(t_{\rm i})$, and
this operator will be found by letting $t$ run backward in Eq.(\ref{A3q}) 
from $+\infty$ to $t_{\rm i}$,
with the final condition $A(+\infty)=I$.
Here again we shall simulate the operator $A(t)$,
solution of the exact equation (\ref{A3q}),
by a trial operator $\AIh(t)$ which will satisfy (\ref{A3q}) approximately.

In order to optimize the generating functional
$\psi\{\xi\} = \ln {\rm Tr}{\boldmath A}(t_{\rm i})\, D $, a variational expression
[Eq. (\ref{MVP5q}) below] will be constructed, which relies on the equations (\ref{BL12q})
for $D$ and (\ref{A3q}) for $A(t)$. These equations are regarded as constraints
with which Lagrange multipliers will be associated. We denote the
Lagrange multiplier accounting for the Bloch equation (\ref{BL12q})
by $\AIh(\tau)$, an operator depending on the "time" $\tau$;
we denote the Lagrange multiplier accounting for Eq.(\ref{A3q})
by $\DIh(t)$, a time-dependent operator.
These notations are inspired by the duality
between observables ${\AIh}$ and states ${\DIh}$ at the root
of the algebraic formulation of quantum mechanics \cite{Emchq,Dav15}
where expectation values are expressed as scalar products ${\rm Tr}\,{\AIh} {\DIh}$.
Here, $\DIh(\tau)$ (for $0 \le \tau \le \beta$) and the multiplier $\DIh(t)$
(for $t \ge t_{\rm i}$) are state-like quantities, whereas the multiplier $\AIh(\tau)$
and the operator $\AIh(t)$ are observable-like quantities.

\subsection{The variational principle} \label{sec2.3}

The implementation of the constraint (\ref{BL12q}) for ${\DIh}(\tau)$
by introducing the Lagrange multiplier ${\AIh}(\tau)$,
and of the constraint (\ref{A3q}) for ${\AIh}(t)$
by introducing of the Lagrange multiplier ${\DIh}(t)$,
results in the {\it variational expression} \cite{BV93q}
\begin{eqnarray}    \label{MVP5q}
&&  \Psi \{\AIh,\DIh \}  =
\ln{\rm Tr}\,{\AIh (t=t_{\rm i})} {\DIh(\tau=\beta)}\\
\nonumber
&&
- \int_{{0}}^{\beta}\,d{\tau}\,
{\rm Tr}\,\AIh(\tau) \left[ \frac{d\DIh(\tau)}{d\tau} +K\DIh(\tau) \right]
[{\rm Tr}\,\AIh(\tau) \DIh(\tau)]^{-1}\\
\nonumber
&&
+ \int_{t_{\rm i}}^{\infty}dt \,{\rm Tr}
\left[ \frac {d\AIh(t)}{dt}
+\frac {1} {i\,\hbar}\,[\AIh(t),\,H]
+i\,\AIh(t)\sum _j \xi _j(t)\,Q_j^{\rm S} \right]
{\DIh (t)} [{\rm Tr}\,\AIh(t) \DIh(t)]^{-1}\,,
\nonumber
\end{eqnarray}
where normalizing denominators are included for convenience. 
Together with the {\it mixed boundary conditions}
\begin{eqnarray}  \label{MVP6q}
      \AIh(t=+\infty)=I\,, \;\;\;\;\;  \DIh(\tau=0)=I\,,
\end{eqnarray}
$\Psi \{\AIh,\DIh \}$ should be made stationary with respect to
the four time-dependent operators
$\DIh(\tau), \AIh(\tau), \AIh(t), \DIh(t)$
(with $0\le \tau\le {\beta}$ and $t_{\rm i}\le t\le +\infty$).
The stationarity conditions include the additional {\it continuity relations}
\begin{equation}  \label{qb10q}
 \DIh(\tau=\beta)=\DIh(t=t_{\rm i})\;\;\; {\rm and}\;\;\;
 \AIh(\tau=\beta)=\AIh(t=t_{\rm i})\,,
\end{equation}
another argument for the notation.
(In view of this continuity one might replace $\tau$ by a complex time
$t = t_{\rm i}+i({\beta}-{\tau})\hbar$ so as to rewrite the two integrals
of (\ref{MVP5q}) as a single integral
on a Keldysh-like contour \cite{BV93q,Ke65q}.)

For unrestricted variations of $\AIh$ and $\DIh$
{\it the stationary value of $\Psi$ is the required
generating functional} $\psi\{\xi\}$.
It is attained for values of $\AIh$ and $\DIh$ that let
the two square brackets of (\ref{MVP5q}) vanish, so that we recover
the evolution equations (\ref{BL12q}) and (\ref{A3q}), the solutions of which are
$\DIh(\tau)=\exp{(-\tau{K})}$ and $\AIh(t)=A(t)$.

 The data of the problem, $K, H$ and the observables $Q_j^{\rm S}$, are operators
in the Hilbert space $\mathscr{H}$. They all enter explicitly
the variational expression $\Psi \{\AIh,\DIh \}$. Simpler
variational principles (VPs) derive from (\ref{MVP5q}) in two
special circumstances. If the initial state $D$ is workable, the
first integral over $\tau$ should be omitted. In this case, the
variational principle (\ref{MVP5q}) can be viewed for $\xi_j(t)=0$
as a transposition of the Lippmann-Schwinger VP \cite{LS50q} from
the Hilbert space, with duality between bras and kets, to the
Liouville space, with duality between observables and states
\cite{BV85q}. For static problems, the last integral over $t$ is
irrelevant \cite{BFV99q}. Classical problems enter the same
framework, with the replacement of the Hilbert space by the phase
space, traces by integrals and commutators by Poisson brackets.

As usual, practical exploitation of the above variational approach relies on  
restricting the trial spaces so that the expression (\ref{MVP5q}) of 
$\Psi \{\AIh,\DIh \}$ can be explicitly worked out. The denominators 
${\rm Tr}\,\AIh \DIh$ have been introduced so as to let the functional 
$\Psi \{\AIh,\DIh\}$ be invariant under time-dependent changes of 
normalization of $\DIh$ and $\AIh$.
This allows us to select a "gauge", that is, to fix at each time the traces of
$\DIh$ and $\AIh$ in such a way that the {\it stationarity conditions} take the form
\begin{eqnarray}
&& \label{MVP10q}
{\rm Tr}\,\delta\AIh(\tau)\,
\left[ \frac{d\DIh(\tau)}{d\tau} + K\DIh(\tau)\right]=0\,, \q\q\q
  ( 0 \le \tau \le \beta)  \\
&& \label{MVP11q}
{\rm Tr} \left[ \frac{d\AIh(\tau)}{d\tau}-\AIh(\tau) K \right]
\delta\DIh(\tau)\q=0\,,
\q\q\q   ( 0 \le \tau \le \beta)  \\
&&  \label{MVP13q}
{\rm Tr}\left[ \frac {d\AIh(t)}{dt}+\frac {1}{i\hbar}\,
[\AIh(t),H]+i\AIh(t)\sum _j \xi _j(t)Q_j^{\rm S} \right]
{\delta \DIh (t)}=0 ,   (t \ge t_{\rm i}) \\
&&   \label{MVP12q}
{\rm Tr}\,{\delta \AIh (t)}\left[ \frac {d\DIh(t)}{dt}
-\frac{1}{i\hbar}\,[H,\DIh(t)]
-i\sum _j\xi_j(t)Q_j^{\rm S}\DIh(t) \right]=0 ,
(t \ge t_{\rm i})
\end{eqnarray}
in the restricted space for $\DIh(\tau)$,  $\AIh(\tau)$,  $\AIh(t)$,  $\DIh(t)$
and the corresponding space for their infinitesimal variations.
In agreement with the boundary conditions (\ref{MVP6q})
and (\ref{qb10q}) relating the sectors $\tau$ and $t$,
equations (\ref{MVP10q}) and (\ref{MVP12q}) for $\DIh$ should be solved forward in time,
with $\tau$ running from  $0$ to $\beta$ and $t$ running from $t_{\rm i}$ to $\infty$,
whereas Eqs.(\ref{MVP13q}) and (\ref{MVP11q}) for $\AIh$ should be solved backward.
We obtain the stationary value of $\Psi$ as
\begin{eqnarray} \label{MVP14q}
  \psi\{\xi\} \simeq \ln \,{\rm Tr}\,\AIh(t)\,\DIh(t) = \ln \,{\rm Tr}\,\AIh(\tau)\,\DIh(\tau)
\end{eqnarray}
for arbitrary $t$ or $\tau$, as a consequence of the stationarity conditions written for
$\delta\AIh \propto \AIh$, $\delta\DIh \propto \DIh$.
In the following the restricted choice of the trial space will imply that the allowed variations
$\delta\AIh$ around $\AIh$ depend on $\AIh$, and likewise for $\delta\DIh$,
so that the {\it forward or backward equations (\ref{MVP10q}-\ref{MVP12q}) are coupled}.
Practical solutions will take advantage of their expansion in powers of the sources $\xi_j(t)$.

The variational procedure has duplicated the dynamical equations,
introducing Eqs.(\ref{MVP11q}) and (\ref{MVP12q}), besides the approximate Bloch equation (\ref{MVP10q}) 
and the approximate equation (\ref{MVP13q}) for the generating operator $A(t)$. While 
the formalism was set up in the Heisenberg picture, the stationarity condition (\ref{MVP12q}) reduces,
in the absence of sources and for unrestricted variations of $\AIh(t)$ and $\DIh(t)$,
to the Liouville-von Neumann equation
\begin{eqnarray}    \label{MVP20q}
 \frac {d\DIh(t)}{dt}  = \frac{1}{i\hbar}\,[H,\DIh(t)] \,.
\end{eqnarray}
The Lagrange multiplier matrix $\DIh(t)$ thus behaves as a time-dependent density operator
in the Schr\"odinger picture. However, such an interpretation does not hold in the presence
of sources, in which case $\DIh(t)$ is not even hermitian.

\section{Lie group as trial space} \label{sec3}

\subsection{Parametrizations and entropy} \label{sec3.1}

From now on, we specialize the trial space for the operators $\AIh$ and $\DIh$
involved in the variational principle (\ref{MVP5q}), assuming it to be endowed with a Lie group structure.
{\it This will be our sole approximation.}
The Lie group is generated by a Lie algebra $\{{\sf M}\}$ of operators acting on the 
Hilbert space $\mathscr {H}$, a basis of which is denoted
as ${\sf M}_{\alpha}$. This algebra is characterized by the structure constants
$\Gamma_{\alpha\beta}^{\gamma}$ entering the commutation relations
\begin{equation} \label{MFA00q}
 [ {\sf M}_{\alpha}, {\sf M}_{\beta}]
   = {i\,\hbar\,} \Gamma_{\alpha\beta}^{\gamma}\,{\sf M}_{\gamma}.
\end{equation}
(We use throughout the convention of summation over repeated indices.)
These constants are antisymmetric and satisfy the Jacobi identity
\begin{eqnarray}  \nonumber
    \Gamma_{\alpha\beta}^{\epsilon} \,\Gamma_{\epsilon\gamma}^{\delta}
  + \Gamma_{\beta\gamma}^{\epsilon} \,\Gamma_{\epsilon\alpha}^{\delta}
  + \Gamma_{\gamma\alpha}^{\epsilon}\,\Gamma_{\epsilon\beta}^{\delta} = 0\,.
\end{eqnarray}
It is convenient to include in the algebra the unit operator ${\sf I}$,
denoted as ${\sf M}_{0}$.
A seminal example is provided, for a many-fermion problem, by the algebra of fermionic
single-particle operators
$ {\sf M}_{\alpha} = a^{\dagger}_{\mu}a_{\nu}$ [with $\alpha \equiv (\nu,\mu) \ne 0$]. Other
examples are given in Sec.\,\ref{sec11.1}, such as, for condensed bosons, the set of creation 
and annihilation operators and of their products in pairs. 

 The operators $\AIh$ and $\DIh$ are then parametrized, at each time $\tau$ or $t$, according to
\begin{eqnarray} \label{bq2q}
\AIh=e^{{L}^{\alpha}{\sf M}_{\alpha}}\,,
     \;\;\;\;\;\;\;\;\;\DIh=e^{{J}^{\alpha}{\sf M}_{\alpha}}\,.
\end{eqnarray}
The parameters $L^{\alpha}$ and $J^{\alpha}$ are functions of the times $\tau$ or $t$;
their sets will be denoted as $\{{L}\}$ and  $\{{J}\}$. They will be complex due to the
presence of the sources ${\xi}_j(t)$, and to a possible non-hermiticity of the
operators $ {\sf M}_{\alpha}$.
[In the fermionic example, the operators ${\DIh}$ have the nature of non-normalized
independent-particle states; for bosonic problems they would encompass coherent states.]

The results will be conveniently expressed by writing $\DIh$ as
${\DIh} = Z\tilde{\DIh}$, where $Z$ denotes the normalization factor
\begin{equation}  \label{MFA8Bq}
 Z\{{J}\} \equiv {\rm Tr}\,\DIh = {\rm Tr}\,e^{{J}^{\alpha}\,{\sf M}_{\alpha}} \,,
\end{equation}
and where $\tilde{\DIh}$ is a normalized operator. Instead of the
set $\{{J}\}$, the operator $\DIh$ can alternatively be parametrized
by $Z$ and by the set $\{{R}\}$, defined by
\begin{equation}  \label{MFA8Cq}
 {R}_{\alpha} \equiv  {\rm Tr}\,{\sf M}_{\alpha}\,\tilde\DIh
= \frac {\partial \ln Z}  {\partial {J}^{\alpha}}
\end{equation}
and including ${R}_{0}=1$. [In the example of fermions, for $\alpha=(\nu,\mu) \ne 0$, the set
${R}_{\alpha} \equiv {R}_{\nu\mu}= {\rm Tr}\,a_{\mu}^{\dagger}a_{\nu}\tilde\DIh$
are the Wick contractions associated with the independent-particle trial operator
$\tilde\DIh$, so that a covariant vector $\{{R}\}$ can also be regarded as 
a single-particle density matrix. Contravariant vectors such as $\{{L}\}$ or $\{{J}\}$
are then regarded as matrices with switched indices, so as to produce the usual
expressions in Hilbert space for operators and scalars, such as
${L}^{\alpha}{\sf M}_{\alpha}=a^{\dagger}_{\mu}{L}^{\mu\nu}a_{\nu}$
or ${J}^{\alpha}{R}_{\alpha} = {J}^{\mu\nu}{R}_{\nu\mu} $.]

The converse equations of (\ref{MFA8Cq}), which relate the set $\{{J}\}$  to the set $\{{R}\},Z$,
involve the {\it von Neumann entropy} $(k_{\rm{B}}=1)$
\begin{equation}  \label{MFA8Dq}
  S\{{R}\} \equiv - {\rm Tr}\,\tilde\DIh \ln \tilde\DIh
     \equiv   \ln{\rm Tr}\,\DIh  - \frac {{\rm Tr}\,\DIh\ln\DIh} {{\rm Tr}\,\DIh }
   = \ln{Z} - {J}^{\alpha}\, {R}_{\alpha} \,,
\end{equation}
and they read
\begin{equation} \label{MFA9A20Bq}
{J}^{\alpha} =  - \frac {\partial S\{{R}\}}
       {\partial {R}_{\alpha}}\;\;({\alpha \ne 0})\,,
\;\;\;\;{J}^0= \ln {Z} - S\{{R}\}
 - \sum_{\alpha \ne 0}{J}^{\alpha}\, {R}_{\alpha}\,.
\end{equation}
 The Legendre transform (\ref{MFA8Dq})-(\ref{MFA9A20Bq}) from $\ln{Z\{{J}\}}$
to the von Neumann entropy $S\{{R}\}$ stems from the exponential
form of $\DIh$ as function of the parameters $\{{J}\}$.
The situation is the same as in thermodynamics
where a Legendre transform relates thermodynamic potentials and entropy
when going from intensive to extensive variables,
due to the the Boltzmann-Gibbs form of the equilibrium states.

The Jacobian matrix of the transformation relating the two parametrizations
of $\tilde\DIh$ is the {\it entropic matrix}
\begin{equation}    \label{MFA10cq}
{\mathbb S}^{\alpha\beta} \equiv
 \frac { \partial ^{2} S\{{R}\} }
       {\partial {R}_{\alpha} \, \partial {R}_{\beta}} =
  - \frac {\partial {J}^{\alpha} }
       {\partial{R}_{\beta}} \,,
 \;\; \;\;
\left( {\mathbb S}^{-1} \right)_{\alpha\beta} =
-\frac {\partial ^{2} \ln {Z}\{{J}\} }
       {\partial J^{\alpha} \, \partial J^{\beta}} =
  - \frac {\partial {R}_{\alpha} }
       {\partial J^{\beta}}
\;\;\;(\alpha,\beta \ne 0)\, ,
\end{equation}
which is the (negative) matrix of second derivatives of $S\{{R}\} $.

 As a remark, we note that a metric $ds^{2}$ can be defined in the full space
of density operators $\tilde{D}$ by $ds^{2}=-d^{2}S\{{\tilde{D}}\}$ \cite{Ba13q}.
The quantity $-\delta{R}_{\alpha}{\mathbb S}^{\alpha\beta}\delta{R}_{\beta}$
can indeed be interpreted  as the square of the distance, within the trial
subset of density operators, between the state $\tilde{\DIh}$ parametrized by
$\{{R}\}$ and the state $\tilde{\DIh}+\delta{\tilde{\DIh}}$ parametrized by
$\{{R}+\delta{R}\}$.
The matrix ${-{\mathbb S}}$ can thus be regarded as a metric tensor,
and the relation
$\delta{J}^{\alpha}= -{\mathbb S}^{\alpha\beta}\delta{R}_{\beta}$\,
$(\alpha,\beta \ne 0)$ as a correspondence between covariant and
contravariant coordinates.

\subsection{Symbols and images} \label{sec3.2}

We shall have to deal with quantities of the form $ {\rm Tr}\,Q\DIh$, where $\DIh$
will be some element of the Lie group and $Q$ some operator in the full Hilbert space
$\mathscr {H}$, not necessarily belonging to the Lie algebra.
In order to take care of such operators
it appears convenient to represent them by means of two useful tools.

Let us first introduce the {\it symbol} $q\{{R}\}$ of $Q$, a {\it scalar} which depends
both on the operator $Q$
(in the  Schr\"odinger picture) and on a normalized {\it running element} $\tilde{\DIh}$
of the Lie group. This symbol is defined by
\begin{equation}  \label{MFA7q}
  q\{{R}\} \equiv  {\rm Tr}\,Q\,\tilde\DIh\,,
\end{equation}
 as function of the parameters ${R}_{\alpha}$ (with $\alpha \ne 0$)
 that characterize $\tilde\DIh$. If $Q$ belongs to the Lie algebra,
this function is linear since ${R}_{\alpha} = {\rm Tr}\,{\sf
M}_{\alpha}\,\tilde\DIh $ itself is the symbol of ${\sf
M}_{\alpha}$. Otherwise $q\{{R}\}$ is non-linear. If $\tilde\DIh$
were a density operator, a symbol would be an expectation value but
here $\tilde\DIh$ is an arbitrary normalized element of the Lie group, not
necessarily hermitian. The symbol $q\{{R}\}$ of the operator $Q$ may
be viewed as a generalization, for any Lie group and for mixed
states, of the expectation value of $Q$ in a coherent state
\cite{Perq}, this value being regarded as a function of the
parameters that characterize this state.

Let us then introduce a second object associated with an operator $Q$,
its {\it image}  ${\sf Q} \{{R}\}$, {\it an element of the Lie algebra}
depending again both on $Q$ and on the running operator $\tilde\DIh$.
The image  ${\sf Q} \{{R}\}$ of $Q$ is constructed by requiring that both operators
${\sf Q} \{{R}\}$ and $Q$ should be equivalent, in the sense that
\begin{eqnarray}    \label{MFA901aq}
&&{\rm Tr}\, {\sf Q}\{{R}\} \, \tilde\DIh  = {\rm Tr}\,{Q}\,\tilde\DIh = q\{{R}\} \, ,
\end{eqnarray}
and that
\begin{eqnarray}     \label{MFA901bq}
&&{\rm Tr}\,{\sf Q}\{{R}\}\,\delta\DIh = {\rm Tr}\,{Q}\,\delta\DIh
\end{eqnarray}
for any infinitesimal variation $\delta\DIh$ around $\tilde\DIh$ within the Lie group, 
in particular for $\delta{\DIh} \propto {\sf M}_{\alpha}\tilde{\DIh}$,
or $\delta{\DIh} \propto \tilde{\DIh}\,{\sf M}_{\alpha}$,
or $\delta{\DIh} \propto \partial\tilde{\DIh}/\partial{R}_{\alpha}$.
Let us show that these conditions are sufficient to determine uniquely the image ${\sf Q} \{{R}\}$
associated with a given $Q$. Since it must belong to the Lie algebra, ${\sf Q} \{{R}\}$
is parametrized by a set of coordinates ${\cal Q}^{\alpha}\{{R}\}$ according to
 \begin{equation}    \label{MFA900q}
 {\sf Q} \{{R}\} = {\cal Q}^{\alpha} \{{R}\} \, {\sf M}_{\alpha}\,.
\end{equation}
For $\alpha \ne 0$, these coordinates ${\cal Q}^{\alpha}\{{R}\}$ are determined by inserting
(\ref{MFA900q}) into (\ref{MFA901bq}), by taking
$\delta{\DIh} \propto (\partial\tilde{\DIh}/\partial{R}_{\beta})\delta{{R}_{\beta}}$
and by using the definition (\ref{MFA7q}) of the symbol of $Q$, which yields
\begin{equation} \label{MFA902q}
  {\cal Q}^{\alpha}\{{R}\}  = \frac { \partial q\{{R}\} } { \partial{R}_{\alpha} } \,,
\;\;\;\;\;\; (\alpha \ne 0)\, .
\end{equation}
The coordinate ${\cal Q}^{0}\{{R}\}$ is obtained from (\ref{MFA901aq}). Altogether,
the image ${\sf Q} \{{R}\}$ of $Q$ defined by their equivalence (\ref{MFA901aq}),(\ref{MFA901bq})
is related to the symbol $q\{{R}\}$ of $Q$ through
\begin{equation} \label{MFA903q}
{\sf Q}\{{R}\}  =  q\{{R}\}\,{\sf M}_{0}
 + ( {\sf M}_{\alpha} - {R}_{\alpha} \,{\sf M}_{0}  )
     \frac {\partial q\{{R}\} }   {\partial{R}_{\alpha}} \;\;\;\; (\alpha \ne 0)\, .
\end{equation}
An operator $Q$ belonging to the Lie algebra coincides with its image. Otherwise,
its coordinates ${\cal Q}^{\beta}$ depend on the $ {R}_{\alpha}$'s, so that ${\sf Q}\{{R}\}$
is an effective operator simulating $Q$ in the Lie algebra, for a state close to $\tilde\DIh$.
The dependence of ${\cal Q}^{\alpha}\{{R}\}$ on $\{{R}\}$ arises from the occurrence
of $\tilde\DIh$ in the equivalence relation (\ref{MFA901aq}),(\ref{MFA901bq}).

\subsection{The commutation matrix ${\mathbb C}$ and the entropic matrix ${\mathbb S}$} \label{sec3.3}

We shall have to handle {\it products of two operators of the Lie algebra}.
When solving the dynamical equations we shall in particular encounter commutators
$[{\sf M}_{\alpha},\,{\sf M}_{\beta}]$.
Their symbol defines the {\it commutation matrix}
\begin{equation}    \label{MFA14q}
    {\mathbb C}_{\alpha\beta} \{{R}\}
   = \frac {1}{i\,\hbar}\,
   {\rm Tr}\, [{\sf M}_{\alpha},\,{\sf M}_{\beta}] \,\tilde\DIh =
   \Gamma^{\gamma}_{\alpha\beta} \,{R}_{\gamma} \, ,
\end{equation}
expressed in terms of the structure constants
$\Gamma^{\gamma}_{\alpha\beta}$ (for ${\gamma} = 0$, we have
${R}_{0}=1$). The matrix $ {\mathbb C}$ will play a crucial role.

Using $\ln\DIh={J}^{\gamma}{\sf M}_{\gamma}$, the vanishing of
 $ {\rm Tr}[{\sf M}_{\beta},\,\ln\DIh]\tilde\DIh = 0$ is expressed by
  \begin{equation}   \label{MFA10dq}
  {\mathbb C}_{\beta\gamma} \, {J}^{\gamma} =
  {\Gamma}^{\delta}_{\beta\gamma}\,{R}_{\delta}\,{J}^{\gamma}=0\,.
 \end{equation}
Taking the derivatives of this identity with respect to ${R}_{\alpha}$ and using
the relation (\ref{MFA10cq}) between the two parametrizations
$\{{J}\}$ and $\{{R}\}$ of $\DIh$, one obtains
for the product ${\mathbb C}\,{\mathbb S}$ the relations
\begin{eqnarray}   \label{MFA10fq}
   &&  {\Gamma}^{\alpha}_{\beta\gamma}\,{J}^{\gamma} =
   {\mathbb C}_{\beta\gamma} \, {\mathbb S}^{\gamma\alpha} \,,
   \;\;\;\; (\alpha \ne 0)
   \\ \nonumber
  && {\Gamma}^{0}_{\beta\gamma}\,{J}^{\gamma} =
   -\, {\mathbb C}_{\beta\gamma} \,{\mathbb S}^{\gamma\delta}\,{R}_{\delta} \,.
\end{eqnarray}
These equations help to show that the automorphism of the Lie algebra
engendered by the element ${\DIh}^{\lambda}$ of the group
can be expressed in the two equivalent forms
 \begin{equation}    \label{MFA10gq}
 {\sf M}_{\alpha}-{R}_{\alpha}  \mapsto
 {\DIh}^{-\,\lambda} ({\sf M}_{\alpha}-{R}_{\alpha}) {\DIh}^{\lambda} =
 \left( e^{ \,i\,{\hbar}\,{\mathbb C}\,{\mathbb S}\,\lambda }
 \right)_{\alpha}^{\,\gamma}  ({\sf M}_{\gamma}-{R}_{\gamma}) \, .
  \end{equation}
 (This is proved by evaluating the derivative with respect to ${\lambda}$
of the left-hand-side, using (\ref{MFA10fq})
then integrating from $0$ to ${\lambda}$.)

The property (\ref{MFA10gq}) will be exploited
later on (Sec.\,\ref{sec5.5} and Appendix A). We use it here to find the expression of
the symbol of the product ${\sf M}_{\alpha}\,{\sf M}_{\beta}$,
or equivalently of the correlation
${\rm Tr}\,{\sf M}_{\alpha}{\sf M}_{\beta} \tilde{\DIh}
-{R}_{\alpha}\,{R}_{\beta} =
{\rm Tr} ({\sf M}_{\alpha}-{R}_{\alpha}){\sf M}_{\beta} \tilde{\DIh}$ 
(with ${R}_{\alpha}={\rm Tr}\,{\sf M}_{\alpha} \tilde{\DIh}$).
To this aim, we start from the expression (\ref{MFA10cq}) of ${\mathbb S}^{-1}$,
and evaluate explicitly therein the derivatives with respect to $\{{J}\}$:
  \begin{eqnarray}   \nonumber
  - \left( {\mathbb S}^{-1} \right)_{\alpha\beta} =
   \frac {\partial ^{2} \ln {Z}\{{J}\} }
       {\partial J^{\alpha} \, \partial J^{\beta}} =
  \frac {\partial} {\partial J^{\alpha}}
  \frac { {\rm Tr}\,e^{\,J^{\gamma}\,{\sf M}_{\gamma}} \,{\sf M}_{\beta}  }
        { {\rm Tr}\,e^{\,J^{\gamma}\,{\sf M}_{\gamma}}  }
 \\ \label{MFA010q}
   =  {\rm Tr}\, \int_{0}^{1} d\lambda \,
    {\tilde\DIh}^{1-\lambda}
    ( {\sf M}_{\alpha}-{R}_{\alpha} )
    {\tilde\DIh}^{\lambda} \, {\sf M}_{\beta} \, ,
  \end{eqnarray}
where we made use of the first-order expansion in the shift $\{\delta{J}\}$ of
 \begin{eqnarray}    \label{MFA011q}
 \exp \left( J^{\gamma}\,{\sf M}_{\gamma}
   + \delta{J}^{\gamma}\,{\sf M}_{\gamma} \right)
  \approx  \DIh  + \delta{J}^{\gamma}  \int_{0}^{1} d\lambda \,
    {\DIh}^{1-\lambda} \,{\sf M}_{\gamma} \, {\DIh}^{\lambda}, \, 
  \end{eqnarray}
with $\DIh =  \exp { J^{\gamma}\,{\sf M}_{\gamma}}  $.
We recognize in the r.h.s. of (\ref{MFA010q}) the Kubo correlation
of ${\sf M}_{\alpha}$ and ${\sf M}_{\beta}$ in the normalized state
$\tilde\DIh$. By means of (\ref{MFA10gq}) the integration over $\lambda$
can be performed explicitly in (\ref{MFA010q}), which yields
 \begin{eqnarray}    \label{MFA012q}
 -\, ({\mathbb S}^{-1})_{\alpha\beta} =
   {\rm Tr} \,\int_{0}^{1} d{\lambda}  \,
  \left( e^ {\, i\,\hbar \,{\mathbb C}\,{\mathbb S}\,{\lambda} }
  \right)_{\alpha}^{\,\gamma}\,
  ( {\sf M}_{\gamma}- {R}_{\gamma} )\, {\sf M}_{\beta}\, {\tilde\DIh} \,
   \\ \nonumber
   =  \left(  \frac { e^{\,i\,\hbar\,{\mathbb C}\,{\mathbb S}}-{\mathbb I} }
           { i\,\hbar\,{\mathbb C}\,{\mathbb S} } \, \right)^{\,\gamma}_{\alpha}
  {\rm Tr}\,({\sf M}_{\gamma} - {R}_{\gamma})\,{\sf M}_{\beta}\,{\tilde\DIh}\,.
  \end{eqnarray}
Hence the ordinary correlations between operators of the Lie algebra
in any element ${\tilde\DIh}$ of the Lie group are found to be given by
 \begin{eqnarray}    \label{MFA10hq}
{\rm Tr}\,{\sf M}_{\alpha}\,{\sf M}_{\beta}\tilde{\DIh}-{R}_{\alpha}\,{R}_{\beta}
 =  \left( -\, \frac  { i\,\hbar\,{\mathbb C}\,{\mathbb S} }
                    {  e^{\,i\,\hbar\,{\mathbb C}\,{\mathbb S} }-  {\mathbb I} }
                  \, {\mathbb S}^{-1}      \right)_{\alpha\beta} \, .
  \end{eqnarray}

\subsection{The variational formalism in the restricted space} \label{sec3.4}

With the above tools in hand, it is possible to implement
the Lie-group form (\ref{bq2q}) of the
trial operators into the variational expression (\ref{MVP5q}) of $\Psi\{ {\AIh},{\DIh} \}$
by taking as variables the parameters $\{{L}\},\{{R}\},Z$ that characterize
${\AIh}$ and ${\DIh}$ at the times $\tau$ or $t$. These operators ${\AIh}$ and ${\DIh}$
appear within $\Psi\{ {\AIh},{\DIh} \}$ through products $\AIh\DIh$ and $\DIh\AIh$.
Such products belong to the Lie group and are characterized by the parameters
${R}_{\alpha}^{\AIh\DIh} =  {{\rm Tr}\,{\sf M}_{\alpha} \AIh\,\DIh} /
                                     {{\rm Tr}\,\AIh\,\DIh}$,
${R}_{\alpha}^{\DIh\AIh} = {{\rm Tr}\, {\sf M}_{\alpha}   \DIh\AIh} /
                                      {{\rm Tr}\,\DIh\AIh} $,
$Z^{\AIh\DIh}=Z^{\DIh\AIh}={\rm Tr}\,\AIh\DIh$, which should be expressed in terms of
our basic variables $\{{L}\},\{{R}\},Z$ by relying on the group properties.

The initial state operator $K=-{\beta}^{-1}\ln{D}$, the Hamiltonian $H$ and the observables
$Q_j^{\rm S}$ enter $\Psi \{\AIh,\DIh \}$ through traces of the form
${{\rm Tr}\,{Q} \AIh\,\DIh} / {{\rm Tr}\,\AIh\,\DIh}$ and
${{\rm Tr}\,{Q} \DIh\,\AIh} / {{\rm Tr}\,\DIh\,\AIh}$,
where $Q$ stands for  $K,H$ or $Q_j^{\rm S}$. We are thus led to introduce, for any
operator $\tilde\DIh$ parametrized by $\{{R}\}$, the symbols
\begin{equation}  \label{MFA9Aq}
  k\{{R}\} \equiv {\rm Tr}\,K\,\tilde\DIh\,, \;\;\;
  h\{{R}\} \equiv {\rm Tr}\,H\,\tilde\DIh\,,  \;\;\;
  q_j\{{R}\} \equiv {\rm Tr}\,Q_j^{\rm S}\,\tilde\DIh\,
\end{equation}
of $K,H$ and $Q_j^{\rm S}$. These symbols occur within $\Psi \{\AIh,\DIh \}$ for values
of $\{{R}\}$ equal to $\{{R}^{\AIh\DIh}\}$ or $\{{R}^{\DIh\AIh}\}$. The {\it variational
expression} $\Psi \{\AIh,\DIh \}$, when specialized to a Lie group, then takes the form
\begin{eqnarray}    \label{GS30q}
&&\Psi \{{\AIh},{\DIh}\} = \ln {\rm Tr}\,\AIh(t=t_{\rm i})\,\DIh (\tau=\beta)
  \\ \nonumber  &&
- \int_{0}^{\beta}\,{d\tau}\,
 \left(\frac {\partial \ln{Z}^{\AIh\DIh}} {\partial{R}_{\alpha}}
\frac {d{R}_{\alpha}} {d\tau} + \frac {d\ln{Z}} {d\tau}
+\frac{i}{\hbar}\, k\{{R}^{\DIh\AIh}\} \right)\\
\nonumber
&&
+ \int_{t_{\rm i}}^{\infty}dt
\left( \frac {\partial \ln{Z}^{\AIh\DIh}} {\partial{L}^{\alpha}}
\frac {d{L}^{\alpha}} {dt}
+\frac {1} {i\,\hbar}\, \left[ h\{{R}^{\DIh\AIh}\}-h\{{R}^{\AIh\DIh}\} \right]
+ i \sum_{j} \xi_{j}(t) \,q_{j} \{{R}^{\DIh\AIh}\}
\right) \, ,
\nonumber
\end{eqnarray}
which should be regarded as a functional of the original trial parameters
$\{{L}\},\{{R}\}$ taken at times $\tau$ and $t$, and
$Z(\tau) = {\rm Tr}\,\DIh (\tau)$.

The {\it stationarity conditions} (\ref{MVP10q}-\ref{MVP12q}), obtained by functional derivation
with respect to these parameters, now read
 \begin{eqnarray}    \label{GS200q}
 &&\frac{d\DIh(\tau)}{d\tau} + {\sf K}\{{R}^{\DIh\AIh}\} \,\DIh(\tau)=0\,, \q\q\q\q
({0} \le {\tau} \le {\beta})\\
&&   \label{GS201q}
 \frac{d\AIh(\tau)}{d\tau} - \AIh(\tau)\, {\sf K}\{{R}^{\DIh\AIh}\} =0\,, \q\q\q\q
  ({0} \le {\tau} \le {\beta})\\
&&  \label{GS202q}
  \frac {d\AIh(t)}{dt} + \frac {1}{i\,\hbar} \left[
 \AIh(t)\,{\sf H}\{{R}^{\DIh\AIh}\} - {\sf H}\{{R}^{\AIh\DIh}\}\,\AIh(t)  \right] \nonumber \\
  &&\q\q\q\q
 + \,i\sum _j \xi _j(t) \,\AIh(t) \, {\sf Q}_{j}\{{R}^{\DIh\AIh}\} =0\,, \q\q\q (t \ge t_{\rm i}) \\
&& \label{GS203q}
\frac {d\DIh(t)}{dt} + \frac {1}{i\,\hbar} \left[
 \DIh(t)\,{\sf H}\{{R}^{\AIh\DIh}\} - {\sf H}\{{R}^{\DIh\AIh}\}\,\DIh(t)  \right] \nonumber \\
  &&\q\q\q\q
 -  \,i\sum _j \xi _j(t)  \, {\sf Q}_{j}\{{R}^{\DIh\AIh}\}\,\DIh(t) =0\,,
  \q\q\q (t \ge t_{\rm i}) \, .
\end{eqnarray}
These equations involve the images $ {\sf K}\{{R}\},{\sf H}\{{R}\}$ and ${\sf Q_j}\{{R}\}$ issued,
according to the general relation (\ref{MFA903q}),
from the derivatives of the corresponding symbols  $ {k}\{{R}\},{h}\{{R}\}$ and ${q_j}\{{R}\}$
defined by (\ref{MFA9Aq}). The stationarity conditions (\ref{GS200q})-(\ref{GS203q}) should
be solved with the boundary conditions (\ref{MVP6q}) and (\ref{qb10q}).
When they are satisfied, ${Z}^{\AIh\DIh}$ is constant
in $\tau$ and $t$, and the optimal estimate for the generating functional $\psi\{\xi\}$
reduces to the first term  $\ln{Z}^{\AIh\DIh}$ of (\ref{GS30q}), in agreement with (\ref{MVP14q}).

For most Lie groups, solving the coupled equations (\ref{GS200q}-\ref{GS203q}) is hindered
by the need of expressing explicitly
the parameters $\{{R}^{\DIh\AIh}\},\{{R}^{\AIh\DIh}\},{Z}^{\AIh\DIh}$ of ${\AIh\DIh}$ and
${\DIh\AIh}$ in terms of those $(\{{L}\},\{{R}\}$ and $Z$) of ${\AIh}$ and  ${\DIh}$. However,
we are interested in the first terms of the expansion of $\psi\{\xi\}$ in powers of the
sources $\xi_j(t)$. Accordingly we shall only need to express,
as functions of the parameters $\{{R}\}$ and $Z$
of an arbitrary element $\DIh$ of the Lie group, the following ingredients:
(i) the symbols (\ref{MFA9Aq}) of the operators $K,H$ and $Q_j^{\rm S}$,
and (ii) the entropy function (\ref{MFA8Dq}) which allows us to relate
the sets $\{{R}\}$ and $\{{J}\}$.
This is feasible for many Lie groups.
[In the example of the algebra of single-fermion operators
$a^{\dagger}_{\mu}a_{\nu}$, this is achieved by Wick's theorem.]
Thus, explicit solutions of the equations of motion will be found
at the first few orders in the sources.

\section{Zeroth and first orders} \label{sec4}

\subsection{Thermodynamic quantities} \label{sec4.1}

At zeroth-order in the sources $\{{\xi}\}$, the quantity of interest
is the partition function ${\rm Tr}\,e^{\,-{\beta}K}$, or the "generalized free energy"
\begin{equation}    \label{MFA11Aq}
  F \equiv -\, {\beta}^{-1} \,\ln{\rm Tr}\,e^{\,-\,{\beta}\,K}
    =-\, {\beta}^{-1} \, \psi\{\xi=0\}
\end{equation}
$(k_{\rm B}=1,\,T={\beta}^{-1})$ which reduces to the standard free energy
for $K=H$, or to the grand potential for $K=H -{\mu} N$. It is variationally approximated by
$ -\, {\beta}^{-1}\,\Psi\{{\AIh}^{(0)}, {\DIh}^{(0)}\}
 = -\, {\beta}^{-1}\,\ln{\rm Tr}\,{\AIh}^{(0)}(t=t_{\rm i})
 {\DIh}^{(0)}({\tau}
 =\beta) $,
where the upper index denotes the order in the sources $\{{\xi}\}$.

For $\{{\xi}\} = 0$, the stationarity condition (\ref{GS202q}) yields
${\AIh}^{(0)}(t)=I$
for $t \ge t_{\rm i}$, hence, from Eq.(\ref{qb10q}), ${\AIh}^{(0)}(\tau=\beta)=I$
and $ F \simeq -\, {\beta}^{-1} \,\ln{\rm Tr}\,{\DIh}^{(0)}({\tau}=\beta)$.
Thus ${\DIh}^{(0)}({\tau}=\beta) \equiv {\DIh}^{(0)}$ appears as an approximation,
variationally suited to the evaluation of thermodynamic quantities,
of the exact state $D=e^{\,-\,{\beta}\,K}$.

To obtain ${\DIh}^{(0)}({\tau}=\beta)$ we have to solve the first two stationarity conditions
(\ref{GS200q}-\ref{GS201q}) with ${\DIh}^{(0)}(0)={\AIh}^{(0)}(\beta)=I$.
We make the Ansatz
    \begin{eqnarray}  \label{qb11q}
       {\DIh}(\tau) {\AIh}(\tau)  = \DIh^{(0)} \, ,
    \end{eqnarray}
where $\DIh^{(0)} \equiv {\DIh}^{(0)}({\tau}=\beta)$ is a constant operator,
still to be determined and characterized by its parameters
${R}_{\alpha}^{(0)} \equiv {\rm Tr}\,{\sf M}_{\alpha}\tilde\DIh^{(0)}$
for $\alpha \ne 0$ and $ {Z}^{(0)} \equiv {\rm Tr}\,\DIh^{(0)}$.
The image ${\sf K}\{{R}^{\DIh\AIh}\}$ is then the constant operator
${\sf K}\{{R}^{(0)}\}$, so that we can solve 
Eqs.(\ref{GS200q}) and (\ref{GS201q}) in the form
${\DIh}^{(0)}(\tau)= [ {\DIh}^{(0)}]^{\tau/\beta}$, 
${\AIh}^{(0)}(\tau)= [ {\DIh}^{(0)}]^{(\beta-\tau)/\beta}$,
where $ {\DIh}^{(0)}$ is determined by the equation
\begin{eqnarray}  \label{qb29q}
  \ln \DIh^{(0)}  =  -\,{\beta} \,{\sf K}\{{R}^{(0)}\} \,.
\end{eqnarray}

 The operator equation (\ref{qb29q}) provides
${\DIh}^{(0)}(\tau=\beta)= {\DIh}^{(0)}$.
More explicitly, in the basis $\{{\sf M}\}$ of the Lie algebra,
the components $\alpha \ne 0$ of (\ref{qb29q}) read
$ {J}^{(0)\alpha}   = -\,{\beta} \, {\cal K}^{\alpha}\{{R}^{(0)}\}$
in terms of the coordinates
$ {J}^{(0)\alpha}$ of $\ln \DIh^{(0)} \equiv {J}^{(0)\alpha} {\sf M}_{\alpha}$
and the coordinates ${\cal K}^{\alpha}\{{R}^{(0)}\} $
of $ {\sf K}\{{R}^{(0)}\} $ [defined as in (\ref{MFA900q}-\ref{MFA903q})].
This yields the self-consistent equations
\begin{eqnarray}  \label{qb30bq}
   \frac {\partial S\{{R}^{(0)}\}} { {\partial {R}}^{(0)}_{\alpha} }
    = \beta\, \frac {\partial k\{{R}^{(0)}\}}  { {\partial{R}}^{(0)}_{\alpha}  } \,
  \end{eqnarray}
which determine the parameters $\{{R}^{(0)}\}$ of
${\tilde{\DIh}}^{(0)}$. The component $\alpha = 0$ of (\ref{qb29q})
yields $ {J}^{(0)0} = k\{{R}^{(0)}\} - {R}^{(0)}_{\alpha}\,
  { \partial k\{{R}^{(0)}\} } /  { {\partial{R}}^{(0)}_{\alpha} }  $,
where Eq.(\ref{MFA903q}) has been used for ${\sf K}$. Together with
(\ref{MFA9A20Bq}) and (\ref{qb30bq}), this equation provides for the
sought {\it free energy} $F$ the alternative expressions
\begin{eqnarray}  \label{qb20Aq}
F \simeq  -\,  {\beta}^{-1} \ln {\rm Tr}\,e^{ \,-\,{\beta} \,{\sf K}\{{R}^{(0)}\} }
  =   -\, {\beta}^{-1}\,\ln{Z}^{(0)} =  f\{{R}^{(0)} \} \, ,
\end{eqnarray}
where we defined the free-energy function $f\{{R}\}$ through
\begin{equation}    \label{MFA12q}
  f\{{R}\}  \equiv  k\{{R}\}-T\,S\{{R}\} \, .
\end{equation}

The above self-consistent equations determine a hermitian operator
${\tilde{\DIh}}^{(0)}$ which can be interpreted as an approximation
for the exact density operator ${\tilde{D}}$ (whereas the trial
operator ${\DIh}$ occurring in the presence of sources is not
hermitian). The general relation (\ref{MFA10dq}) entails
\begin{equation}    \label{MFA30q}
   \Gamma^{\gamma}_{\alpha\beta} \,
   {\cal K}^{\beta}\{{R}^{(0)}\} \, {R}_{\gamma}^{(0)} = 0 \, ,
\end{equation}
an equation which is equivalent to ${\rm Tr}\,{\sf M}_{\gamma}[\ln
{\DIh}^{(0)},\, \tilde\DIh^{(0)}] = 0$. [In the Hartree-Fock
example, (\ref{MFA30q}) expresses the commutation of the
single-particle density matrix with the effective hamiltonian
matrix.]

 Our variational principle provides solutions for $\psi\{\xi=0\}= -\,{\beta}\,F$
that are not maxima but only stationary values of $\Psi\{
\AIh,\DIh\}$; it relies on the Bloch equation rather than on the
maximization of the entropy under constraints. However, it turns out
that the above result (\ref{qb30bq})
coincides with the outcome of the standard maximum entropy
procedure. Indeed the latter amounts to minimizing the left-hand
side of the Bogoliubov inequality
\begin{eqnarray}  \label{qbABCq}
   {\rm Tr}\,{K}\,\tilde\DIh  +  {\beta}^{-1}\,{\rm Tr}\,\tilde\DIh \ln \tilde\DIh
 \ge     -\, {\beta}^{-1} \ln{\rm Tr}\,e^{\,-\,{\beta}\,K}  \equiv  F
\end{eqnarray}
with respect to the normalized trial density operator $\tilde\DIh$.
When $\tilde\DIh$ in (\ref{qbABCq}) is restricted to the Lie group,
the left-hand side reduces to $f\{{R}\}$, where $\{{R}\}$ is the set
characterizing $\tilde{\DIh}$. The best estimate for $F$ is thereby
the minimum of the free-energy function $f\{{R}\}$, which requires that the 
equations (\ref{qb30bq}) are satisfied. The equivalence
between the two variational approaches provides a criterion for
selecting the best solution $\{{R}^{(0)}\}$ of the self-consistent stationarity
conditions (\ref{qb30bq}) when they have several solutions, namely,
the one for which $f\{{R}^{(0)}\}$ is the {\it absolute minimum} of $f\{{R}\}$.

The standard relations
$ S= - \partial F/\partial T \simeq
 - {\rm Tr}\,\tilde\DIh^{(0)}  \ln \tilde\DIh^{(0)} =  S\{{R}^{(0)} \}  $ and
$ \langle K \rangle = F + TS \simeq  k\{{R}^{(0)}\} $ are satisfied as usual, so that
the approximation is thermodynamically consistent.
Thermodynamic coefficients are obtained by derivation, which introduces
the matrix $(\alpha,\beta \ne 0)$
 \begin{equation} \label{MFA20q}
 {\mathbb F}^{\alpha\beta} =
 \frac {\partial^{\,2} f\{{R}^{(0)}\}} {\partial{R}^{(0)}_{\alpha}\,
   \partial{R}^{(0)}_{\beta}}
 = {\mathbb K}^{\alpha\beta} - T\, {\mathbb S}^{\alpha\beta}
 =   \frac {\partial^{\,2} k\{{R}^{(0)}\}} {\partial{R}^{(0)}_{\alpha}\,
   \partial{R}^{(0)}_{\beta}}
    -  T \,
 \frac {\partial^{\,2} S\{{R}^{(0)}\}} {\partial{R}^{(0)}_{\alpha}\,
  \partial{R}^{(0)}_{\beta}}\,.
 \end{equation}
In particular, the heat capacity is found for $K=H$ as
\begin{eqnarray}  \label{qbCTq}
C \simeq \beta \,{\cal K}^{\alpha}\{{R}^{(0)}\} ({\mathbb
F}^{-1})_{\alpha\beta}\,{\cal K}^{\beta}\{{R}^{(0)}\} \,.
\end{eqnarray}
 The positivity of the {\it stability matrix} ${\mathbb F}$ at the minimum
of $f\{{R}\}$, entailed by the inequality (\ref{qbABCq}),
will play an important role below.
[For the fermionic single-particle Lie algebra $ a^{\dagger}_{\mu}a_{\nu}$,
we recover the thermal HF approximation, either under the form (\ref{qb29q})
or through the minimization of $f\{{R}\}$.]

\subsection{Expectation values} \label{sec4.2}

Expectation values are obtained by expanding  the generating functional $\psi\{\xi\}$
at first order in the sources $\xi_j(t)$ [Eq.(\ref{GO3q})].
These sources occur both directly in $\Psi\{ \AIh,\DIh\}$,
as exhibited by the last term of (\ref{GS30q}),
and indirectly through the values of $\AIh$ and $\DIh$ at the stationarity point
where $ \psi\{{\xi}\} = \Psi\{ \AIh,\DIh\} $.
At this point, however, the partial derivatives of
$\Psi\{{\AIh},{\DIh}\}$ with respect to $\AIh$ and $\DIh$ vanish;
one is left with the explicit derivative
 $\partial\Psi\{\AIh^{(0)},\DIh^{(0)}\} / \partial\xi_j(t )$
taken at the zeroth-order point $\{\AIh^{(0)}$,$\DIh^{(0)}\}$, so that
\begin{equation}    \label{DVP21q}
 \langle Q_j \rangle_{t} =
 \left.\frac{1}{i}\frac {\partial\psi\{\xi\}  }{\partial \xi_j(t)}\right|_{\xi=0}
 \simeq \frac { {\rm Tr}\,Q_j^{\rm S}\,\DIh^{(0)}(t) \,\AIh ^{(0)}(t) }
      { {\rm Tr}\,\AIh^{(0)}(t)\, \DIh^{(0)}(t) } \, 
\end{equation}
involves only the zeroth-order approximations 
$\AIh ^{(0)}(t)$ and $\DIh ^{(0)}(t)$.

We have seen that $ \AIh ^{(0)}(t)=I$ for $t \ge t_{\rm i}$, and that
$\tilde\DIh^{(0)}(t_{\rm i})$, equal to
$\tilde\DIh^{(0)}(\tau = \beta) \equiv \tilde\DIh^{(0)}$, is given self-consistently
by (\ref{qb29q}). Hence, the static expectation value ${\langle Q_j \rangle}_{t_{\rm i}}$
in the state $\tilde{D}$ is variationally expressed by
\begin{equation} \label{MFA130q}
{\rm Tr}\, {Q}^{\rm S}_j \,\tilde{D} = \frac { {\rm Tr}\,Q^{\rm
S}_j\, e^{\,-\,{\beta}K} }  { {\rm Tr}\, e^{\,-\,{\beta}K} }
   \simeq  {\rm Tr}\, {Q}_j^{\rm S}\,\tilde\DIh^{(0)}
\equiv  q_{j}\{{R}^{(0)}\}  =   {\cal Q}^{\alpha}_{j} \{ {R}^{(0)}
\}  \, {R}^{(0)}_{\alpha} \, ,
  \end{equation}
that is, by the symbol (\ref{MFA7q}) of the observable ${Q}^{\rm
S}_j$ of interest.

For dynamical problems, the expectation value (\ref{DVP21q}) involves $\tilde{\DIh}^{(0)}(t)$, 
provided by the last stationarity condition (\ref{GS203q}) taken for $\xi_j(t)=0$
and for ${\AIh}^{(0)}(t)=I$. Thus, $\tilde\DIh^{(0)}(t)$ is determined by the zeroth order self-consistent equation
\begin{eqnarray}   \label{qb23q}
   \frac {d\tilde\DIh^{(0)}(t)} {dt}=
   \frac {1} {i\,\hbar}\, \left[ {\sf H}\{{R}^{(0)}(t)\},
  \,\tilde\DIh^{(0)}(t) \right]\, ,
\end{eqnarray}
with the initial condition $\tilde\DIh^{(0)}(t_{\rm i}) = \tilde\DIh^{(0)}$.
[The norm $Z^{(0)}(t)$ is constant and equal to $Z^{(0)}$.] In (\ref{qb23q})
the operator ${\sf H}\{{R}^{(0)}(t)\}$ is the image taken for
${R}_{\alpha}^{(0)}(t) =  {\rm Tr}\,{\sf M}_{\alpha}\tilde\DIh^{(0)}(t)$ of the Hamiltonian $H$.
The {\it time-dependent expectation values} (\ref{DVP21q}) are therefore found as
\begin{equation}    \label{MFA125q}
  \langle Q_j \rangle_t \equiv   {\rm Tr}\,Q^{\rm H}_j(t,t_{\rm i})\tilde{D}
      \simeq   {\rm Tr}\,Q^{\rm S}_j \, \tilde{\DIh}^{(0)}(t)
      \equiv   q_{j}\{ {R}^{(0)}(t)\} \,.
 \end{equation}
As noted at the end of Sec.\,\ref{sec3}, the variational equation (\ref{qb23q})
for the Lagrange multiplier $\DIh^{(0)}(t)$ comes out
as an approximation for the Liouville-von Neumann equation (\ref{MVP20q}).

In coordinate form, the equations of motion for the variables
${R}^{(0)}_{\alpha}(t) = \langle {\sf M}_{\alpha} \rangle_{t}$
that parametrize $\tilde\DIh^{(0)}(t)$ are found from (\ref{qb23q}) as
 \begin{equation}   \label{MFA13q}
\frac { d{R}^{(0)}_{\alpha}(t) } {dt} =
 {\mathbb C}_{\alpha\beta} \{ {R}^{(0)}(t) \} \,
   \frac  {  \partial {h}  \{ {R}^{(0)}(t) \}   }
          {  \partial {R}^{(0)}_{\beta}(t)   }  \, ,
 \end{equation}
with the initial conditions $\{ {R}^{(0)}(t_{\rm i}) \}= \{{R}^{(0)}\}$.
[For the single-fermion Lie algebra,
we recover the time-dependent Hartree-Fock (TDHF) approximation
for the single-particle density matrix ${R}^{(0)}_{\nu\mu}(t)$, with the static HF solution
as initial condition. Indeed, the multiplication in (\ref{MFA13q})
by ${\mathbb C}_{\alpha\beta} \{{R}\}$
produces for this algebra a commutation with 
${R}_{\nu\mu}^{(0)}(t)$. 
The usual single-particle effective HF Hamiltonian comes out as
the image ${\sf H}\{{R}^{(0)}(t)\}$. The current use of TDHF to evaluate
expectation values is thus given a variational status.]

Equations (\ref{MFA13q}) can be rewritten in the alternative form
\begin{equation}    \label{MFA119q}
\frac {d{R}^{(0)}_{\alpha}} {dt}= {\mathbb
L}_{\alpha}^{\;{\gamma}}\, {{R}^{(0)}_{\gamma}(t)} \, ,
\end{equation}
where the matrix
\begin{equation}   \label{MFA118q}
{\mathbb L}_{\alpha}^{\;\gamma}\{{R}^{(0)}(t)\} =
\Gamma_{\alpha\beta}^{\;\gamma}\,
   {\cal H} ^{\beta}\{{R}^{(0)}(t)\}     \,
\end{equation}
plays the role of an effective Liouvillian acting in the Lie
algebra; the quantities ${\cal H}^{\beta}\{{R}\}$ are the
coordinates ${\cal H}^{\beta}\{{R}\} \equiv
\partial{h\{{R}\}}/\partial{R}_{\beta}$ [Eq.(\ref{MFA902q})] of the
image ${\sf H}\{{R}\}$ of $H$.

The dynamics of the density operator $\tilde\DIh^{(0)}(t)$ takes therefore
the classical form (\ref{MFA119q}) in terms of the scalar variables
${{R}^{(0)}_{\alpha}}$ parametrizing $\tilde\DIh^{(0)}(t)$. This classical structure 
will be analyzed in Sec.\,\ref{sec10}. Non-linearity occurs through the restriction of the trial space:
the image ${\sf H}\{{R}\}$ depends on the variables $\{{R}^{(0)}(t)\}$
when $H$ does not belong to the Lie algebra.

\section{Correlation functions} \label{sec5}

Variational approximations for the two-time correlation functions $C_{jk}(t',t'')$
are provided by the second-order term of the expansion (\ref{GO3q})
of the generating functional $\psi\{\xi\}$ in powers of the sources $\xi_j(t) $.
However, we have seen that, owing to the stationarity of $\Psi$, zeroth order was
sufficient to determine the variational approximation for expectation values.
Likewise, it is sufficient here to expand up to first order the first derivative
of the generating functional $\psi\{{\xi}\}$ according to
  \begin{eqnarray}    \label{DVP21Aq}
  && \frac{1}{i}\frac {\partial \psi\{\xi\}  }{\partial \xi_k(t'')}
  \approx   \langle Q_k \rangle_{t''}
  + i\int_{t_{\rm i}}^\infty dt' \sum _{j} \xi_j(t') \, C_{jk}(t',t'')
  \\  \nonumber
  && \simeq    \frac{1}{i}\frac {\partial \Psi\{ \AIh,\DIh\}  }{\partial \xi_k(t'')}
  =  \frac {{\rm Tr}\,Q_k^{\rm S} \,\DIh (t'') \,\AIh(t'')}
             {{\rm Tr}\,{\DIh(t'')}\,{\AIh (t'')}} \, ,
  \end{eqnarray}
where the stationarity of $\Psi\{\AIh,\DIh\}$ with respect of $\AIh$ and $\DIh$ has been used. We can thus obtain $C_{jk}(t',t'')$ from the first-order contribution to the r.h.s. of (\ref{DVP21Aq}), and only ${\AIh}^{(0)}$, 
${\DIh}^{(0)}$, ${\AIh}^{(1)}$ and ${\DIh}^{(1)}$ have to be determined from the variational equations (\ref{GS200q}-\ref{GS203q}).

\subsection{The approximate backward Heisenberg equation} \label{sec5.1}

One building block for correlation functions will be the quantity
${\sf Q}_k^{\rm H}(t'',t)$ defined for $t'' > t$ by expanding
the trial operator $\AIh(t)$, which enters (\ref{DVP21Aq}), up to first order as
\begin{eqnarray} \label{MFA01q}
\AIh(t) \approx \AIh^{(0)}(t) +  \AIh^{(1)}(t)  \equiv  I
 + i\int_{t}^{\infty}dt''\sum_{k} \xi_{k}(t'')\,{\sf Q}_k^{\rm H}(t'',t)\,.
\end{eqnarray}
Comparison with the expansion of the exact generating operator $A(t)$ defined by (\ref{A1q})
shows that ${\sf Q}_k^{\rm H}(t'',t)$ simulates the Heisenberg observable
${Q}_k^{\rm H}(t'',t)$. Since $\AIh(t)$ belongs to the Lie group,
${\sf Q}_k^{\rm H}(t'',t)$ belongs to the Lie algebra and can be expressed as
 ${\cal Q}_k^{{\rm H}\alpha}(t'',t)  {\sf M}_{\alpha}$.

The stationarity condition (\ref{GS202q}) with respect to $\AIh(t)$, expanded up to first order,
determines the coordinates ${\cal Q}_k^{{\rm H}\alpha}(t'',t)$.
For $\alpha \ne 0$ these obey for $t \le t''$ the equations
 \begin{equation} \label{MFA16Bq}
\frac {d {\cal Q}^{{\rm H}\alpha}_k(t'',t)}{dt}= -\,{\cal Q}^{{\rm
H}\beta}_k(t'',t) \,
 \left(  {\mathbb L} + {\mathbb C}\,{\mathbb H} \right)_{\beta}^{\;\alpha}
\;\;\;\;(\alpha,\beta \ne 0)\, ,
\end{equation}
which appear as the {\it reduction in the Lie algebra of the backward
Heisenberg equation} (\ref{BH5q}) for ${Q}^{{\rm H}}_k(t'',t)$. The
matrix $ {\mathbb L} $ is the effective Liouvillian (\ref{MFA118q}),
${\mathbb C}$ is the commutation matrix defined by (\ref{MFA14q})
and ${\mathbb H}$ is the second-derivative matrix of the symbol
$h\{{R}\}$ of the Hamiltonian $H$,
 \begin{equation} \label{MFA905q}
{\mathbb H}^{\alpha\beta}\{{R}\} =
\frac{ \partial^{\,2} h\{{R}\} }
     { \partial{R}_{\alpha}\,\partial{R}_{\beta} } \,.
\end{equation}
These three matrices are functions of $\{{R}\}$; in (\ref{MFA16Bq}) they are
taken at the point $ {R}_{\alpha} = {R}^{(0)}_{\alpha}(t)$ determined
by the zeroth-order equations (\ref{MFA119q}).
Equations (\ref{MFA16Bq}) should be solved backward in time
from the final boundary condition
  \begin{equation}   \label{MFA16Cq}
 {\cal Q}^{{\rm H}\alpha}_k(t'',t'') =
  {\cal Q}^{\alpha}_k\{ {R}^{(0)}(t'')\}
   =  \frac { \partial q_k\{{R}^{(0)}(t'')\} } { \partial{R}_{\alpha}^{(0)}(t'') } \,,
\end{equation}
where ${\cal Q}^{\alpha}_k\{ {R}\}{\sf M}_{\alpha} = {\sf Q}_k\{
{R}\}$ is the image of the observable $Q_k$ while $q_k\{{R}\}$
is its symbol.

The {\it exact} Ehrenfest equation for ${\rm Tr}\,{\sf M}_{\alpha}\tilde D(t)$
[issued from the Liouville-von Neumann Eq.(\ref{MVP20q})]
and the {\it exact} Heisenberg equation for
${Q}^{{\rm H}}_k(t'',t'')$ involve the same Liouvillian.
However, in the variational treatment, the corresponding {\it approximate}
Ehrenfest equation (\ref{MFA119q}) and the {\it approximate} Heisenberg equation
(\ref{MFA16Bq}) differ; the latter contains the corrective term
${\mathbb C}{\mathbb H}$ in addition to the effective Liouvillian ${\mathbb L}$.
This difference arises because the variational equations for the set
${R}_{\alpha}^{(0)}(t)$ occur at zeroth order in the sources
whereas those for the set ${\cal Q}^{{\rm H}\alpha}_k(t'',t)$ occur at first order.

From the stationarity condition (\ref{GS202q}) for $\AIh(t)$ one also finds the component
${\cal Q}^{{\rm H}0}_k(t'',t)$ of $\AIh^{(1)}$. This gives an alternative expression
for the time-dependent expectation value of $Q_k$ at a time $t$,
\begin{eqnarray}  \label{MFA02q}
 \langle Q_k \rangle_{t} \simeq  q_{k}\{ {R}^{(0)}(t)\}
   =  {\cal Q}^{{\rm H}\alpha}_{k}(t,t'')  \, {R}^{(0)}_{\alpha}(t'')
   = {\rm Tr}\,{\sf Q}^{\rm H}_{k}(t,t'') \, \tilde{\DIh}^{(0)}(t'')
 \end{eqnarray}
(including $\alpha = 0$ with ${R}^{(0)}_{0}=1$), which holds
for any intermediate time $t''$, and thus interpolates the Schr\"odinger picture
for $t''=t$ [Eq.(\ref{MFA125q})] and the Heisenberg picture for $t''=t_{\rm i}$,
as does the exact expression
\begin{eqnarray}    \label{MFA03q}
 \langle Q_k \rangle_{t} = {\rm Tr}\,Q^{\rm H}_{k}(t,t'')
 [ U(t'',t_{\rm i}) \,\tilde{D} \, U^{\dagger}(t'',t_{\rm i})]\,.
 \end{eqnarray}

\subsection{Bypassing ${\DIh}^{(1)}(t)$ for real times $t$} \label{sec5.2}

The correlation functions that we want to determine through (\ref{DVP21Aq})
depend on the combination 
${\DIh}^{(1)}(t'')+{\DIh}^{(0)}(t'')\,{\AIh}^{(1)}(t'')$,
since ${\AIh}^{(0)}={\cal I}$.
The time-dependence of ${\AIh}^{(1)}(t'')$ has been expressed
through Eqs.(\ref{MFA01q}) and (\ref{MFA16Bq}).
We still need the other ingredient, the first-order operator ${\DIh}^{(1)}(t'')$
whose evolution is not simple.
However, using the coupled equations of motion
for ${\DIh}(t)$ and ${\AIh}(t)$, we will show below that the time $t''$ occurring
in (\ref{DVP21Aq}) can be shifted down to $t_{\rm i}$.

From Eqs.(\ref{GS202q}) and (\ref{GS203q}), one can derive
for the products ${\DIh}(t){\AIh}(t)$ and ${\AIh}(t){\DIh}(t)$
the uncoupled equations
\begin{eqnarray}
 &&  \label{MFB10q}
  \frac {d({\DIh}{\AIh})}{dt} =
  \frac {1}{i\,\hbar} \left[ {\sf H}\{{R}^{\DIh\AIh}\},\, \DIh\AIh \right]
 +\,i\sum _j \xi _j(t)\,\left[ {\sf Q}_{j}\{{R}^{\DIh\AIh}\},\,\DIh\AIh \right] \,,
 \\
 && \label{MFB11q}
 \frac {d({\AIh}{\DIh})}{dt} =
  \frac {1}{i\,\hbar} \left[ {\sf H}\{{R}^{\AIh\DIh}\},\, \AIh\DIh \right] \, .
 \end{eqnarray}
These equations cannot be fully solved because the boundary conditions
on $\AIh$ and $\DIh$ occur at different times [Eq.(\ref{MVP6q})].
Nevertheless, Eq.(\ref{MFB10q}) will happen to be sufficient for our purpose.

In parallel with Eq.(\ref{MFA01q}), we parametrize at first order
the product $\DIh(t)\AIh(t)$ by ${R}^{(1)}_{j\beta}(t,t')$ according to
 \begin{eqnarray} \label{MFB12q}
  {R}^{\DIh\AIh}_{\beta}(t) =
      \frac {{\rm Tr}\,{\sf M}_{\beta}\,\DIh (t) \,\AIh(t)}
            {{\rm Tr}\,{\DIh(t)}\,{\AIh (t)}}
 \approx {\rm Tr}\,{\sf M}_{\beta} [{\tilde\DIh}^{(0)}(t) + \delta{\tilde\DIh}(t)]
 \\  \nonumber
={R}^{(0)}_{\beta}(t) +
   i\int_{t_{\rm i}}^{\infty} dt'\sum_{j} \xi_{j}(t')\,{R}^{(1)}_{j\beta}(t,t').
 \end{eqnarray}
We have in (\ref{MFB12q}) denoted as $\delta{\tilde\DIh}(t)$
the first-order variation
of $\DIh\AIh /{\rm Tr}\,{\DIh}{\AIh}$ around ${\tilde{\DIh}}^{(0)}(t)$. 
The quantity $C_{jk}(t',t'')$ that we wish to evaluate now satisfies
from (\ref{DVP21Aq}) and (\ref{MFB12q}) the relation
  \begin{eqnarray}    \label{DVP22Bq}
   + \, i\int_{t_{\rm i}}^\infty dt' \sum _{j} \xi_j(t') \, C_{jk}(t',t'')
   =  {\rm Tr}\,Q_k^{\rm S} \,\delta{\tilde\DIh}(t'')  \, .
  \end{eqnarray}
Since $\delta{\tilde\DIh}(t'')$ is a variation around ${\tilde{\DIh}}^{(0)}(t'')$
within the Lie group, the image property (\ref{MFA901bq}) allows us to replace
the operator $Q_k^{\rm S}$ by its image
${\sf Q}_k\{{R}^{(0)}(t'')\}
  = {\cal Q}_k^{\beta}\{{R}^{(0)}(t'')\} {\sf M}_{\beta}$.
Comparing Eqs.(\ref{MFB12q}) and (\ref{DVP22Bq}), we can get rid
of the sources $ \xi_j(t')$, so that
 \begin{eqnarray}   \label{MFB13q}
  C_{jk}(t',t'') =  {R}^{(1)}_{j\beta}(t'',t') \,
    {\cal Q}^{\beta}_k\{ {R}^{(0)}(t'')\} \, .
 \end{eqnarray}
Remember that the coordinates ${\cal Q}_k^{\beta}\{{R}^{(0)}(t'')\}$
of the image of $Q_k^{\rm S}$ are related to the symbol
$q_k\{{R}^{(0)}(t'')\} \equiv {\rm Tr}\, Q_k^{\rm S}\,{\tilde{\DIh}}^{(0)}(t'')$
by
  \begin{eqnarray}   \label{MFB13Bq}
  {\cal Q}^{\beta}_k\{ {R}^{(0)}(t'')\}
   =  \frac { \partial q_k\{{R}^{(0)}(t'')\} } { \partial{R}_{\beta}^{(0)}(t'') } \,.
 \end{eqnarray}

The time-dependence of ${R}^{(1)}_{j\beta}(t,t')$ is found by expanding
Eq.(\ref{MFB10q}) at first order. Noting that
${{\rm Tr}\DIh \AIh}$ does not depend on time, one obtains
   \begin{eqnarray}   \label{MFB14q}
   \frac {d {R}^{(1)}_{j\beta}(t,t')}{dt} =
  \left(  {\mathbb L} + {\mathbb C}\,{\mathbb H} \right)_{\beta}^{\;\gamma}
   {R}^{(1)}_{j\gamma}(t,t')
  +i\,{\hbar} \, {\mathbb C}_{\beta\gamma} \,
    {\cal Q}^{\gamma}_j\{ {R}^{(0)}(t)\} \, \delta(t-t') ,
  \end{eqnarray}
 where again ${\mathbb L}$, ${\mathbb C}$ and ${\mathbb H}$ depend on time through
$\{{R}^{(0)}(t)\}$. The same kernel
${\mathbb L} + {\mathbb C}\,{\mathbb H}$ as in the approximate
backward Heisenberg equation (\ref{MFA16Bq}) is recovered; it will also be encountered
in the context of small deviations (Sec.\,\ref{sec9.2}).
[For fermionic systems, ${\mathbb L} + {\mathbb C}\, {\mathbb H}$
is the time-dependent RPA matrix
issued from the kernel ${\mathbb L}$ of the TDHF equation (\ref{MFA119q}).]

As a consequence of the duality between the kernels of Eqs.(\ref{MFA16Bq})
and (\ref{MFB14q}), we obtain the identity
 \begin{eqnarray}   \label{MFB16q}
 \frac {d}{dt} [  {R}^{(1)}_{j\beta}(t,t') \,{\cal Q}^{{\rm H}\beta}_k(t'',t) ]
 = - \, i\,{\hbar} \,  {\cal Q}^{\gamma}_j\{ {R}^{(0)}(t)\} \,
       {\mathbb C}_{\gamma\beta}\,
    {\cal Q}^{{\rm H}\beta}_k(t'',t)\,    \delta(t-t') \, .
 \end{eqnarray}
Since $ {\cal Q}^{{\rm H}\beta}_k(t'',t) $ vanishes for $t''< t$,
the r.h.s. of (\ref{MFB16q}) disappears if $t''< t'$.
One can therefore evaluate $C_{jk}(t',t'')$ for $t' \ge t''$ from (\ref{MFB13q})
by noting that the product
$ {R}^{(1)}_{j\beta}(t,t')\,{\cal Q}^{{\rm H}\beta}_k(t'',t) $
does not depend on $t$ for $t' \ge t'' > t > t_{\rm i}$.
Using the boundary condition
${\cal Q}^{{\rm H}\beta}_k(t'',t'') = {\cal Q}^{\beta}_k\{ {R}^{(0)}(t'')\} $,
we then shift $t''$ down to $t_{\rm i}$ in (\ref{MFB13q}), which yields
 \begin{eqnarray}   \label{MFB17q}
    C_{jk}(t',t'') =   {R}^{(1)}_{j\beta}(t_{\rm i},t')
     \,{\cal Q}^{{\rm H}\beta}_k(t'',t_{\rm i})  \,,
   \;\;\;\; (t' \ge t'')\,.
 \end{eqnarray}

The continuity condition
$\DIh(t_{\rm i})\AIh(t_{\rm i})=\DIh(\beta)\AIh(\beta)$ entails that
this correlation function (\ref{MFB17q}) is equal to
 \begin{eqnarray}   \label{MFB17Bq}
    C_{jk}(t',t'') =   {R}^{(1)}_{j\beta}(\beta,t') \,
  {\cal Q}^{{\rm H}\beta}_k(t'',t_{\rm i})\,
   \;\;\;\; (t' \ge t'')\,
 \end{eqnarray}
in terms of the boundary value at $\tau = \beta$ of $\{{R}^{(1)}_{j}({\tau},t')\}$. 
One needs therefore to determine, only in the interval $0 \le \tau \le \beta$,
the set $\{{R}^{(1)}_{j}({\tau},t')\}$ defined through the first-order
parametrization of $\DIh(\tau)\AIh(\tau)$ by
  \begin{eqnarray} \label{MFB30q}
  {R}^{\DIh\AIh}_{\beta}(\tau) =
      \frac { {\rm Tr}\,{\sf M}_{\beta}\,\DIh (\tau) \,\AIh(\tau)}
            {{\rm Tr}\,{\DIh(\tau)}\,{\AIh (\tau)}}
 \approx {R}^{(0)}_{\beta} +
   i\int_{t_{\rm i}}^{\infty} dt'\sum_{j} \xi_{j}(t')
   {R}^{(1)}_{j\beta}(\tau,t') \, ,
  \end{eqnarray}
where ${R}^{(0)}_{\beta} = {\rm Tr}\,{\sf M}_{\beta}\,{\tilde\DIh}^{(0)}$.

The evaluation of ${\DIh(t)}$ at times $t > t_{\rm i}$ has been bypassed,
and it remains to solve the coupled equations (\ref{GS200q})
and (\ref{GS201q}) for ${\AIh}^{(1)}(\tau)$ and ${\DIh}^{(1)}(\tau)$
in the range $0 \le \tau \le \beta$,
with the boundary conditions $\DIh^{(1)}({\tau}={0}) = 0$ and
 \begin{eqnarray}   \label{MFB22q}
 \AIh^{(1)}({\tau}=\beta)  =
    i\int_{t_{\rm i}}^{\infty} dt'\sum_{j} \xi_{j}(t') \,
 {\cal Q}_j^{\rm H\alpha}(t',t_{\rm i}) \, {\sf M}_{\alpha}
 \end{eqnarray}
 issued from (\ref{MVP6q}) and (\ref{MFA01q}), respectively.
For $t''\ge t'$, the symmetry $C_{jk}(t',t'') = C_{kj}(t'',t') $ can be checked
by means of the r.h.s. of Eq.(\ref{MFB16q}).

\subsection{Two-time correlation functions} \label{sec5.3}

The linear structure of the boundary condition (\ref{MFB22q})
for ${\AIh}^{(1)}(\tau)$, together with the boundary condition
${\DIh}^{(1)}(\tau=0)=0$ for ${\DIh}^{(1)}(\tau)$, entail the occurrence
of an overall factor ${\cal Q}^{{\rm H}\alpha}_j(t',t_{\rm i})$
in ${\AIh}^{(1)}(\tau)$ and ${\DIh}^{(1)}(\tau)$,
hence in ${R}^{(1)}_{j\beta}(\tau,t')$,
and finally in the correlation function $C_{jk}(t',t'')$.
We had already acknowledged the explicit dependence of $C_{jk}(t',t'')$
on ${\cal Q}^{{\rm H}\beta}_k(t'',t_{\rm i})$.
Therefore, for any trial Lie group,
the two-time correlation functions have the general form $(\alpha,\beta \ne 0)$
\begin{eqnarray}  \label{MFA18q}
  C_{jk}(t',t'') \simeq   {\cal Q}_{j}^{{\rm H}\alpha}(t',t_{\rm i})\,
 {\mathbb B}_{\alpha\beta} \, {\cal Q}_{k}^{{\rm H}\beta}(t'',t_{\rm i}) \, ,
   \;\;\;\;(t'>t'') \, .
  \end{eqnarray}
Beside the approximate Heisenberg observables given by the dynamical
equations (\ref{MFA16Bq}) and the boundary condition  (\ref{MFA16Cq}),
a matrix ${\mathbb B}_{\alpha\beta}$ appears,
which depends only on the zeroth and first order contributions
to ${\DIh}(\beta)$ and ${\AIh}(\beta)$.

 Noticeably, the expression (\ref{MFA18q}) has the same {\it factorized
structure} as the exact formula (\ref{Q4q}).
In spite of the coupling between the stationarity conditions in the sectors
$t \ge t_{\rm i}$ and $ 0 \le \tau \le \beta $, Eq.(\ref{MFA18q})
[as Eq.(\ref{Q4q})] displays separately two types of ingredients:
The result ${\mathbb B}$ of the optimization of the initial state is disentangled
from that of the dynamics, embedded in the approximate Lie-algebra
Heisenberg observables
${\sf Q}_{j}^{\rm H}(t',t_{\rm i})$ and ${\sf Q}_{k}^{\rm H}(t'',t_{\rm i})$.

\subsection {The correlation matrix ${\mathbb B}$} \label{sec5.4}

The last task consists in determining explicitly the correlation matrix
${\mathbb B}_{\alpha\beta}$.
We have noted above that, through the boundary condition (\ref{MFB22q}),
the factor ${\cal Q}^{{\rm H}\alpha}_j(t',t_{\rm i})$
occurs in ${R}^{(1)}_{j\beta}(\tau,t')$. By introducing
new ($\alpha$-indexed) operators
${\AIh}^{(1)}_{\alpha}(\tau)$ and ${\DIh}^{(1)}_{\alpha}(\tau)$,
again determined by the coupled equations  (\ref{GS200q}) and (\ref{GS201q})
but with the boundary conditions
${\AIh}^{(1)}_{\alpha}(\beta) = {\sf M}_{\alpha}$
and ${\DIh}^{(1)}_{\alpha}(\beta) =0$,
one can explicitly factorize ${R}^{(1)}_{j\beta}(\tau,t')$ as
    \begin{eqnarray}   \label{MFB25q}
     {R}^{(1)}_{j\beta}(\tau,t') =
      {\cal Q}^{{\rm H}\alpha}_j(t',t_{\rm i})\,
       {R}^{(1)}_{\alpha\beta}(\tau) \,.
    \end{eqnarray}
The quantities $ {R}^{(1)}_{\alpha\beta}(\tau)$
defined by (\ref{MFB30q}) and (\ref{MFB25q}) will then be given by
 \begin{eqnarray}   \label{MFB20q}
 {R}^{(1)}_{\alpha\beta}({\tau}) =
   {\rm Tr}\,
   [  {\DIh}^{(0)}({\tau})\, {\AIh}^{(1)}_{\alpha}({\tau})
  +  {\DIh}^{(1)}_{\alpha}({\tau})\, {\AIh}^{(0)}({\tau})  ]
 ( {\sf M}_{\beta} - {R}^{(0)}_{\beta} )
 \, / \,  {\rm Tr}\,{\DIh}^{(0)} \,,
 \end{eqnarray}
and the correlation matrix ${\mathbb B}_{\alpha\beta}$ will be found as
${\mathbb B}_{\alpha\beta} = {R}^{(1)}_{\alpha\beta}(\tau=\beta)$, that is,
  \begin{eqnarray}   \label{MFB23q}
 {\mathbb B}_{\alpha\beta}     =
 {\rm Tr}\,{\sf M}_{\alpha} ( {\sf M}_{\beta}
  - {R}^{(0)}_{\beta} ) {\tilde\DIh}^{(0)}
  +\frac  { {\rm Tr}\,{\DIh}^{(1)}_{\alpha}(\beta)
    ({\sf M}_{\beta}- {R}^{(0)}_{\beta})  }
          {  {\rm Tr}\,{\DIh}^{(0)}  } \,.
 \end{eqnarray}

It remains to determine the coupled operators
${\DIh}^{(1)}_{\alpha}(\tau)$ and ${\AIh}^{(1)}_{\alpha}(\tau)$.
The solution is explicitly worked out in Appendix A.1.
It had already been given in special cases,
for fermionic systems \cite{BV93q}, extended BCS theory \cite{BFV99q},
${\phi}^{4}$ quantum field \cite{CMq}.

The result thus found for the {\it correlation matrix} ${\mathbb B}$ is
\begin{eqnarray} \label{MFA19Aq}
  && {\mathbb B} =
      \frac      { i\,\hbar\,{\mathbb C}\, {\mathbb F} }
                 { \Id -\exp( {-i\,\hbar}\,\beta\,{\mathbb C}\,{\mathbb F} )  }
         \,{\mathbb F}^{-1}
  \\ \nonumber
    &&  \;  =   \left[ \frac{1}{2}\,i\,\hbar\,{\mathbb C}\,{\mathbb F}\,
\coth \left(
\frac{1}{2}\,i\,\hbar\,\beta\,{\mathbb C}\,{\mathbb F}
\right) \right] {\mathbb F}^{-1}
   + \frac{1}{2}\,i\,\hbar\,{\mathbb C}\,.
\end{eqnarray}
It involves the commutation matrix ${\mathbb C}$ defined by (\ref{MFA14q})
and the positive matrix ${\mathbb F}={\mathbb K}-T\,{\mathbb S}$
of second derivatives (\ref{MFA20q})
of the trial free energy. Both are taken for the parameters $\{{R}^{(0)}\}$ that were
determined at zeroth order.
For vanishing eigenvalues of $i{\mathbb C}{\mathbb F}$,
the coefficient of ${\mathbb F}^{-1}$ is meant as ${\beta}^{-1}$.
[For the single-particle fermionic Lie group, the matrix
$i{\mathbb C}{\mathbb F}$ is the static RPA kernel.]

By letting $t'-0 = t'' = t_{\rm i}$ and
$Q_j^{\rm S} = {\sf M}_{\alpha}$, $Q_k^{\rm S} = {\sf M}_{\beta}$
in (\ref{MFA18q}), one identifies ${\mathbb B}_{\alpha\beta}$
as the variational approximation
for the correlations, in the initial state $\tilde{D}$,
of the operators ${\sf M}_{\alpha}$ that span the Lie algebra:
 \begin{equation}  \label{MFA18Aq}
 {\rm Tr}\, {\sf M}_{\alpha}{\sf M}_{\beta} \tilde{D}
  - {\rm Tr}\,{\sf M}_{\alpha}\tilde{D} \,
  {\rm Tr}\,{\sf M}_{\beta} \tilde{D}
   \simeq    {\mathbb B}_{\alpha\beta} \,,
 \;\;\;\;  (\alpha,\beta \ne 0)\,.
 \end{equation}

The variational expressions (\ref{MFA18q}), (\ref{MFA19Aq})
[together with Eqs.(\ref{MFA16Bq})]
for the correlation functions issued from the use of a Lie-group trial space
are the most important outcomes of the present variational approach.
We comment below the features of this expression, and work out its consequences
in the forthcoming sections.

\subsection  {Status of the result for static correlations; Kubo correlations} 
 \label{sec5.5}

The optimization of thermodynamic quantities and expectation values 
(Sec.\,\ref{sec4})
has resulted in a mean-field type of approximation, with the mere replacement
of the exact state
$D$ by the zeroth-order contribution $ {\DIh}^{(0)}$
to the trial object ${\DIh}$, as in
${\rm Tr}\,{\sf M}_{\alpha}\tilde{D}
\simeq {\rm Tr}\,{\sf M}_{\alpha}{{\tilde\DIh}}^{(0)}={R}^{(0)}_{\alpha} $.
However, the optimized approximation (\ref{MFA19Aq}),(\ref{MFA18Aq}) that we found
for the correlations ${\mathbb B}_{\alpha\beta}$
of ${\sf M}_{\alpha}$ and ${\sf M}_{\beta}$
does not follow from such a simple replacement.
Evaluated in the state ${\tilde{\DIh}}^{(0)}$
by the formula (\ref{MFA10hq}), such correlations, instead of (\ref{MFA19Aq}),
would be given by
\begin{eqnarray} \label{MFA19Bq}
     {\rm Tr}\, {\sf M}_{\alpha} \,{\sf M}_{\beta} \,{\tilde\DIh}^{(0)}
                -  {R}^{(0)}_{\alpha}\, {R}^{(0)}_{\beta}
    = - \left(   \frac   { i\,\hbar\,{\mathbb C}\, {\mathbb S} }
                        {  \exp[ {i\,\hbar}\,{\mathbb C}\,{\mathbb S} ] -\Id }
    \,{\mathbb S}^{-1}   \right)_{\alpha\beta} \, ,
\end{eqnarray}
where ${\mathbb S}$ and ${\mathbb C}$ are evaluated from (\ref{MFA10cq})
and (\ref{MFA14q}) for $\{{R}\}=\{{R}^{(0)}\}$.
Contrary to ${\mathbb B}$, the naive expression (\ref{MFA19Bq}) is not 
variationally optimized. While the first term of the variational
expression (\ref{MFB23q}) of ${\mathbb B}$
provides a contribution equal to (\ref{MFA19Bq}),
its second term, arising from ${{\DIh}}^{(1)}_{\alpha}(\beta)$, introduces
a correction which substitutes
${\mathbb S} - {\beta}{\mathbb K}= - {\beta}{\mathbb F} $ to ${\mathbb S}$ .
The matrix ${\mathbb K}$ takes into account effects coming from the part
of the operator $K$ that lies outside the Lie algebra.
[When $\{{\sf M}\}$ is the fermionic single-particle algebra,
the left-hand-side of (\ref{MFA19Bq}) is the Fock term from
${\rm Tr}\,(a^{\dag}_{\mu}a_{\nu})(a^{\dag}_{\sigma}a_{\tau}){\tilde{\DIh}}^{(0)}$ 
since $ {R}^{(0)}_{\alpha}\,{R}^{(0)}_{\beta}$ is the Hartree term.
The full matrix ${\mathbb B}$ involves an RPA kernel,
in which the matrix ${\mathbb K}$ is the effective two-body interaction.]

 Although the approximations for expectation values and correlations functions
stem from the same variational principle, they appear intrinsically different.
Not only ${\mathbb B}$ cannot be expressed from ${\DIh}^{(0)}$,
but moreover there is {\it no density operator approximating} $D$ in the original
Hilbert space $\mathscr{H}$ that would
produce the optimized correlations ${\mathbb B}$ in the same form as (\ref{MFA18Aq}).
While ${\tilde{\DIh}}^{(0)}$ can be interpreted as a state,
the trial operator ${\DIh}$ has no perturbative status and is just a calculational
tool involving the sources.
In the expression (\ref{MFB23q}) of ${\mathbb B}$, the operator
${\DIh}^{(1)}_{\alpha}$, which depends on ${\sf M}_{\alpha}$,
is not a correction to ${\tilde{\DIh}}^{(0)}$.

Nevertheless, we will show in Sec.\,\ref{sec7} that, through a mapping of the original
Hilbert space ${\mathscr{H}}$ into a {\it new space} $\underline{\mathscr{H}}$ and
of the Lie algebra $\{{\sf M}\}$ into a reduced algebra $\{\underline{\sf M}\}$,
the matrix elements of ${\mathbb B}$ can be interpreted
as {\it exact} correlations between the operators $\{\underline{\sf M}\}$
in an effective state $\tilde{\underline{D}}$.
The quantities $\{{R}^{(0)}\}$ will also appear as exact
expectation values of $\{\underline{\sf M}\}$ in $\tilde{\underline{D}}$.
The variational results, for both expectation values and correlations,
will thus be unified through a modification of the Lie algebra.

One may also wonder about the origin of the Bose-like factor exhibited,
for arbitrary Lie groups, by the expression (\ref{MFA19Aq}) of ${\mathbb B}$.
A clue will be given in Secs.\,\ref{sec7} and \ref{sec8} where an algebra of Bose 
operators arises from the mapped Lie algebra $\{\underline{\sf M}\}$.

An alternative understanding of the structure of the matrix
${\mathbb B}$ can be reached by relating it to the matrix
of {\it Kubo correlations}, even though the latter have less direct physical relevance
than ordinary correlations. Kubo correlations between the operators
${\sf M}_{\alpha}$ and ${\sf M}_{\beta}$ are defined
for the exact state $\tilde{D}$ by
 \begin{eqnarray}   \label{MOC50q}
     {\rm Tr} \, \frac{1}{\beta}   \int_{{0}}^{\beta}\,d{\tau} \,
       e^{ \,   {\tau}\,{K} } \, {\sf M}_{\alpha}\,
       e^{ \,-\,{\tau}\,{K} }  \,{\sf M}_{\beta}  \, \tilde{D}
  -  {\rm Tr} \,{\sf M}_{\alpha}  \,\tilde{D} \,
     {\rm Tr} \,{\sf M}_{\beta}  \, \tilde{D} \, .
  \end{eqnarray}
We have seen in Sec.\,\ref{sec3.3} that, if the state $\tilde{D}$ were bluntly replaced
in (\ref{MOC50q}) by the element $\tilde{\DIh}^{(0)}$ of the Lie group,
Kubo correlations would be directly related by (\ref{MFA010q}) to the
matrix ${\mathbb S}$. We show in Appendix A.2 that a variational approximation
${\mathbb B}^{\rm K}$ for the Kubo correlations (\ref{MOC50q}) reads
   \begin{eqnarray}   \label{MOC54q}
 {\mathbb B}^{\rm K} = \frac{1}{\beta} ({\mathbb F}^{-1}) = ({\beta}\,{\mathbb K}-{\mathbb S})^{-1} \,.
  \end{eqnarray}
The naive approximation (\ref{MFA010q}), namely ${\mathbb B}^{\rm K}\simeq -{\mathbb S}^{-1}$, 
obtained by replacing $\tilde{D}$ by $\tilde{\DIh}^{(0)}$ in (\ref{MOC50q}), is thus 
variationally corrected by inclusion in (\ref{MOC54q}) of the term
${\beta}\,{\mathbb K}$. Likewise, the naive approximation (\ref{MFA19Bq}) for the ordinary 
correlations in the state $\tilde{D}$ results from the variational expression
(\ref{MFA19Aq}) for ${\mathbb B}$ by omission of ${\mathbb K}$ within ${\mathbb F}={\mathbb K}-T\, {\mathbb S}$.

Once ${\mathbb B}^{\rm K}$ has been obtained in the form (\ref{MOC54q}), 
it is possible to recover from it the expression (\ref{MFA19Aq}) for
 ${\mathbb B}$. 
In the special case ${\mathbb K}=0$, for which
${\beta}\,{\mathbb F}=-{\mathbb S}$, this was achieved in Sec.\,\ref{sec3.3}.
We saw there that the factor
$[{\DIh}^{(0)}]^{-\tau/{\beta}}({\sf M}_{\alpha}-{R}_{\alpha})
 [{\DIh}^{(0)}]^{\tau/{\beta}}$
entering the Kubo correlation (\ref{MOC50q}) is expressed as
$\left( e^ {\,i\,\hbar\,{\mathbb C}\,{\mathbb S}\,{\tau}/{\beta} }
\right)_{\alpha}^{\,\gamma}\,( {\sf M}_{\gamma}- {R}_{\gamma} ) $
through the automorphism (\ref{MFA10gq}) of the Lie algebra. Integration of 
$ e^{ \,i\,\hbar\,{\mathbb C}\,{\mathbb S}\,\tau / \beta }$
between $0$ and $\beta$ yields
  \begin{eqnarray}   \label{MOC58q}
       {\mathbb B}^{\rm K} =                      \frac
   { e^{  \,i\,\hbar\,{\mathbb C}\,{\mathbb S}  } - {\mathbb I} }
   { i\,\hbar\,{\mathbb C}\,{\mathbb S} } \,{\mathbb B} \,,
   \end{eqnarray}
in agreement with (\ref{MFA10hq}). In the general case ${\mathbb K} \ne 0$,
it is shown likewise in Appendix B that
  \begin{eqnarray}   \label{MOC56q}
 {\mathbb B}({\mathbb B}^{\rm K})^{-1} =
   \frac      { i\,\hbar\,{\beta}\,{\mathbb C}\, {\mathbb F} }
              { \Id -\exp( {-i\,\hbar}\,\beta\,{\mathbb C}\,{\mathbb F} )  } \,.
 \end{eqnarray}
This ratio stems therefore from a property inherent to the Lie group
underlying our variational
approach, namely the exponentional form of the automorphism (\ref{MFA10gq}).
The Bose-like factor that enters ${\mathbb B}$ arises from this property,
and from the simplicity of ${\mathbb B}^{\rm K}$.
[The quasi-boson structure of the static thermal RPA
for fermions \cite{BFV99q} appears as a special case, see Sec.\,\ref{sec8}.]
\vspace{.35cm}

Thermodynamic quantities and expectation values were determined in 
Sec.\,\ref{sec4};
they depend on the mean-field images of $K$ and $H$ within the Lie algebra, namely,
${\sf K}^{(0)}$ which self-consistently (Sec.\,\ref{sec4.1}) determines $\{{R}^{(0)}\}$
and ${\sf H}$ which governs (Sec.\,\ref{sec4.2}) the time-dependence of $\{{R}^{(0)}(t)\}$.
More elaborate ingredients are required for the evaluation of static and dynamic correlation functions:
The matrix ${\mathbb K}$ enters ${\mathbb B}$ through ${\mathbb F}$
while the matrix ${\mathbb H}$ (together with the mean-field Liouvillian ${\mathbb L}$)
governs the time-dependence of the approximate Heisenberg observables
${\sf Q}^{\rm H}_{j}(t',t)$.

In the rest of this article we will analyse
the properties of the approximate correlation functions
[expressed by Eqs.(\ref{MFA18q}),(\ref{MFA19Aq}) and (\ref{MFA16Bq})]
and of the dynamics [expressed by Eqs.(\ref{MFA13q})].

\section{Properties of approximate correlation functions and fluctuations} \label{sec6}

In this section we review some consequences of the variational expressions found
above for the correlation functions, encompassing special cases and conservation laws.

\subsection{Time-dependent correlation functions in an equilibrium initial state}
 \label{sec6.1}
 We first consider the special case of an equilibrium initial state
$D=\exp(-{\beta}K)$ for which the operator $K$ is equal to the Hamiltonian $H$,
plus possibly some constants of motion such as the
particle number for a grand-canonical equilibrium. One can then generate simply
the dynamics by using $K$ instead of $H$ in the backward Heisenberg
equation (\ref{BH5q}).

The exact expectation values ${\langle Q_j \rangle}_t$ then do not
evolve. Their approximation (\ref{MFA125q}) depends on time through 
$\{{R}^{(0)}(t) \}$ governed by (\ref{MFA119q}). In the effective
Liouvillian $ {\mathbb L}$ defined by (\ref{MFA118q}), the
coordinates ${\cal H}^{\beta}$ of the image of the Hamiltonian are
replaced by those of the image ${\sf K}$ of $K$, that is, ${\cal
K}^{\gamma}\{{R}\} =
\partial{k\{{R}\}}/\partial{R}_{\gamma}$.
Using the self-consistent equilibrium condition (\ref{qb29q}) for
$\tilde{\DIh}^{(0)}$ and the identity (\ref{MFA10fq}) for the
product ${\mathbb C}{\mathbb S}$, we find
 \begin{equation}   \label{MFA120q}
 {\mathbb L}_{\beta}^{\;\alpha}\{{R}^{(0)}\} =
  \Gamma_{\beta\gamma}^{\;\alpha}\, {\cal K} ^{\gamma} \{{R}^{(0)}\}
  =  -\, T \, \Gamma_{\beta\gamma}^{\;\alpha}\, {J}^{(0)\gamma}
  =   -\, T \, {\mathbb C}_{\beta\gamma} \, {\mathbb S}^{\gamma\alpha} \, .
 \end{equation}
The identity (\ref{MFA30q}) then entails, as expected, that
${R}^{(0)}_{\alpha}(t) = {R}^{(0)}_{\alpha}$, and hence
that the approximation ${\langle{Q_j}\rangle}_t$ does not
depend on time.

For two observables $Q_j$ and $Q_k$ that do not commute with $K$,
the exact two-time correlation function $C_{jk}(t',t'')$ defined by
(\ref{Q4q}) {\it depends only on the time difference} $t'-t''$. This 
property is not obvious for the approximation (\ref{MFA18q}), which 
involves two factors depending separately on $t'$ and $t''$. To elucidate 
this point, let us solve for $ {\mathbb H} = {\mathbb K}$ the
approximate backward Heisenberg equation (\ref{MFA16Bq}) for the
image ${\sf Q}_{j}^{\rm H}(t',t) = {\sf Q}_{j}^{\rm H}(t'-t)$ with
the boundary condition ${\sf Q}_{j}^{\rm H}(0) =  {\sf
Q}_j\{{R}^{(0)}\}$. This
equation involves the kernel ${\mathbb L}+{\mathbb C}{\mathbb H}$
which, according to (\ref{MFA120q}), takes the simple form ${\mathbb
C}({\mathbb K}-T\,{\mathbb S}) = {\mathbb C}{\mathbb F}$. Hence, the
Heisenberg equation (\ref{MFA16Bq}) simplifies into ${d{\cal
Q}^{{\rm H}\alpha}_j(t',t)}/{dt}=  -\,{\cal Q}^{{\rm
H}\beta}_j(t',t) \,
 \left(   {\mathbb C}\,{\mathbb F} \right)_{\beta}^{\;\alpha} $,
where $ {\mathbb C}$ and ${\mathbb F}$ are evaluated for $\{{R}^{(0)}\}$; it is readily solved as
 \begin{equation}   \label{MFA140q}
   {\cal Q}^{{\rm H}\alpha}_j(t',t_{\rm i}) =
   {\cal Q}_j^{\beta} \{{R}^{(0)}\}   \,
    \left[  e^{\,{\mathbb C}\,{\mathbb F}\,(t'-t_{\rm i})}\right]^{\,\alpha}_{\beta}
 \end{equation}
(with $\alpha,\beta \ne 0$) in terms of the boundary condition
${\cal Q}_j^{\beta} \{{R}^{(0)}\} = { \partial q_j\{{R}^{(0)}\} } /
{ \partial{R}_{\beta}^{(0)} }$.

In order to evaluate the correlation function (\ref{MFA18q}), use is also made of
\begin{equation}   \label{MFA142q}
   {\cal Q}^{{\rm H}\alpha}_k(t'',t_{\rm i}) =
     \left[  e^{\,-\,{\mathbb F}\,{\mathbb C}\,(t''-t_{\rm i})}  \right]_{\;\beta}^{\alpha}
  \, {\cal Q}_k^{\beta} \{{R}^{(0)}\}   \,  ,
  \end{equation}
 and of the relation
 \begin{equation}   \label{MFA143q}
    {\mathbb F}^{-1} \,  e^{\,-\,{\mathbb F}\,{\mathbb C} \,(t''-t_{\rm i}) }=
              e^{\,-\,{\mathbb C}\,{\mathbb F}  \,(t''-t_{\rm i})} \,{\mathbb F}^{-1} \,.
  \end{equation}
The explicit dependence on time of the correlation functions is then
given for $t' > t''$ by
  \begin{equation}   \label{MFA141q}
    C_{jk}(t',t'')  \simeq
   {\cal Q}_j^{\alpha} \{{R}^{(0)}\}   \,
    \left[  e^{\, {\mathbb C}\,{\mathbb F}\,(t'-t'')} \,
             \frac { i\,\hbar\,{\mathbb C}\, {\mathbb F}}
                 {\Id -\exp({-i\,\hbar}\,\beta\,{\mathbb C}\,{\mathbb F})}
         \,{\mathbb F}^{-1}          \right]_{\alpha\beta} \,
   {\cal Q}_k^{\beta} \{{R}^{(0)}\}  \, .
   \end{equation}

The identity of $H$ and $K$ has led to the occurrence of the same matrix
${\mathbb C}\,{\mathbb F}$ in the dynamical equation (\ref{MFA140q})
and in the matrix ${\mathbb B}$ [Eq.(\ref{MFA19Aq})] that accounts
for the correlations in the initial state. As a consequence, only $t'-t''$ 
appears in (\ref{MFA141q}), as it should.
This property would not have been satisfied if the correlation
matrix ${\mathbb{B}}$ were naively evaluated, as in Eq.(\ref{MFA19Bq}),
by replacing in (\ref{MFA18Aq})
the exact state $\tilde{D}$ by ${\tilde{\DIh}}^{(0)}$.

For $j=k$, $ C_{jj}(t'-t'')$ provides the variational approximation
for the autocorrelation function of the observable $Q_j$ in the
equilibrium state $D \propto \exp(-{\beta}K)$.

\subsection{Other special cases} \label{sec6.2}

\subsubsection{Commutators and linear responses} \label{sec6.2.1}

The antisymmetric part of ${\mathbb B}$, namely  $i\hbar{\mathbb C}/2$,
is simple and depends only on the zeroth order in the sources, not on 
${\mathbb F}$.
As a consequence the approximation ${\mathbb B} _{\alpha\beta} - {\mathbb B} _{\beta\alpha}$
for the expectation value
${\rm Tr}\,[{\sf M}_{\alpha},\,{\sf M}_{\beta}] \tilde{D} = i\hbar
  {\rm Tr}\, \Gamma^{\gamma}_{\alpha\beta} {\sf M}_{\gamma} \tilde{D} $
of the commutator $[{\sf M}_{\alpha},\,{\sf M}_{\beta}]$
is obtained as $i\hbar\,{\mathbb C} _{\alpha\beta}\{{R}^{(0)}\}$,
a property in agreement with the expectation value ${R}^{(0)}_{\gamma}$
of $ {\sf M}_{\gamma}$ found at first order in the sources.

Linear responses are expectation values of commutators, and therefore involve
only this antisymmetric part of $C_{jk}(t',t'')$. Alternatively, they can
be evaluated directly from the variational expression
$\Psi\{\AIh,\DIh\}$ by including a time-dependent perturbation in the Hamiltonian.
Then, the response of $Q_j$ to a perturbation $Q_k$ appears as an expectation value,
and is hence directly obtained at first order in the sources, without the occurrence
of ${\mathbb K}$ which enters the symmetric part of $C_{jk}(t',t'')$.
Both approaches yield
\begin{eqnarray}
   \chi_{jk}(t',t'')  \!\!\!\!\!\!\!\! &&=
 ({1}/{i\hbar}) \theta(t'-t'') [C_{jk}(t',t'') - C_{kj}(t'',t') ]   \nonumber \\
                     &&\simeq      \label{MFA22q}  \theta(t'-t'') \,
{\cal Q}_j^{{\rm H}\alpha}(t',t)\, {\mathbb
C}_{\alpha\beta}\{{R}^{(0)}(t)\}\, {\cal Q}_k^{{\rm
H}\beta}(t'',t)\, ,
\end{eqnarray}
where $t$ is an arbitrary time in the interval $t_{\rm i} \le t \le
t''$ and $\theta$ the usual step function. In particular, by letting
$t=t_{\rm i}$ in (\ref{MFA22q}), the responses are variationally 
expressed in the Heisenberg picture
in terms of the matrix ${\mathbb C} \{{R}^{(0)}\}$ and of the
Heisenberg observables ${\sf Q}_j^{{\rm H}}(t',t_{\rm i})$ and ${\sf
Q}_k^{{\rm H}}(t'',t_{\rm i})$ given by the approximate backward
Heisenberg equations (\ref{MFA16Bq}).

For $H=K,\,D\propto\exp(-{\beta}K)$, the response (\ref{MFA22q}) in
an equilibrium state depends only on the time difference and  takes
the form
 \begin{eqnarray}   \label{MFA22Eq}
   \chi_{jk}(t',t'')  \!\!\!\!\!\!\!\! &&=
 ({1}/{i\hbar}) \, \theta(t'-t'') \,{\rm Tr}\, {\tilde{D}}
  \left[  e^{\,i\,H\,(t'-t'') / \hbar} \, Q_j \,  e^{\,-\,i\,H\,(t'-t'')},\,Q_k \right]
 \nonumber \\
      &&\simeq      
        \theta(t'-t'') \,
       {\cal Q}_j^{\alpha}\{{R}^{(0)}\}
      \left[    e^{\, {\mathbb C}\,{\mathbb F}\,(t'-t'')} \,{\mathbb C} \right]_{\alpha\beta} \,
       {\cal Q}_k^{\beta}\{{R}^{(0)}\} \,  .
\end{eqnarray}

\subsubsection{Static correlations; classical limit} \label{sec6.2.2}

 Static correlations between observables $Q_j$ and $Q_k$ in the state
$\tilde{D} \propto e^{-{\beta}K}$ are variationally obtained by
letting $t'-0 = t'' = t_{\rm i}$ in (\ref{MFA18q}), which yields
\begin{equation}   \label{MFA180q}
  C_{jk}(t_{\rm i}+0,t_{\rm i})  =
  {\cal Q}_j^{\alpha}\{{R}^{(0)}\}
  \,{\mathbb B}_{\alpha\beta}\,
   {\cal Q}_k^{\beta}\{{R}^{(0)}\}\, .
\end{equation}
The  matrix $ {\mathbb B}$ of correlations between the operators $\{{\sf M}\}$ is here saturated
by the coordinates
${\cal Q}_j^{\alpha} \{{R}^{(0)}\} = { \partial q_j\{{R}^{(0)}\} } / { \partial{R}_{\alpha}^{(0)} }$
of the images in the Lie algebra of the considered observables.

 For an Abelian algebra, the matrix ${\mathbb C}$ vanishes
and the ratio (\ref{MOC56q}) reduces to unity.
The ordinary and Kubo correlations are identical,
and the matrix ${\mathbb B}$ simplifies into
${\mathbb B}^{\rm K} = (\beta\,{\mathbb F})^{-1}$. In the high-temperature 
limit $\beta \to 0$ and in the classical limit $\hbar \to 0$, the ratio 
(\ref{MOC56q}) also tends to ${\mathbb I}$, so that
 \begin{eqnarray}   \nonumber
  {\mathbb B}  \to ({\beta}{\mathbb F})^{-1} .
 \end{eqnarray}
If $Q_j$ and $Q_k$ are commuting conserved observables,
the occurrence of the commutation matrix ${\mathbb C}$ in this ratio
also reduces it to ${\mathbb I}$.

For a Curie-Weiss model of interacting spins ${\sigma}_j= \pm 1$ at equilibrium,
the Weiss mean-field expressions for
the thermodynamic properties and the expectation values $\langle {\sigma}_j \rangle$ are recovered,
while the Ornstein-Zernike approximation \cite{OZq} for correlations
is recovered from ${\mathbb B}=({\beta}\,{\mathbb F})^{-1}$.

For the fermionic single-particle Lie algebra, the matrix ${\mathbb B}$ is the variational
approximation for the correlations of the operators
$a^{\dag}_{\mu}a_{\nu}$ and $a^{\dag}_{\sigma}a_{\tau}$:
\begin{equation}  \label{HF40q}
{\rm Tr}\, (a^{\dag}_{\mu}a_{\nu})(a^{\dag}_{\sigma}a_{\tau}) \tilde{D}
 - {\rm Tr}\, a^{\dag}_{\mu}a_{\nu} \tilde{D} \,
   {\rm Tr}\,  a^{\dag}_{\sigma}a_{\tau} \tilde{D}
      \simeq   {\mathbb B}_{\nu\mu,\tau\sigma} \,.
\end{equation}
If $\tilde{D}$ were replaced by an independent-particle state $\tilde{\DIh}$,
the expectation value
${\rm Tr}\,(a^{\dag}_{\mu}a_{\nu})(a^{\dag}_{\sigma}a_{\tau})\tilde{\DIh}$
would be given by Wick's theorem and
${\rm Tr}\, a^{\dag}_{\mu}a_{\nu} \tilde{\DIh}\,
 {\rm Tr}\,a^{\dag}_{\sigma}a_{\tau} \tilde{\DIh}$
would be the Hartree term so that
the left-hand side of (\ref{HF40q}) would reduce to the Fock term
${\rm Tr}(a^{\dag}_{\mu}a_{\tau}\tilde{\DIh})\,
 {\rm Tr}(a_{\nu}a_{\sigma}^{\dag}\tilde{\DIh})$.
For a more general state $\tilde{D}$,
the expression (\ref{MFA19Aq}) of ${\mathbb B}$ involves
the static RPA kernel  $i{\mathbb C}{\mathbb F}$,
which takes into account not only the Fock term
but also, through ${\mathbb K}$, effects of the interactions present in $K$.

 \subsubsection{Fluctuations} \label{sec6.2.3}

The static fluctuation $\Delta {Q}_j$ of the observable $Q_j$ in the
state $\tilde{D}$ is variationally given by (\ref{MFA180q}) where
$k=j$, that is,
\begin{equation}  \label{MFA184q}
      \Delta {Q}_j^{2}   =
    {\cal Q}_j^{\alpha}\{{R}^{(0)}\}
  \,{\mathbb B}_{\alpha\beta}\,
    {\cal Q}_j^{\beta}\{{R}^{(0)}\}\, .
\end{equation}

This fluctuation $\Delta {Q}_j(t)$ evolves in time according to
(\ref{MFA18q}) with $k=j,\, t'=t''=t$, that is,
\begin{equation}  \label{MFA185q}
 \Delta{Q}_j^{2}(t)  \simeq
   {\cal Q}_{j}^{{\rm H}\alpha}(t,t_{\rm i})\,
{\mathbb B}_{\alpha\beta} \, {\cal Q}_{j}^{{\rm H}\beta}(t,t_{\rm i})  \, ,
\end{equation}
which involves only the symmetric part of (\ref{MFA19Aq}).
For $H=K$, the fluctuation is time-independent.
For $H \ne K$, examples of time-dependences that are not 
properly accounted for by non-variational mean-field approximations
are given at the end of Sec.\,\ref{sec6.1} and in Sec.\,\ref{sec6.3}.

\subsubsection{Initial state in the Lie group} \label{sec6.2.4}

 In case the exact density operator $\tilde{D}$ belongs to the trial Lie group,
the operator $K$ belongs to the Lie algebra and coincides with its image,
$\tilde{\DIh}^{(0)}$ equals $\tilde{D}$;
the matrix ${\mathbb K}$ vanishes, ${\mathbb F}$ reduces to $-T\,{\mathbb S}$
and ${\mathbb B}$ to the trivial form (\ref{MFA19Bq}).
[For fermions, ${\mathbb B}$ reduces to the Fock terms.]
The quantities $ {\cal Q}_{j}^{{\rm H}\alpha}(t',t_{\rm i})$
and ${\cal Q}_{k}^{{\rm H}\beta}(t'',t_{\rm i})$
are in this case the only ingredients, apart from $\tilde{\DIh}^{(0)}$,
that enter $C_{jk}(t',t'')$.

\subsubsection{Zero-temperature limit} \label{sec6.2.5}

The present formalism encompasses ground-state properties, found by letting
$\beta \to \infty$. This limit entails simplifications.
[For instance, in the fermionic
case, the parameters $\{{R}^{(0)}\}$ of $\tilde{\DIh}^{(0)}$ constitute a matrix satisfying
$[{R}^{(0)}]^{2}={R}^{(0)}$.]
The number of vanishing eigenvalues of the commutation matrix ${\mathbb C}$ increases. 
While $S\{{R}^{(0)}\}$ tends to $0$ as $T \to 0$, the quantities
${\beta}^{-1}{\partial{S}}/{\partial{R}^{(0)}_{\alpha}}$, ${\beta}^{-1}{\mathbb S}$
and ${\mathbb C}{\mathbb F}$ remain finite, due to the singularity
of the von Neumann entropy (\ref{MFA8Dq}) for vanishing eigenvalues of $\tilde{\DIh}$.
The resulting simplifications of the correlation matrix ${\mathbb B}$
will be exhibited below in the diagonalized form (\ref{MFA50q}) of ${\mathbb B}$.

A further simplification occurs if the initial state lies in the Lie group. 
For fermions (possibly with pairing) the initial state is then a Slater determinant
(or a BCS state). In this case, it has been shown \cite{BV84q} that one 
can {\it by-pass the solution of the equations} (\ref{MFA16Bq}) for 
$ {\cal Q}_{j}^{{\rm H}\alpha}(t',t_{\rm i})$, of the RPA type. The proof 
relies on the fact that these equations (\ref{MFA16Bq}) for 
$ {\cal Q}_{j}^{{\rm H}\alpha}(t',t_{\rm i})$ involve the same kernel as 
the dynamical equations (\ref{MFA12Aq}) for small deviations 
$\delta {R}_{\alpha}^{(0)}(t)$, and that the latter equations can 
in practice be worked out by expansion of the simpler time-dependent 
mean-field equations (\ref{MFA13q}) for ${R}_{\alpha}^{(0)}(t)$. 
Two-time correlation functions $C_{jk}(t',t'')$
and time-dependent fluctuations $\Delta{Q}_j(t)$
can thus be evaluated by running the existing TDHF (or TDHFB) codes alternatively
forward and backward, with appropriate shifts in the boundary conditions. (For 
another derivation, see \cite{BV92q,Bro09,Si12q}.) This technique, variationally 
consistent, has been successfully
applied \cite{MK85q,BF85q,BS08q,Si11q} to describe,
in nuclear systems, the fluctuations
of single-particle observables which were severely underestimated by the conventional
use of TDHF; for a review, see \cite{Si12q}.

\subsection{Images of Heisenberg operators; conservation laws} \label{sec6.3}

It has already been noted that the equations of motion (\ref{MFA16Bq}) for the coordinates
${\cal Q}_{j}^{{\rm H}\alpha}(t',t)$ of ${\AIh}^{(1)}(t)$ appear as
a variational counterpart of the backward Heisenberg equation (\ref{BH5q}).
To be more precise, let us write the time dependence of the Lie algebra operator
${\sf Q}_{j}^{{\rm H}}(t',t)={\cal Q}_{j}^{{\rm H}\alpha}(t',t)\,{\sf M}_{\alpha}$. 
Using the relation
$\left(  {\mathbb L} + {\mathbb C}\,{\mathbb H} \right)_{\beta}^{\;\alpha}  =
{\Gamma}^{\,\alpha}_{\beta\gamma}\,{\cal H}^{\gamma} \{{R}^{(0)}(t)\}  +
 {\mathbb C}_{\beta\gamma}\{{R}^{(0)}(t)\}\,{\mathbb H}^{\gamma\alpha}\{{R}^{(0)}(t)\}
=\partial ({\mathbb C}_{\beta\gamma}{\cal H}^{\gamma}) / \partial{R}^{(0)}_{\alpha}(t)$
and the definition (\ref{MFA903q}) of images, one can rewrite the equations of motion
for ${\cal Q}_{j}^{{\rm H}\alpha}(t',t)$ (including ${\alpha} = 0$) as
 \begin{equation} \label{Im01q}
  \frac {d {\sf Q}^{{\rm H}}_j(t',t)}{dt}= {\rm Image \,of}
  \left\{  -\, \frac {1}{i\,\hbar} \,  [ {\sf Q}^{{\rm H}}_j(t',t), \, H ] \right\}
  =  -\, \frac {1}{i\,\hbar} \,  [ {\sf Q}^{{\rm H}}_j(t',t), \, {\sf H} ] \, ,
\end{equation}
the image ${\sf H}$ of the Hamiltonian $H$ being evaluated with respect to $\{{R}^{(0)}(t)\}$.
While the equations (\ref{MFA16Bq}) were written in terms of the coordinates
${\cal Q}_{j}^{{\rm H}\alpha}$ of ${\sf Q_j^{\rm H}}$,
the introduction of images gives these equations a simple operator form:
The right-hand side is simply the image of the r.h.s. 
of the exact backward Heisenberg equation (\ref{BH5q}) for time-dependent observables.
The boundary condition ${\sf Q}^{{\rm H}}_j(t',t') = {\sf Q}^{{\rm S}}_j$ is also the image
of the boundary condition for the Heisenberg observable ${Q}^{{\rm H}}_j(t',t)$.

Up to now, no specific assumptions have been made about the data.
Let us now consider an observable ${Q}^{{\rm S}}_j$ that belongs to the Lie algebra
and commutes with $H$. This conservation law, together with the equation of motion
(\ref{Im01q}), shows that the operator $ {\sf Q}^{{\rm H}}_j(t',t)$ is constant and
equal to ${\sf Q}^{{\rm S}}_j$. Hence, the approximate expectation value
$ \langle Q_j \rangle_{t}$, evaluated through (\ref{MFA02q}) for $t'= t_{\rm i}$, is constant.
Less trivially, the fluctuation $\Delta{{Q}_j(t)}$,
evaluated through (\ref{MFA185q}), is also constant as it should.
This property, which arises naturally through the use of the approximate
backward Heisenberg equation, was not granted.
[For instance, in the time-dependent Hartree-Fock approximation
for fermions, fluctuations evaluated from ${\DIh}^{(0)}(t)$ through Wick's 
 theorem are not constant for a conserved single-particle observable.]

Another conservation property holds for two observables $Q_{j}^{\rm
S}$ and $Q_{k}^{\rm S}$ of the Lie algebra such that $Q_{j}^{\rm S}$
is conserved, $[Q_{j}^{\rm S},\,H]=0$, and that $[Q_{k}^{\rm
S},\,H]= i{\hbar}\,Q_{j}^{\rm S}$. This occurs for instance if
$Q_{k}^{\rm S}={\sf X}$ is a coordinate of the center of mass of the
system, and $Q_{j}^{\rm S}={\sf P}/m$ the associated velocity
operator (${m}$ is the total mass). Then, Eq.(\ref{Im01q}) implies
that, as the exact Heisenberg operators, the approximate ones
satisfy ${\sf P}^{\rm H}(t',t) = {\sf P}^{\rm S}$ and ${\sf X}^{\rm
H}(t',t) = {\sf X}^{\rm S} + (t'-t){\sf P}^{\rm S}/{m}$ (see
Sec. 5.5 of \cite{BV92q}). Hence, not only ${\langle{\sf
X}\rangle}_t$ and ${\langle{\sf P}\rangle}_t$, but also the
fluctuations and correlations of ${\sf X}$ and ${\sf P}$ produced by
the general formula (\ref{MFA18q}) have the proper time dependence.
In particular, $\Delta{\sf P}$ is constant and $\Delta{{{\sf
X}}^{2}}(t)$ satisfies $d\,\Delta{{{\sf X}}^{2}}(t)/dt= \Delta{{{\sf
P}}^{2}}/{m}^{2}$; we thus acknowledge that the present
approximation for the fluctuations accounts for the exact spreading
of the wave packet [whereas the TDHF approximation produces a
constant width].

\section{Mapped Lie algebra and mapped Hilbert space} \label{sec7}

The purpose of this section is to rewrite in a unified form the
variational approximations (\ref{MFA130q}),(\ref{MFA125q}) for
expectation values and
(\ref{MFA18q}),(\ref{MFA19Aq}),(\ref{MFA141q}) for correlation
functions . To this aim, we will rely on a correspondence that
associates with the Lie algebra $\{{\sf M}\}$ a simpler Lie algebra
$\{\underline{\sf M}\}$. It will turn out that the various
expressions found above can be re-expressed as traces over a single
effective density operator $\tilde{\underline{D}}$ acting in a new
mapped space $\underline{\mathscr{H}}$ rather than in the original
Hilbert space $\mathscr{H}$.

 \subsection{Unifying the approximate expectation values
and correlations in the initial state} \label{sec7.1}

The optimization of the expectation value
${ \langle {\sf M}_{\alpha} \rangle }  = {\rm Tr}\,\tilde{D}\,{\sf M}_{\alpha}$
has yielded the approximation
${ \langle {\sf M}_{\alpha} \rangle }_{\rm app} = {R}^{(0)}_{\alpha}=
{\rm Tr}\,\tilde{\DIh}^{(0)}{\sf M}_{\alpha}$; in contrast the optimization of
$ {  \langle {{\sf M}}_{\alpha}\, {{\sf M}}_{\beta}  \rangle  } =
    {\rm Tr}\,\tilde{{D}} \,{{\sf M}}_{\alpha}\,{{\sf M}}_{\beta}$
has yielded ${\langle {\sf M}_{\alpha} {\sf M}_{\beta} \rangle}_{\rm app}
 = {\mathbb B}_{\alpha\beta} + {R}^{(0)}_{\alpha}\,{R}^{(0)}_{\beta} $
that cannot be expressed as a trace over a density operator
in the Hilbert space ${\mathscr{H}}$.
We wish to map the set $\{{{\sf M}}\}$ acting in ${\mathscr{H}}$
onto a new set  $\{{\underline{\sf M}}\}$
acting in a new space $\underline{\mathscr{H}}$,
and to introduce in $\underline{\mathscr{H}}$
an effective density operator $\tilde{\underline{D}}$ so as to re-express our approximations in terms of $\tilde{\underline{D}}$.
Namely, we wish the exact expectation values over $\underline{\tilde{D}}$
in the mapped space $\underline{\mathscr{H}}$,
denoted as $ \langle {\underline{\sf M}}_{\alpha} \rangle_{\rm map}$ and
$\langle{\underline{\sf M}}_{\alpha}\,
{\underline{\sf M}}_{\beta}\rangle_{\rm map}$,
to coincide with the corresponding variational approximations
in the original space ${\mathscr{H}}$,
denoted as ${\langle {\sf M}_{\alpha} \rangle}_{\rm app}$ and
${\langle {\sf M}_{\alpha} {\sf M}_{\beta} \rangle}_{\rm app} $, according to
 \begin{eqnarray}   \label{MOC13aq}
 &&  \langle {\underline{\sf M}}_{\alpha} \rangle_{\rm map}  \equiv
     \underline {\rm Tr}  \,{\underline{\sf M}}_{\alpha}\, \tilde{\underline{D}}
  = { \langle {\sf M}_{\alpha} \rangle }_{\rm app}  = {R}^{(0)}_{\alpha} \, ,
  \\       \label{MOC13bq}
 &&  \langle   {\underline{\sf M}}_{\alpha} \,
  {\underline{\sf M}}_{\beta}  \rangle_{\rm map}
      \equiv
 \underline {\rm Tr} \,{\underline{\sf M}}_{\alpha}\,{\underline{\sf M}}_{\beta}\,
 \tilde   {\underline{D}}
  = {  \langle {{\sf M}}_{\alpha}\, {{\sf M}}_{\beta}  \rangle  }_{\rm app}
   = {\mathbb B}_{\alpha\beta} + {R}^{(0)}_{\alpha}\,{R}^{(0)}_{\beta}  \, .
\end{eqnarray}
Going from the space ${\mathscr{H}}$ to $\underline{\mathscr{H}}$ will be a price 
to pay for expressing both expectation values and correlations
in terms of a unique effective state $\underline{\tilde{D}}$.

The first step consists in replacing, in the original Lie structure
$[ {\sf M}_{\alpha}, {\sf M}_{\beta}] =
{i\,\hbar\,} \Gamma_{\alpha\beta}^{\gamma}\,{\sf M}_{\gamma}$,
the operator ${\sf M}_{\gamma}$ on the right side by the c-number
${R}^{(0)}_{\gamma} ={\rm Tr} \, {\sf M}_{\gamma} {\tilde\DIh}^{(0)}$,
the expectation value
of ${\sf M}_{\gamma}$ in the state ${\tilde\DIh}^{(0)}$ of ${\mathscr{H}}$.
This procedure associates with the original Lie algebra
$\{{\sf M}\}$ in ${\mathscr{H}}$
a {\it reduced Lie algebra} $\{{\underline{\sf M}}\}$ characterized by the simpler commutation relations
\begin{eqnarray} \label{MOC10q}
 [ {\underline{\sf M}}_{\alpha}, \,{\underline{\sf M}}_{\beta} ]
  =  {i\,\hbar\,} \Gamma_{\alpha\beta}^{\,\gamma}\,{R}^{(0)}_{\gamma} \, {\underline{\sf M}}_{0}
     = i\,\hbar\,{\mathbb C}_{\alpha\beta}\, .
\end{eqnarray}
(From now on we shall most often drop, as in the end of (\ref{MOC10q}),
the unit operator ${\underline{\sf M}}_{0}$.)

Multiplication of any number of operators ${\underline{\sf M}}_{\alpha}$ generates
an enveloping algebra, the space of representation of which
defines $\underline{\mathscr{H}}$.
The structure of this space will be cleared up in Sec.\,\ref{sec8} by setting
${\mathbb C}_{\alpha\beta}$ into a canonical form.

In order to satisfy the conditions (\ref{MOC13aq}) on expectation values
${ \langle {\sf M}_{\alpha} \rangle }_{\rm app} $
the sought effective density operator $\tilde{\underline{D}}$ should depend only
on the differences ${\underline{\sf M}}_{\alpha} -  {R}^{(0)}_{\alpha}$.
As regards the conditions (\ref{MOC13bq}), we remember that correlations
are often generated in statistical mechanics by a probability distribution
having the form of an exponential of the free energy
regarded as a function of the running variables.
For instance an energy distribution is the product
$ e^{-{\beta}\,F({E})} = e^{-{\beta}\,{E}+{S}({E}) }$
of the Boltzmann-Gibbs exponential $e^{-{\beta}\,{E}}$
by the level density, the exponential $e^{\,{S}({E})}$ of the entropy, 
which accounts for the discarded variables.
Another example was provided in our approximation
by the thermodynamic coefficients given
by the second derivative (\ref{MFA20q}) of the free energy
$f\{{R}\}$ at $\{{R}\}=\{{R}^{(0)}\}$.
Here, likewise, the reduction from $\{{{\sf M}}\}$ to $\{{\underline{\sf M}}\}$
suggests to rely on
an effective free-energy operator rather than on an effective Hamiltonian.
We therefore replace, in the free-energy function
$f\{{R}\} \equiv k\{{R}\} - TS\{{R}\} $ defined by (\ref{MFA12q}),
the variables $\{{R}\}$ by the corresponding new operators $\{{\underline{\sf M}}\}$.
The operators ${\underline{\sf M}}_{\alpha}$ fluctuate around ${R}^{(0)}_{\alpha}$,
and this leads us to expand the operator $f\{{\underline{\sf M}}\}$ as
 \begin{eqnarray} \label{MOC11q}
 f\{{\underline{\sf M}}\} \approx
   f\{ {R}^{(0)} \}  + \frac{1}{2}
 ({\underline{\sf M}}_{\alpha} - {R}^{(0)}_{\alpha})
  \,{\mathbb F}^{\alpha\beta}\,
 ({\underline{\sf M}}_{\beta} - {R}^{(0)}_{\beta}) + ... \, ,
 \end{eqnarray}
where the first-order term is absent thanks to the stationarity
of $f\{{R}\}$ at $\{{R}^{(0)}\}$.

We thus guess that the distribution $\tilde{\underline{D}}$
governing the operators $\{{\underline{\sf M}}\}$, which is expected
to yield the identities (\ref{MOC13aq})-(\ref{MOC13bq}), should have {\it in
the mapped space} $\underline{\mathscr{H}}$ the exponential form
\begin{eqnarray}   \label{MOC12q}
\tilde{\underline{D}} \equiv  \frac {  e^{ \,-\,{\beta}\,\underline{F} }  }
                { {\underline{\rm Tr}}\, e^{ \,-\,{\beta}\,\underline{F} }  } \, ,
\;\;\;\;\;\;\;
  \underline{F} \equiv \frac{1}{2} \,
 ( {\underline{\sf M}}_{\alpha} - {R}^{(0)}_{\alpha} ) \,
  {\mathbb F}^{\alpha\beta} \,
 ( {\underline{\sf M}}_{\beta} -  {R}^{(0)}_{\beta}  )  \, ,
\end{eqnarray}
where $\underline{F}$ behaves as a kind of free-energy operator. The inclusion of an 
{\it entropic contribution} in the exponent of $\tilde{\underline{D}}$ accounts 
for the {\it elimination of degrees of freedom} associated with the replacement of 
the original Hilbert space ${\mathscr{H}}$ by the space $\underline{\mathscr{H}}$ 
that involves a smaller set of observables, those 
generated by the operators $\{{\underline{\sf M}}\}$ and their products. The 
non-negativity of the operator $\underline{F}$ is ensured by that of the matrix
${\mathbb F}$. More concrete interpretations of $\underline{F}$ will
be given in Sec.\,\ref{sec8} by Eqs.(\ref{MFA82q}) or (\ref{MFA74q}).

While $\langle {\underline{\sf M}}_{\alpha} \rangle_{\rm map}= {R}^{(0)}_{\alpha}$
is evident, the surmise (\ref{MOC13bq}) is proved in Appendix B.
It is first shown there that the Kubo correlations of the operators
$\{{\underline{\sf M}}\}$ in the state $\tilde{\underline{D}}$ are given by
 \begin{eqnarray}   \label{MOC20q}
    \left \langle \frac{1}{\beta} \,  \int_{{0}}^{\beta}\,d{\tau} \,
     e^{ \,{\tau}\underline{F} }
 \, ( {\underline{\sf M}}_{\alpha} - {R}^{(0)}_{\alpha} )
        \,   e^{ \,-\,{\tau}\underline{F} }  \,\,
    ( {\underline{\sf M}}_{\beta} - {R}^{(0)}_{\beta} ) \right \rangle_{\rm map}
 = \frac {1}{\beta} \, \left( {\mathbb F}^{-1} \right)_{\alpha\beta}  \,.
\end{eqnarray}
The ordinary correlation matrix of the operators $\{{\underline{\sf M}}\}$
is then derived from (\ref{MOC20q})
and shown to coincide with the matrix $\mathbb{B}$ defined by (\ref{MFA19Aq}):
 \begin{eqnarray}   \label{MOC24q}
  \langle  ( {\underline{\sf M}}_{\alpha} -  {R}^{(0)}_{\alpha} )
            ( {\underline{\sf M}}_{\beta} - {R}^{(0)}_{\beta} )  \rangle _{\rm map}
 =   \left(   \frac { i\,\hbar\,{\mathbb C}\,{\mathbb F}}
                 {\Id -\exp[{-i\,\hbar}\,\beta\,{\mathbb C}\,{\mathbb F}]}
 \,{\mathbb F}^{-1}    \right)_{\alpha\beta}
  =  {\mathbb B}_{\alpha\beta}  \, .
 \end{eqnarray}
Thus, the operator $\tilde{\underline{D}}$ can be regarded as a substitute
in $\underline{\mathscr{H}}$
to the exact state $\tilde{D} \propto \exp(-{\beta}K)$
 that satisfies the anticipated identities
(\ref{MOC13aq}) and (\ref{MOC13bq}):
We can interpret the matrix elements of ${\mathbb{B}}$ as exact correlations
of the operators $\{{\underline{\sf M}}\}$ in the state $\tilde{\underline{D}}$.

At first order in the sources, the variational treatment amounted to replace
the operators $\{{\sf M}\}$ by their expectation values $\{{R}^{(0)} \}$.
Here, at second order, we reproduce the variational approximation
$\mathbb{B}$ for correlations by keeping contributions of lowest order
in the deviation $\{ {\underline{\sf M}} - {R}^{(0)} \}$,
both as regards the reduction (\ref{MOC10q})
of the algebra of $\{{\sf M}\}$ into that of $\{{\underline{\sf M}}\}$
and as regards the expansion (\ref{MOC11q}) of the free energy operator.

\subsection{Heisenberg dynamics of the mapped Lie algebra} \label{sec7.2}

Let us extend the above results to the time-dependent correlation functions.
We restrict here to the case where $H=K$, as in Sec.\,\ref{sec6.1}. As already seen,
the solution of the approximate backward Heisenberg equation is then generated
by the kernel $i{\mathbb C}{\mathbb F}$ according
to Eqs.(\ref{MFA140q}) for the coordinates $\alpha \ne 0$
[and to Eq.(\ref{MFA02q}) for $\alpha = 0$].
This time-dependence keeps the Heisenberg operators ${\sf M}^{\rm H}_{\alpha}(t',t)$
in the Lie algebra $\{{\sf M}\}$ since they are given by
 \begin{equation}   \label{MOC30q}
  {\sf M}^{\rm H}_{\alpha}(t',t) - {R}^{(0)}_{\alpha}  =
    \left[  e^{\,{\mathbb C}\,{\mathbb F}\,(t'-t)} \right]_{\alpha}^{\;\beta}
    (  {\sf M}_{\beta} - {R}^{(0)}_{\beta}  ) \, ,
 \end{equation}
or equivalently by
\begin{equation}   \label{MOC31q}
    \frac {d\,{\sf M}^{\rm H}_{\alpha}(t',t)} {d\,t}   =
   - ( {\mathbb C}\,{\mathbb F} )_{\alpha}^{\;\beta} \,
  ( {\sf M}^{\rm H}_{\beta}(t',t) - {R}^{(0)}_{\beta}  ) \, .
 \end{equation}

The equations (\ref{MOC31q}) constitute a linear set. We are thus led to define,
in the mapped space $\underline{\mathscr{H}}$, the time-dependent operators
${\underline{\sf M}}^{\rm H}_{\alpha}(t',t)$ by the corresponding equation
\begin{equation}   \label{MOC32q}
    \frac { d\,{\underline{\sf M}}^{\rm H}_{\alpha}(t',t) } {d\,t}   =
   - ( {\mathbb C}\,{\mathbb F} )_{\alpha}^{\;\beta} \,
  ( {\underline {\sf M}}^{\rm H}_{\beta}(t',t)  - {R}^{(0)}_{\beta} ) \, ,
 \end{equation}
with the boundary condition
${\underline{\sf M}}^{\rm H}_{\alpha}(t,t) = {\underline{\sf M}}_{\alpha}$.
Moreover, from the definition (\ref{MOC12q}) of the operator
${\underline{F}}$ and from the mapped algebra (\ref{MOC10q}), it follows that
     \begin{equation}   \label{MOC33q}
    [ {\underline{\sf M}}_{\alpha},\, {\underline{F}} ] =
     i\,\hbar\, ( {\mathbb C}\,{\mathbb F} )_{\alpha}^{\;\beta} \,
  (  {\underline {\sf M}}_{\beta} - {R}^{(0)}_{\beta} ) \, .
 \end{equation}
Hence the equations of motion of the set $\{  {\underline {\sf M}}^{\rm H}(t',t) \}$
defined by (\ref{MOC32q}) read alternatively
 \begin{equation}   \label{MOC34q}
 \frac { d\,{\underline{\sf M}}^{\rm H}_{\alpha}(t',t) } {d\,t}  =
  -\, \frac{1} {i\,\hbar}  \left[  {\underline {\sf M}}^{\rm H}_{\alpha}(t',t), \,
{\underline{F}} \right] \, .
  \end{equation}

One recognizes the structure of an {\it exact backward Heisenberg equation} (\ref{BH5q})
in the space $\underline{\mathscr{H}}$, where the free-energy operator 
$ {\underline{F}} $ plays the role of a Hamiltonian. As 
in the static case, the mapping of the operators $\{{{\sf M}}\}$
onto $\{{\underline{\sf M}}\}$
replaces approximate properties in ${\mathscr{H}}$ by exact ones in $\underline{\mathscr{H}}$,
here the approximate dynamical equation (\ref{MOC31q}) in the space ${\mathscr{H}}$
by (\ref{MOC34q}) where the commutator is restored
in the space $\underline{\mathscr{H}}$. Accordingly,
the mapped Heisenberg operators ${\underline {\sf M}}^{\rm H}_{\alpha}(t',t)$
are given by
 \begin{equation}   \label{MOC35q}
 {\underline {\sf M}}^{\rm H}_{\alpha}(t',t)  =
    e^{ \,    i\, {\underline{F}}(t'-t) / \hbar }   \, {\underline {\sf M}}_{\alpha} \,
    e^{ \,- \,i\, {\underline{F}}(t'-t) / \hbar } \, ,
 \end{equation}
a mere unitary transformation in the space $\underline{\mathscr{H}}$, whereas the 
transformation (\ref{MOC30q}) of the set $\{{\sf M}\}$ in the space ${\mathscr{H}}$ is not unitary.

The linearity of Eqs.(\ref{MOC32q}), or equivalently the quadratic nature
of ${\underline{F}}$, exhibits harmonic-oscillator dynamics in the mapped space.
This will be made more precise in Secs.\,\ref{sec8.4} and \ref{sec8.5}.
In Eq.(\ref{MOC34q}) the effective "Hamiltonian" $ {\underline{F}} $
in the space $\underline{\mathscr{H}}$ should not be confused with an
approximation for the Hamiltonian $H=K$ of the original problem.
Its expressions (\ref{MOC12q}) and (\ref{MFA20q})
are related to the free-energy function $f\{{R}\}$ and its deviations, rather 
than to the original Hamiltonian $H$. The contribution to ${\underline{F}}$ 
of the entropic term $-{\beta}^{-1}{\mathbb S}$,
through ${\mathbb F} = {\mathbb K} - {\beta}^{-1}{\mathbb S}$, is essential.
This contribution is the only one left if $K$ belongs to the Lie algebra,
and it remains {\it finite at zero temperature}.

\subsection{Unified formulation of the variational expressions} \label{sec7.3}

 Our mapping has provided a formalism (Sec.\,\ref{sec7.1}) in which both
optimized expectation values and correlations in the state $\tilde{D}$
are generated, in the mapped space $\underline{\mathscr{H}}$,
as traces over the effective density
operator $\underline{\tilde{D}} \propto \exp( -{\beta} {\underline{F}} )$. This operator depends on the
temperature both explicitly and through the kernel ${\mathbb F}$ of the operator ${\underline{F}}$.

An arbitrary observable $Q_j$ acting in the original Hilbert space is now
represented by its mapped image given, according to (\ref{MFA903q}), by
   \begin{eqnarray} \label{MOC40q}
   {\underline{\sf Q}}_{j}    =   {q}_{j}\{{R}^{(0)}\} \, {\underline{\sf M}}_{0}
  + ( {\underline{\sf M}}_{\alpha} - {R}^{(0)}_{\alpha}\, {\underline{\sf M}}_{0} )\,
   \frac   { \partial q_j\{{R}^{(0)}\} }
 { \partial{R}_{\alpha}^{(0)} }  \;\;\;(\alpha \ne 0)\,.
 \end{eqnarray}
Then, the optimized expectation value of a single operator $Q_j$ is
   \begin{eqnarray} \label{MOC41q}
  { \langle Q_j \rangle }_{\rm app}  =  {q}_{j}\{{R}^{(0)}\}  =  {\langle  {\underline{\sf Q}}_{j}  \rangle}_{\rm map}
    \equiv  {\underline{\rm Tr}} \, {\underline{\sf Q}}_{j} \, { \underline{\tilde{D}} } \, ,
   \end{eqnarray}
and that of the product $Q_j\,Q_k$ is
   \begin{eqnarray} \label{MOC42q}
  { \langle Q_j\,Q_k \rangle }_{\rm app}  =
   {\langle  {\underline{\sf Q}}_{j} \,  {\underline{\sf Q}}_{k}  \rangle}_{\rm map}
    \equiv  {\underline{\rm Tr}} \, {\underline{\sf Q}}_{j} \, {\underline{\sf Q}}_{k}\,
            { \underline{\tilde{D}} } \, ,
   \end{eqnarray}
 both being directly obtained from the effective state $ { \underline{\tilde{D}} }$ in
the space $\underline{\mathscr{H}}$.

In the case $H=K$ considered in Secs.\,\ref{sec6.1} and \ref{sec7.2}, the 
time-dependence of the Heisenberg operator associated with 
${\underline{\sf Q}}_{j}$
is governed by the backward Heisenberg equation (\ref{MOC34q}), which yields
 \begin{equation}  \label{MOC44q}
  {\underline {\sf Q}}^{\rm H}_j(t)  =
    e^{ \,    i\, {\underline{F}}\,t / \hbar }   \, {\underline {\sf Q}}_j \,
    e^{ \,- \,i\, {\underline{F}}\,t / \hbar } \, .
 \end{equation}
Hence, the variational {\it two-time correlation functions} take in the mapped space
$\underline{\mathscr{H}}$ the simple form
 \begin{equation}  \label{MOC43q}
C_{jk}(t',t'')  = {  \langle \,{T}\, \underline {\sf{Q}}_j^{\rm H}(t'- t'')  \,
     \underline{\sf{Q}}_k \rangle  }_{\rm map}
  - { \langle  \underline{\sf{Q}}_j \rangle  }_{\rm map}  \,
  { \langle  \underline{\sf{Q}}_j \rangle  }_{\rm map} \, .
  \end{equation}

The same operator
${\underline{F}}$ [Eq.(\ref{MOC12q})] occurs both in the density operator
$ \underline{\tilde{D}}$
and as an effective Hamiltonian in (\ref{MOC44q}).
It depends on the following ingredients:
${R}_{\alpha}^{(0)}$ defined self-consistently by Eq.(\ref{qb30bq}),
${\mathbb C}_{\alpha\beta} \{{R}^{(0)}\}$ given by
$\Gamma^{\gamma}_{\alpha\beta} \,{R}_{\gamma}^{(0)} $,
${\mathbb F}$ given by (\ref{MFA20q})
and ${ \underline{\tilde{D}} }$ by (\ref{MOC12q}).

\section{The eigenmodes and their interpretation} \label{sec8}

In this section, we consider only initial states at equilibrium,
in which case the dynamics is governed by the Hamiltonian $H=K$.
The variational expressions of correlation functions obtained in 
Sec.\,\ref{sec5} 
then involve functions of the sole matrices $\mathbb{C}$ and $\mathbb{F}$.
Practical evaluations rely on their diagonalization.
The eigenvalues and eigenvectors thus obtained will
enable us to interpret the above results.

\subsection{Diagonalization of the evolution kernel and the correlation matrix}
 \label{sec8.1}

The product $i{\mathbb C}{\mathbb F}$ of the {\it commutation matrix} ${\mathbb C}$
and the {\it stability matrix} ${\mathbb F}$ is the kernel which governs the evolution,
as exhibited by the dynamical equations (\ref{MFA140q}) of Sec.\,\ref{sec6.1}.
It also occurs in the expression (\ref{MFA19Aq})
of the correlation matrix ${\mathbb B}$.
The diagonalization of the matrix $i{\mathbb C}{\mathbb F}$,
and the study of its properties, appear therefore appropriate.

 The Lie algebra is globally hermitian, namely,
if the operators ${\sf M}_{\alpha}$ of the basis are not individually
hermitian, they come in conjugate pairs (e.g., $a^{\dagger}_{\mu}a_{\nu}$ and
$a^{\dagger}_{\nu}a_{\mu}$ in the fermionic example). We denote as $\overline{\alpha}$
the index of the operator ${\sf M}_{\overline{\alpha}} \equiv {\sf M}^{\dagger}_{\alpha}$,
which may or may not differ from ${\sf M}_{\alpha}$. The change
${\alpha} \mapsto \overline{\alpha}$ of all the indices in the quantities
${R}_{\alpha},\,{\cal Q}^{\alpha},\,{\mathbb C}_{\alpha\beta}$
or ${\mathbb F}^{\alpha\beta}$ transforms them into their complex conjugates.
Hence,  $i{\mathbb C}$, which is antisymmetric, and ${\mathbb F}$, which is symmetric,
are equivalent to hermitian matrices. Moreover, ${\mathbb F}$ is non negative.
(Only the case of a strictly positive matrix ${\mathbb F}$ is examined here;
vanishing eigenvalues are considered in the end of Sec.\,\ref{sec9.3}.)
Altogether, the matrix $i{\mathbb C}{\mathbb F}$ is
equivalent to the antisymmetric hermitian matrix
$i{\mathbb F}^{1/2}{\mathbb C}{\mathbb F}^{1/2}$,
so that its right and left eigenvectors, respectively denoted
as ${\psi}$  and  ${\phi}$,
constitute a complete biorthonormal basis in the space $\alpha$,
while its eigenvalues are real and either vanish or come in opposite pairs.

Using the above properties,
one can classify into three subsets the eigenvectors of $i{\mathbb C}{\mathbb F}$,
denoted with indices $n,\,-n$ and $p$, respectively:

(i) To the positive eigenvalues $\Omega_n >0$ are associated the right and left
eigenvectors ${\psi}^n$ and ${\phi}_n$ defined by $(\alpha,\beta \ne 0)$
\begin{eqnarray} \label{MFA30Aq}
 && i\,{\mathbb C}_{\alpha\gamma}\, {\mathbb F}^{\gamma\beta} \,{{\psi}}_{\beta}^{n}
=\Omega_{n}\,  {{\psi}}_{\alpha}^{n}\,,
 \\ \label{MFA30Bq}
&&  {\phi}^{\alpha}_{n} \, i\,{\mathbb C}_{\alpha\gamma}\, {\mathbb F}^{\gamma\beta} \,
  =  {\phi}^{\beta}_{n}\, \Omega_{n}\, ,
\;\;\;\;\;\;\;\;
  {\phi}^{\beta}_{n} =   {({\psi}_{\overline\alpha}^{n})}^{*} \, {\mathbb F}^{\alpha\beta}\,.
\end{eqnarray}

(ii) Taking the complex conjugate of (\ref{MFA30Aq}) provides the right and left
eigenvectors ${\psi}^{-n}$ and ${\phi}_{-n}$ associated with the negative eigenvalue
$-{\Omega}_{n}$:
\begin{equation} \label{MFA34q}
 {{\psi}}_{\alpha}^{-n} =  {( {\psi}_{\overline{\alpha}}^{n} )}^{*} \,,
\;\;\;\;\;\;\;\;
 {\phi}^{\alpha}_{-n} =    {( {\phi}^{\overline{\alpha}}_{n} )}^{*} \,.
 \end{equation}

(iii) The eigenvectors $ {\psi}^{p}$ and ${\phi}_{p}$ associated with vanishing
eigenvalues are given by
\begin{eqnarray}   \label{MFA44q}
i\,{\mathbb C}_{\alpha\gamma}\, {\mathbb F}^{\gamma\beta} \,{\psi}_{\beta}^{p}  = 0\,,
\;\;\;\;\;\;\;
{\phi}^{\alpha}_{p}\, {\mathbb C}_{\alpha\beta} = 0\,,
   \\   \label{MFA44Bq}
 {\psi}_{\alpha}^{p} =  {( {\psi}_{\overline{\alpha}}^{p} )}^{*} \, ,
\;\;\;\;\;\;
 {\phi}^{\beta}_{p} = {( {\phi}^{\overline{\beta}}_{p} )}^{*}
=   {( {\psi}_{\overline{\alpha}}^{p} )}^{*} \,  {\mathbb F}^{\alpha\beta}\,.
\end{eqnarray}

The biorthonormality of the sets $ \{{{\psi}}\}$ and $\{{{\phi}}\}$ is equivalent
to the {\it orthonormalization relations} for the right eigenvectors
$ \{{{\psi}}\}$ expressed by
\begin{eqnarray}    \nonumber
&& {\phi}^{\beta}_{n}\,{{\psi}}_{\beta}^{n'} = {( {\psi}_{\overline{\alpha}}^{n} )}^{*}\,{\mathbb F}^{\alpha\beta}
  \,{\psi}_{\beta}^{n'} = \delta^{nn'}\,,
\;\;\;\;
 {\phi}^{\beta}_{p}\,{{\psi}}_{\beta}^{n'} = { ( {\psi}_{\overline{\alpha}}^{p} ) }^{*} \, {\mathbb F}^{\alpha\beta}\,
 {\psi}_{\beta}^{p'} = {\delta}^{pp'}\,,
\\   \label{MFA45Aq}
&&{\psi}_{\alpha}^{n}\,{\mathbb F}^{\alpha\beta}\,{\psi}_{\beta}^{n'}  =
  {\psi}_{\alpha}^{n}\,{\mathbb F}^{\alpha\beta}\,{\psi}_{\beta}^{p}  =
 {({\psi}_{\overline{\alpha}}^{n})}^{*}\,{\mathbb F}^{\alpha\beta} \,{\psi}_{\beta}^{p} =0\,.
 \end{eqnarray}
 Likewise, the {\it closure property} is equivalent to
\begin{eqnarray} \label{MFA46q}
 {({\mathbb F}^{-1})}_{\alpha\beta}  =
\sum_{n} [ {\psi}_{\alpha}^{n} \,  {({\psi}_{\overline{\beta}}^{n})}^{*}   +
           {({\psi}_{\overline{\alpha}}^{n})}^{*}  \, {\psi}_{\beta}^{n}  ]
  +  \sum_{p}  {\psi}_{\alpha}^{p}\, {\psi}_{\beta}^{p} \,.
 \end{eqnarray}

The matrix $i{\mathbb C}{\mathbb F}$ is thus diagonalized as
\begin{eqnarray} \label{MFA49q}
  i ({\mathbb C}\,{\mathbb F})_{\alpha}^{\beta} =
 \sum_{n} \Omega_{n} [ {\psi}_{\alpha}^{n} \, {\phi}^{\beta}_{n}
   -  {({\psi}_{\overline{\alpha}}^{n})}^{*} \, {({\phi}^{\overline{\beta}}_{n})}^{*}   ]\ ,
  \end{eqnarray}
a sum involving only the modes $n$ associated with $\Omega_{n} > 0$.
The matrices ${\mathbb F}$ and ${\mathbb C}$ can then be expressed in the form
 \begin{eqnarray}    \label{MFA48q}
  &&  {\mathbb F}^{\alpha\beta} =
    \sum_{n}  [  ({\phi}^{\overline{\alpha}}_{n})^{*} \, {\phi}^{\beta}_{n}
                +  {\phi}^{\alpha}_{n} \, ({\phi}^{\overline{\beta}}_{n})^{*} ]
     +  \sum_{p}  {\phi}_{p}^{\alpha}\,  {\phi}_{p}^{\beta}\,,
   \\    \label{MFA47q}
 &&  i\,{\mathbb C}_{\alpha\beta} =
 \sum_{n} \Omega_{n} [ {\psi}_{\alpha}^{n} \,  {({\psi}_{\overline{\beta}}^{n})}^{*}
                      -  {({\psi}_{\overline{\alpha}}^{n})}^{*}  \,  {{\psi}}_{\beta}^{n} ]\,,
 \end{eqnarray}
which for ${\mathbb F}$ includes also terms associated with the vanishing eigenvalues
of $i{\mathbb C}{\mathbb F}$.

The above diagonalization appears as a generalization, for an arbitrary Lie group and at
non-zero temperature, of the standard diagonalization of the RPA matrix
for fermion systems \cite{RS80q,BR86q}.
The reality of the eigenvalues
${\Omega}_{n}$, which follows, as shown above, from the positivity of the matrix
${\mathbb F}$  is well known in that case \cite{TH60q,dClq,Mer63q}.
The present diagonalization is also similar to the diagonalization of a quadratic
Hamiltonian of boson operators \cite{BR86q}.

Inserting the expansions (\ref{MFA48q}) and (\ref{MFA47q}) of ${\mathbb F}$ and ${\mathbb C}$
into the expression (\ref{MFA19Aq}) of the {\it correlation matrix} ${\mathbb B}$ yields
 \begin{eqnarray}  \label{MFA50q}
 {\mathbb B}_{\alpha\beta} =
\sum_{n} {\psi}_{\alpha}^{n}\,
  \frac {\hbar\,{\Omega}_{n}} {1-\exp(-\beta\,\hbar\,{\Omega}_{n})}
   \,  {({\psi}_{\overline{\beta}}^{n})}^{*}
  \\ \nonumber
   + \sum_{n}  {({\psi}_{\overline{\alpha}}^{n})}^{*}  \,
  \frac {\hbar\, {\Omega}_{n}} {\exp(\beta\,\hbar\,{\Omega}_{n}) - 1}
   \,    {\psi}_{\beta}^{n}\,
    +  \frac {1} {\beta} \sum_{p}  {\psi}_{\alpha}^{p}\,  {\psi}_{\beta}^{p} \,.
 \end{eqnarray}
This expression involves only the eigenvalues and right eigenvectors of $i{\mathbb{C}}{\mathbb{F}}$,
normalized according to (\ref{MFA45Aq}).

In the zero-temperature limit, only the first term of (\ref{MFA50q}) survives, and $ {\mathbb B}$ reduces to
\begin{eqnarray}  \label{MFA51q}
 {\mathbb B}_{\alpha\beta} =
\sum_{n} {\psi}_{\alpha}^{n}\, {\hbar\,{\Omega}_{n}} \, {({\psi}_{\overline{\beta}}^{n})}^{*}  \, ,
 \end{eqnarray}
where the parameters ${\Omega}_{n}$ and $ {\psi}_{\alpha}^{n}$ are found from the limit
$\beta \to \infty $ of $i{\mathbb C}{\mathbb F}$. The quasi-scalars do not contribute.

In the high-temperature limit or in the classical limit, ${\mathbb B}={({\beta}{\mathbb F})}^{-1}$ is 
given in diagonalized form by (\ref{MFA46q}). The same holds for the Kubo correlations.

\subsection{Diagonalization of the effective free-energy operator}
 \label{sec8.2}
 
The above diagonalization of the matrix ${\mathbb F}$ will help us to give
an interpretation of the effective Hamiltonian
${\underline{F}}$ defined
in the mapped space $\underline{\mathscr{H}}$ [Eq.(\ref{MOC12q}) of
 Sec.\,\ref{sec7.1}].
Let us introduce, besides the unit operator
${\underline{\sf M}}_{0} = {\underline{\sf I}}$, the following new basis
for the mapped algebra $\{{\underline{\sf M}}\}$ $(\alpha \ne 0)$:
\begin{eqnarray} \label{MFA80q}
&& {\underline{\sf A}}_{n} \equiv
 \frac { {\phi}_n^{\alpha}\, ({\underline{\sf M}}_{\alpha}  - {R}^{(0)}_{\alpha}) }
{ \sqrt{\hbar\,\Omega_n} }\, ,
\;\;\;\;  {\underline{\sf A}}_{-n} = {\underline{\sf A}}_{n}^{\dagger} \, ,
   \\ \nonumber
&& {\underline{\sf Y}}_{p} \equiv {{\phi}}_p^{\alpha}\,
 ({\underline{\sf M}}_{\alpha} - {R}^{(0)}_{\alpha})\sqrt\beta
=  {({\underline{\sf Y}}_{p})}^{\dagger} \,.
\end{eqnarray}
The pair of operators ${\underline{\sf A}}_{n}$, ${\underline{\sf A}}_{-n}$ is associated with each mode
${\Omega_n}$, and the single operator ${\underline{\sf Y}}_{p}$ with each vanishing eigenvalue
of $i{\mathbb C}{\mathbb F}$. Conversely, the original operators of the mapped Lie algebra are decomposed over the modes $n$ and $p$ according to
  \begin{eqnarray}   \label{MFA81q}
 {\underline{\sf M}}_{\alpha}  - {R}^{(0)}_{\alpha}  = \sum_{n}{ \sqrt{\hbar\,\Omega_n} } \,
   [ {\psi}_{\alpha}^{n}\,{\underline{\sf A}}_{n}
     + {({\psi}^{n}_{\overline\alpha})}^{*} \, {\underline{\sf A}}_{n}^{\dagger}   ]
+ \sum_p {{\psi}}_{\alpha}^{p}\, {\underline{\sf Y}}_{p} / \sqrt{\beta} \,.
\end{eqnarray}
The eigenvectors $\psi$ appear as the amplitudes of ${\underline{\sf M}}_{\alpha}$ on the different modes.

The {\it free-energy operator} ${\underline{F}}$ takes in the new basis the diagonal form
   \begin{eqnarray} \label{MFA82q}
  {\underline{{F}}}  =  \sum_{n}  \hbar\,{\Omega}_{n} \, \frac{1}{2}   \left(
     {\underline{\sf{A}}_{n}^{\dagger}} \,   {\underline {\sf{A}}_{n}}
  +  {\underline {\sf{A}}_{n}} \,  {\underline{\sf{A}}_{n}^{\dagger}}  \right)
   +  \sum_{p}\frac {1}{2\beta} \, \underline{\sf{Y}}_{p}^{2} \,.
 \end{eqnarray}
The {\it commutation relations} of the operators
$ {\underline {\sf{A}}_{n}},\,{\underline{\sf{A}}_{n}^{\dagger}}$
and ${\underline{\sf{Y}}_p}$ follow from those ($ [ {\underline{\sf
M}}_{\alpha}, \,{\underline{\sf M}}_{\beta} ] = i\,\hbar\,{\mathbb
C}_{\alpha\beta}$) of the operators $\{{\underline{\sf M}}\}$, from
the diagonal form (\ref{MFA47q}) of the commutation matrix ${\mathbb
C}_{\alpha\beta}$ and from the biorthogonality (\ref{MFA45Aq}) of
the amplitudes $\psi$ and $\phi$. The transformation (\ref{MFA80q})
thus implies
\begin{eqnarray} \label{MFA46mq}
&&  [{\underline{\sf{A}}_n,}\,{\underline{\sf{A}}_{n'}^{\dagger}}] = \delta_{nn'}\, ,
   \\ \nonumber
 &&  [ {\underline{\sf{A}}_n},\,  \, {\underline{\sf{A}}_{n'}}  ]
 = [{\underline{\sf{A}}_n},\,  \,{\underline{\sf{Y}}_p} ]
  = [ {\underline{\sf{Y}}_p} ,  \, {\underline{\sf{Y}}_{p'}}  ]  =0 \, .
\end{eqnarray}

\subsection{A bosonic and scalar algebra} \label{sec8.3}

The mapped algebra $\{\underline{\sf M}\}$ is spanned, according to 
(\ref{MFA81q}), by two sets of operators.
On the one hand, the commutation relations (\ref{MFA46mq}) express that this algebra has a symplectic sector $n$: the set
$\{ \underline {\sf{A}},{\underline{\sf{A}}}^{\dagger}  \}$ is simply an algebra
of bosonic operators,
with single-boson states labelled by the index $n$.
On the other hand, in the sector $p$, the set $\{{\underline{\sf{Y}}}\}$
commute with all other operators of the mapped algebra,
and hence each ${\underline{\sf{Y}}}_{p}$ can be regarded as a scalar random variable.

The form (\ref{MFA82q}) of the free-energy operator $\underline{F}$
in the space $\underline{\mathscr{H}}$ is suggestive. It can be
identified with a Hamiltonian of {\it non-interacting bosons}, with
single-particle states $n$ having the energy $\hbar\,{\Omega}_n$,
plus a {\it classical quadratic part} with a coefficient $1/2{\beta}$ for
each variable ${\underline{\sf{Y}}_p}$.

The evaluation of expectation values ${\langle ... \rangle}_{\rm map}$ involves
traces over the space
$\underline{\mathscr{H}}$ with
the weight $\tilde{\underline{D}} \propto \exp(-{\beta}{\underline{F}})$.
As regards the bosonic part of (\ref{MFA82q}),
$\underline{\mathscr{H}}$ contains a Fock space and the trace is as usual
a summation over the occupation numbers of each single-boson state.
As regards the scalar part,
the trace is meant as the integration
${\prod}_{p} \int d{\underline{\sf{Y}}_p}$. Thus, the expectation values over
$\tilde{\underline{D}} \propto \exp(-{\beta}{\underline{F}})$ of single operators
$ {\underline {\sf{A}}}_{n},{\underline{\sf{A}}}^{\dagger}_{n}$
or $\underline{\sf{Y}}_{p}$ vanish,
while we find for pairs
 \begin{eqnarray} \label{MFA46cq}
 && { \langle   \underline{\sf A}_{n} ^{\dagger} \,
   \underline{\sf A}_{n} \rangle }_{\rm map}
     = \frac {1} { e^{\,\beta\, \hbar\,{\Omega}_{n}} - 1 }
  = { \langle  \underline{\sf A}_{n}\,
     \underline{\sf A}_{n} ^{\dagger}  \rangle}_{\rm map} - 1\, ,
  \\  \nonumber
 &&  {\langle { \underline{\sf Y}_{p} }^{2} \rangle}_{\rm map}
    = {\langle  \underline{\sf M}_{0} ^{2} \rangle}_{\rm map}  = 1 \, ;
 \end{eqnarray}
all other expectation values of pairs vanish. Each mode $n$ yields a Bose factor associated
in canonical equilibrium with the energy $\hbar\,{\Omega}_{n}$; no chemical potential occurs for the distribution
$\underline{\tilde{D}} \propto \exp(-{\beta}\underline{F})$.
For the scalar variables,
the variance $1$ arising from the Gaussian weight
$\exp(- {\underline{\sf Y}_{p} }^{2} / {2} )$
within $\tilde{\underline{D}}$ provides an expectation value
$1/2{\beta}$ for the term
$ { { \underline{\sf{Y}}  }_{p}   }^{2} / {2\beta} $ of $\underline{F}$,
in agreement with the equipartition theorem of classical statistical mechanics.

We are considering in this section the dynamics (\ref{MOC35q}) for $H=K$,
and we have seen that $\underline{F}$ then plays the role of a
Hamiltonian in the space $\underline{\mathscr{H}}$. The diagonal
form (\ref{MFA82q}) of $\underline{F}$ indicates that the mapped
operators $ {\underline {\sf A}}_{n} $ oscillate as
\begin{eqnarray}   \label{MFA42CDq}
 {\underline {\sf A}}_{n}^{\rm H}(t',t) =
 e ^{-i{{\Omega}_n}(t'-t) }{\underline {\sf A}}_{n} \, ,
\end{eqnarray}
while the scalars $ {\underline {\sf Y}}_{p}$ remain constant. Thus, the operators
$ {\underline {\sf{A}}_{n}} \,,  {\underline{\sf{A}}_{n}^{\dagger}} $
behave as bosonic annihilation and creation operators in all respects:
{\it commutation relations} (\ref{MFA46mq}), average {\it occupation in canonical equilibrium} (\ref{MFA46cq})
and {\it dynamics} (\ref{MFA42CDq}).

\subsection{Interpretation in terms of oscillators} \label{sec8.4}

 The mapped bosonic Fock space can equivalently be regarded as a space of oscillators,
each single bosonic state $n$ corresponding to an oscillator mode
with frequency ${\Omega}_n$. The operators $\underline{\sf{A}}_{n},
\underline{\sf{A}}_{n} ^{\dagger}$ are thus replaced by the position
and momentum operators
  \begin{eqnarray} \label{MFA73q}
 \underline{\sf{X}}_{n}= \sqrt{\hbar/2}\,( \underline{\sf{A}}_{n}
     + \underline {\sf{A}}_{n} ^{\dagger})\, ,
   \;\;\;\;\;\;
  \underline{\sf{P}}_{n}= \sqrt{\hbar/2}\,( \underline{\sf{A}}_{n}
     - \underline{\sf {A}}_{n} ^{\dagger} )  / i \,,
  \end{eqnarray}
satisfying the canonical commutation relations $[
\underline{\sf{X}}_{n},\,\underline{\sf{P}}_{n'}] = i\hbar
\delta_{nn'}$. The effective Hamiltonian $\underline{\sf{F}}$ then
takes the form
\begin{eqnarray} \label{MFA74q}
  \underline{\sf{F}}  =  \sum_{n}  {\Omega}_{n} \,
  ( {\underline{\sf {P}}_{n}}^{2} +  {\underline{\sf {X}}_{n}}^{2} )
   + \frac {1}{2\beta}  \sum_{p} \, {\underline{\sf{Y}}_{p}}^{2} \,.
 \end{eqnarray}
It now describes {\it uncoupled harmonic oscillators, plus a quadratic
energy} associated with the classical variables
$\underline{\sf{Y}}_{p}$. In this alternative interpretation, we
have
  \begin{eqnarray}   \label{MFA73Aq}
   {\langle   {\underline{\sf{X}}}_{n}^{2} \rangle}_{\rm map}
  ={\langle   {\underline{\sf{P}}}_{n}^{2} \rangle}_{\rm map}
  = \frac {\hbar} {2}\, \coth \frac { {\beta}\,\hbar\,{\Omega}_n } {2} \,,
  \end{eqnarray}
 and the dynamics describes harmonic oscillators with frequency $ {\Omega}_{n}$.

\subsection{Quasi-bosons and quasi-scalars} \label{sec8.5}

From now on, we return to the original space ${\mathscr{H}}$.
The correspondence between the algebras $\{{\sf M}\}$ and $\{\underline{\sf M}\}$
leads us to perform on the original Lie algebra the same transformation
as (\ref{MFA80q})-(\ref{MFA81q}),
which introduces {\it for a given} ${\DIh}^{(0)}$ {\it a new basis}
$ {{\sf{A}}}_{n}$, \,${{\sf{A}}}_{n}^{\dagger}$,\, ${{\sf{Y}}}_{p}$
(besides ${\sf M}_{0}$) {\it in the algebra} $\{{\sf M}\}$.

The commutators between these operators, issued
from the Lie structure (\ref{MFA00q}), are not simple.
However, the expectation values of these commutators,
evaluated as traces over ${\tilde\DIh}^{(0)}$, have the same structure
as the commutators (\ref{MFA46mq}) of the corresponding mapped operators
${\underline{\sf A}}_{n}$, ${\underline{\sf A}}_{n}^{\dagger}$, ${\underline{\sf Y}}_{p}$,
that is,
\begin{eqnarray} \label{MFA46bq}
&& { \langle [ {\sf A}_{n},\,
    {\sf A}_{n'} ^{\dagger} ]  \rangle  }_{ \rm app }  = \delta_{nn'}\, ,
   \\ \nonumber
 && { \langle [ {\sf A}_{n},\,  \, {\sf A}_{n'}  ] \rangle  }_{\rm app}
 = { \langle [ {\sf A}_{n},\,  \, {\sf Y}_{p}  ] \rangle  }_{\rm app}
  = { \langle [ {\sf Y}_{p} ,\,  \, {\sf Y}_{p'}  ] \rangle   }_{\rm app}=0 \, .
\end{eqnarray}
 The operators ${\sf A}_{n}$, ${\sf A}_{n}^{\dagger}$ can thus be termed
"quasi-boson" annihilation and creation operators,
while the operators ${\sf Y}_{p}$ can be termed "quasi-scalars".

The matrix $i\hbar\,{\mathbb C}_{\alpha\beta}$ was defined as the expectation value
$ {\langle [ {{\sf M}}_{\alpha},\,{{\sf M}}_{\beta} ] \rangle}_{\rm app} $
in the original basis of the Lie algebra. Going to the new basis 
amounts to diagonalize ${\mathbb C}$ according to (\ref{MFA47q}).
In this new basis, $i\hbar\,{\mathbb{C}}$
takes a form involving only $2 \times 2$ diagonal blocks
  $ {\bf  \left(  \begin{array} {cc}
  0 & 1 \\
  -1 & 0
  \end{array} \right)}$ and zeros, in agreement with the above relations.

The variational expectation values in the state ${\tilde\DIh}^{(0)}$ of the 
 operators ${\sf{A}}_{n},\,{{\sf{A}}}^{\dagger}_{n}$ and ${\sf{Y}}_p$ vanish,
\begin{eqnarray}   \label{MFA42Cq}
 { \langle {\sf A}_{n} \rangle }_{{\rm app}}=
 {\langle {\sf A}_{n}^{\dagger} \rangle}_{{\rm app}}
    = {\langle {\sf Y}_{p} \rangle }_{{\rm app}} =  0 \, ,
  \end{eqnarray}
 while ${\langle {\sf M}_{0} \rangle}_{{\rm app}} =1$.
The approximate expectation values of pairs of operators,
 \begin{eqnarray} \label{MFA46aq}
 && { \langle   {\sf A}_{n} ^{\dagger} \, {\sf A}_{n} \rangle }_{\rm app}
     = \frac {1} { e^{\,\beta\, \hbar\,{\Omega}_{n}} - 1 }
  = { \langle  {\sf A}_{n}\, {\sf A}_{n} ^{\dagger}  \rangle}_{\rm app} - 1\, ,
  \\  \nonumber
 &&  {\langle { {\sf Y}_{p} }^{2} \rangle}_{\rm app}
    = {\langle  {\sf M}_{0} ^{2} \rangle}_{\rm app}  = 1 \, ,
 \end{eqnarray}
are the same as the equations (\ref{MFA46cq}) which were exact
in the space $\underline{\mathscr{H}}$;
again all other expectation values of pairs vanish.
We recognize, as in the mapped space,
the Bose factor of a mode $n$ for the quasi-boson operators
${\sf A}_{n},\, {\sf A}_{n} ^{\dagger}$, and the unit variance
for the Gaussian quasi-scalar variables ${\sf Y}_{p} $.

The change of basis
$\{ {\sf M}\}$ $\mapsto$ $\{ {\sf A}, {{\sf A}}^{\dagger}, {\sf Y}\}$
is issued from the diagonalization of Sec.\,\ref{sec8.1}. This change leads 
to re-express the correlation matrix ${\mathbb{B}}$ as
 \begin{eqnarray}  \label{MFA52q}
 {\mathbb B}_{\alpha\beta} =
  {\langle {{\sf M}}_{\alpha}\,{{\sf M}}_{\beta} \rangle}_{\rm app}
    - {R}^{(0)}_{\alpha} \, {R}^{(0)}_{\beta}
   = \sum_{n} {\psi}_{\alpha}^{n}\, \hbar\,{\Omega}_{n} \,
  { \langle  {\sf A}_{n} \, {\sf A}_{n} ^{\dagger}  \rangle}_{\rm app}
   \,  { ( {\psi}_{\overline{\beta}}^{n} ) }^{*}
  \\ \nonumber
    + \sum_{n}  {({\psi}_{\overline{\alpha}}^{n})}^{*}  \, \hbar\,{\Omega}_{n} \,
    { \langle   {\sf A}_{n} ^{\dagger} \, {\sf A}_{n} \rangle }_{\rm app}
   \,    {\psi}_{\beta}^{n}\,
+  \frac {1} {\beta} \sum_{p}  {\psi}_{\alpha}^{p}
{\langle { {\sf Y}_{p} }^{2} \rangle}_{\rm app}
 {\psi}_{\beta}^{p} \,.
 \end{eqnarray}
The Bose-like factors that appeared in the expression (\ref{MFA50q})
of ${\mathbb B}$ now come out as expectation values (\ref{MFA46aq})
of pairs of quasi-boson operators
$ {\sf A}_{n}$ and ${\sf A}_{n'} ^{\dagger}$,
while the last term of (\ref{MFA50q}) is consistent
with the variance equal to $1$ of the quasi-scalar operators ${\sf Y}_{p} $.
The coefficients arise from the change of basis (\ref{MFA81q}) that expresses the operators
$ {{\sf M}}_{\alpha}  - {R}^{(0)}_{\alpha} $ on the basis
$\{ {\sf A}, {{\sf A}}^{\dagger}, {\sf Y} \} $.

 More generally, to write the optimized correlation of arbitrary observables
$Q_j$ and $Q_k$, one should express their images on the basis
$ {{\sf{A}}},{{\sf{A}}}^{\dagger}, {\sf{Y}}$ according to
 \begin{eqnarray} \label{MFA57q}
   {\sf Q}_{j} =
     {q}_{j}\{{R}^{(0)}\} \, {\sf M}_{0}
    +  \sum_{n} [  {\cal Q}^{n}_{j{\rm b}}   \, { {\sf A}}_{n}
           + {( {\cal Q}^{n}_{j{\rm b}} )}^{*}   \,  {{\sf A}}_{n}^{\dagger}  ]
           +   \sum_{p}  {\cal Q}^{p}_{j{\rm s}}   \,  { {\sf Y}}_{p} \, ,
  \end{eqnarray}
 where the new, bosonic and scalar, coordinates are $(\alpha \ne 0)$
 \begin{eqnarray} \label{MFA58aq}
  && {\cal Q}^{n}_{j{\rm b}} = \sqrt {\hbar\,{\Omega}_{n}} \,
   {{\psi}}_{\alpha}^{n}\,  {\cal Q}^{\alpha}_j\{ {R}^{(0)}\}
    \\  \nonumber
  &&   {\cal Q}_{j{\rm s}}^{p}  =  \frac {1} {\sqrt \beta } \,
         {{\psi}}_{\alpha}^{p}\, {\cal Q}^{\alpha}_j\{ {R}^{(0)}\}\, ,
 \end{eqnarray}
(we recall that $ {\cal Q}^{\alpha}_j\{ {R}^{(0)}\} \equiv {\partial
q_{j}\{{R}^{(0)}\} } / { \partial{R}^{(0)}_{\alpha} }$). One thus finds for the {\it correlations}
\begin{eqnarray} \label{MFA60q}
&&  {\langle {Q}_j {Q}_k  \rangle }_{\rm app}
    - {\langle {Q}_j \rangle}_{\rm app}  { \langle {Q}_k  \rangle }_{\rm app}
      = C_{jk}(t_{\rm i},t_{\rm i})
 \\ \nonumber
 &&   =   \sum_{n} \left[
    \coth  \, \frac { \beta\,\hbar\,\Omega_n } {2}
    \, {\rm {Re}}  \, {\cal Q}^{n}_{j{\rm b}} \,  {( {\cal Q}^{n}_{k{\rm b}} )}^{*}
    +  i\,  {\rm {Im}} \,  {\cal Q}^{n}_{j{\rm b}} \,
  {( {\cal Q}^{n}_{k{\rm b}} )}^{*}  \right]
   \\  \nonumber
  && +  \sum_{p}  {\cal Q}^{p}_{j{\rm s}} \,{\cal Q}^{p}_{k{\rm s}} \,.
 \end{eqnarray}
The antisymmetric part of the correlations,
 \begin{eqnarray}  \label{MFA61q}
   \frac {1}{2} {\langle {Q}_j {Q}_k - {Q}_k {Q}_j \rangle}_{\rm app}   =
   \sum_{n} i\, {\rm {Im}}\,{\cal Q}^{n}_{j{\rm b}}\,{( {\cal Q}^{n}_{k{\rm b}} )}^{*} \,,
 \end{eqnarray}
agrees with $ [{\sf Q_j},{\sf Q_k}] =  i\,\hbar\,
    {\cal Q}_j^{\alpha}\{{R}^{(0)}\}
  \,{\mathbb C}_{\alpha\beta}\,
   {\cal Q}_k^{\beta}\{{R}^{(0)}\}$.
An alternative interpretation of the symmetric part is obtained by replacing,
as in Sec.\,\ref{sec8.4},
quasi-bosons by quasi-oscillators with variables ${\sf X}_{n}$, ${\sf P}_{n}$.
Writing the image of $Q_j$ in the basis $\{{\sf{X}},{\sf{P}},{\sf{Y}}\}$
then produces the factors
$({\hbar}/ {2}) \coth ({ {\beta}\,\hbar\,{\Omega}_n }/ {2})$
which are the expectation values
$ {\langle {\sf{X}}_{n}^{2} \rangle}_{\rm app}
 = {\langle {\sf{P}}_{n}^{2} \rangle}_{\rm app}$.

Since $H=K$ here, the approximate Heisenberg operators
in the space ${\mathscr{H}}$ follow the same evolution as (\ref{MFA42CDq})
in the mapped space $\underline{\mathscr{H}}$, that is,
 \begin{eqnarray}  \label{MFA75q}
{{\sf A}}_{n}^{\rm H}(t',t) =  e ^{-i{{\Omega}_n}(t'-t) }{{\sf A}}_{n} \,,
  \;\;\;\;
 {{\sf Y}}_{p}^{\rm H}(t',t) = {{\sf Y}}_{p} \,.
 \end{eqnarray}
Hence, the {\it two-time correlation function}
$ C_{jk}(t',t'') $ is given, when the initial state is at equilibrium, by
 \begin{eqnarray}  \label{MFA126q}
 C_{jk}(t',t'') &= & \sum_{n} \left[{\cal Q}^{n}_{j{\rm b}}  \,
  \frac {e^{\,-\,i\,{\Omega}_{n}(t'-t'')}} {1- e^{\,-\,\beta\,\hbar\,\Omega_n}}\,
                       {( {\cal Q}^{n}_{k{\rm b}} )}^{*}      \right.
  \\ \nonumber
 && + \left.   {( {\cal Q}^{n}_{j{\rm b}} )}^{*}   \,
  \frac { e^{\,i\,{\Omega}_{n}(t'-t'')} } { e^{\,\beta\,\hbar\,\Omega_n} - 1} \,
    {\cal Q}^{n}_{k{\rm b}}  \right]
 +  \sum_{p}  {\cal Q}^{p}_{j{\rm s}} \,{\cal Q}^{p}_{k{\rm s}} \, ,
  \;\;\;\;(t'>t'')\, ,
\end{eqnarray}
which is the explicit form for the general equation (\ref{MOC43q}).
It exhibits the coefficients of the images ${\sf Q_j}$ and ${\sf Q}_k$
on the quasi-boson and quasi-scalar basis,
the Bose factors and the boson dynamics.

\subsection{Linear response and excitation energies} \label{sec8.6}

 The linear response (\ref{MFA22q}) is directly found from (\ref{MFA22Eq})
and (\ref{MFA57q});
it does not involve the quasi-scalar contributions nor the Bose factor.
Its dissipative part, defined through a Fourier transform with respect to $t'-t''$, comes out as
\begin{equation}   \label{MFA128q}
 {\chi}_{jk}^{''}(\omega)   = \frac{\pi}{\hbar}   \sum_{n}          \left[
    {\cal Q}^{n}_{j{\rm b}} \,\delta (\omega - {\Omega}_{n})\,  {( {\cal Q}^{n}_{k{\rm b}} )}^{*}
   -  {( {\cal Q}^{n}_{j{\rm b}} )}^{*}  \,\delta (\omega + {\Omega}_{n})\, {\cal Q}^{n}_{k{\rm b}}
                                                             \right] \,.
\end{equation}
Thus, the present approximation yields the ${\Omega}_n$'s as {\it resonance frequencies}.
Consistency properties such as the
Kramers-Kronig dispersion relations and the Kubo fluctuation-dissipation relations are satisfied.

At the zero-temperature limit, the exact expression of ${\chi}_{jk}^{''}(\omega) $ is given,
in terms of the ground state $\vert 0 \rangle$ of $K$ and of the excited states $\vert {\rm exc} \rangle$
with excitation energies $E_{{\rm exc}}$, as
 \begin{eqnarray} \label{MFA136q}
    {\chi}_{jk}^{''}(\omega)
    =  \pi \sum _{{\rm exc}} [
   \langle 0 \vert Q_{j} \vert {\rm exc} \rangle
   \langle {\rm exc} \vert Q_{k} \vert 0 \rangle  \, \delta(\hbar\omega - E_{{\rm exc}})
  \\    \nonumber
  -   \langle 0 \vert Q_{k} \vert {\rm exc} \rangle  \, \langle {\rm exc} \vert Q_{j} \vert {0} \rangle \,
      \delta(\hbar\omega + E_{{\rm exc}})  ] \,.
 \end{eqnarray}
  Comparison with (\ref{MFA128q}) shows that the positive eigenvalues $\hbar{\Omega}_n$
of $i\hbar{\mathbb{C}}{\mathbb{F}}$ can be identified as approximations
for the excitation energies $E_n$ of some set of states, labelled as $\vert {n} \rangle$.
The amplitude $ {\cal Q}^{n}_{j{\rm b}}$ of the image
${{\sf Q}}_{j}$ over the quasi-boson annihilation operator ${{\sf A}}_{n}$ appears as
an approximation for the matrix element $\langle 0 \vert Q_{j} \vert {n} \rangle $.
In particular, the quasi-boson operators ${{\sf A}}_{n}$ satisfy approximately
 \begin{eqnarray}
   \langle 0 \vert  {{\sf A}}_{n} \vert {n'} \rangle  = {\delta}_{nn'} \, ,
 \end{eqnarray}
as if the state $ \vert {n} \rangle$ were obtained by creating a quasi-boson.

\section{Small deviations} \label{sec9}

We consider in this section static and dynamic changes brought in by a shift of the initial conditions. 

\subsection{Static deviations} \label{sec9.1}

The change in the equilibrium properties arising from a variation $\delta K$
of the operator $K$
that defines the state $\tilde{D} {\propto} \exp(-{\beta}K)$
is accounted for in the variational treatment by the shift
$\delta\tilde\DIh^{(0)}$ of $\tilde\DIh^{(0)}$, parametrized by the set
\begin{equation} \label{qb40q}
\delta{R}_{\alpha}^{(0)}   = {\rm Tr}\, \delta\tilde\DIh^{(0)}\,{\sf M}_{\alpha}
  ={\rm Tr} [ \tilde\DIh^{(0)} + \delta\tilde\DIh^{(0)} ]
            ( {\sf M}_{\alpha} - {R}_{\alpha}^{(0)} ) \, .
\end{equation}
 Denoting by $\delta {\cal K}$ the image of $\delta {K}$, the first-order change
in the self-consistent equations (\ref{qb30bq}) provides
\begin{equation} \label{qb42q}
 {\mathbb S}^{\alpha\beta} \, \delta{R}^{(0)}_{\beta} =
    \beta \, \delta {\cal K}^{\alpha}
  + \beta \, {\mathbb K}^{\alpha\beta} \, \delta{R}^{(0)}_{\beta}  \,,
\end{equation}
that is, using the definition (\ref{MFA20q}) of the matrix ${\mathbb F}$,
\begin{equation} \label{MFA250q}
\delta{R}_{\alpha}^{(0)}  \equiv
 -  ({\mathbb F}^{-1})_{\alpha\beta} \,\delta {\cal K}^{\beta}
\;\;\;\;\;(\alpha,\beta \ne 0) \, .
\end{equation}
The resulting {\it change} $\Delta F$ {\it of the free energy}
$F \simeq {\rm min}f\{{R}\} = {\rm min}[k\{{R}\}-T\,{S}\{{R}\}]$,
expanded up to second order in $\delta{K}$, is found as
 \begin{eqnarray} \label{qb44q}
 &&  \Delta F  \approx    \delta {k}  + \frac {1}{2}\, \delta{R}^{(0)}_{\alpha} \,
                       {\mathbb F}^{\alpha\beta} \, \delta{R}^{(0)}_{\beta}
                   + \delta {\cal K}^{\alpha} \, \delta{R}^{(0)}_{\alpha}
  \\ \nonumber
 && = \delta k
    - \frac {1}{2}\, \delta{R}^{(0)}_{\alpha} \,
                       {\mathbb F}^{\alpha\beta} \, \delta{R}^{(0)}_{\beta}
               =  \delta k - \frac {1}{2} \delta {\cal K}^{\alpha} \,
  ( {\mathbb F}^{-1})_{\alpha\beta} \, \delta {\cal K}^{\beta}  \, .
  \end{eqnarray}
In (\ref{qb42q})-(\ref{qb44q}), the matrices
${\mathbb K}$, ${\mathbb S}$, $ {\mathbb F}$,
the image $\delta {\cal K}$ and the symbol $\delta{k}$ of $\delta{K}$
are taken at $\{{R}^{(0)}\}$.
In particular, the expression (\ref{qbCTq}) of the heat capacity
is recovered by taking $\delta{\sf K}={\sf K}\delta\beta/\beta$.

This shift takes a suggestive form if $ {\mathbb F}^{\alpha\beta}$ is replaced
by its diagonalized form (\ref{MFA48q}). Using (\ref{qb40q}) and (\ref{MFA42Cq}), and denoting as
 \begin{eqnarray} \label{qb48q}
 && \delta{R}_{n{\rm b}}^{(0 )}
    = {\rm Tr}\,[ \tilde\DIh^{(0)} + \delta\tilde\DIh^{(0 )} ] \, {\sf A}_n
                =  {\rm Tr}\, \delta\tilde\DIh^{(0 )} \, {\sf A}_n\, ,
                 \\ \nonumber
 && \delta{R}_{n{\rm b}}^{(0)*} =  {\rm Tr}\, \delta\tilde\DIh^{(0 )} \, {\sf A}_n^{\dagger}\, ,
    \;\;\;\;
    \delta{R}_{p{\rm s}}^{(0 )}   =  {\rm Tr}\, \delta\tilde\DIh^{(0 )} \, {\sf Y}_p
\end{eqnarray}
the variations of the symbols of the quasi-boson and quasi-scalar operators
${\sf A}_n, {\sf A}_n^{\dagger}$ and $ {\sf Y}_p$,
one finds from the  variation of $f\{{R}\}$ the form
 \begin{eqnarray} \label{qb50q}
 &&\Delta{F}  \approx  \delta{k}
  - \frac {1}{2}\,\delta{R}^{(0)}_{\alpha}\,{\mathbb F}^{\alpha\beta}\,\delta{R}^{(0)}_{\beta}
  \\ \nonumber
 && =  \delta{k}  -   \sum_{n}  \hbar\,\Omega_{n} \, \vert\delta   {R}_{n{\rm b}}^{(0 )}   \vert^{2}
  - \sum_{p} \frac{1}{2\beta} \, {( \delta{R}_{p{\rm s}}^{(0 )}  )} ^{2}\,.
\end{eqnarray}
Each oscillator mode provides a contribution
$\hbar\,\Omega_{n}$ weighted by the amplitude
${\vert \delta{R}_{n{\rm b}}^{(0 )}  \vert}^{2}$, while each quasi-scalar
mode brings in the equipartition contribution ${(2\beta)}^{-1}$.

\subsection{Dynamic deviations} \label{sec9.2}

Let us turn to the change in the time-dependent expectation values induced
by a shift $\tilde\DIh^{(0)} \mapsto \tilde\DIh^{(0)} + \delta\tilde\DIh^{(0)}$
around the equilibrium state $\tilde\DIh^{(0)}$.
The resulting small deviations
$\delta {R}_{\alpha}^{(0)}(t)$ around ${R}_{\alpha}^{(0)}(t)$
are governed by the set of equations
\begin{eqnarray}    \label{MFA12Aq}
\frac {d\,  \delta {R}_{\alpha}^{(0)}(t)  } {dt}  =
  \left(  {\mathbb L} + {\mathbb C}\, {\mathbb H}  \right) _{\alpha}^{\,\beta}
   \delta {R}_{\beta}^{(0)}(t) \, ,
\end{eqnarray}
obtained by varying ${R}_{\alpha}^{(0)}(t)$ in Eq.(\ref{MFA13q});
the initial conditions $\delta{R}_{\alpha}^{(0)}(t_{\rm i})$ are given
by (\ref{qb40q}).
[For the fermionic single-particle Lie-algebra, (\ref{MFA12Aq})
is the dynamical RPA equation issued from the TDHF equation.]
The kernel ${\mathbb L}+ {\mathbb C}{\mathbb H}$ governing the forward
equation (\ref{MFA12Aq}) for
the small deviations turns out to be the dual of the kernel governing
the backward equation
(\ref{MFA16Bq}) for the approximate Heisenberg observables
${\cal Q}^{{\rm H}}_{j}(t',t)$.
The shift $\delta {\langle{Q_j}\rangle}_t$, variationally given by
$\delta {\langle{Q_j}\rangle}_t=
{\cal Q}^{\alpha}_{j}\{{R}^{(0)}(t)\}\delta{R}^{(0)}_{\alpha}(t)$,
is therefore equal to
\begin{eqnarray}    \label{qb51q}
\delta {\langle{Q_j}\rangle}_t  \simeq
{\cal Q}^{{\rm H}\alpha}_{j}(t,t'')\,\delta{R}^{(0)}_{\alpha}(t'')
\end{eqnarray}
for any intermediate time $t''$.
This independence on $t''$ agrees with the expression (\ref{MFA02q})
of ${\langle{Q_j}\rangle}_t$.

\subsection{Dynamic and static stability} \label{sec9.3}

We now specialize to the case $H=K$, for which an initial shift
$\{\delta{R}^{(0)}\}$ generates a deviation $\{\delta{R}^{(0)}(t)\}$
of $\{{R}^{(0)}(t)\}$ around the fixed value $\{{R}^{(0)}\}$.
The linearized equations (\ref{MFA12Aq}) then involve the constant kernel
$ {\mathbb L} + {\mathbb C}\, {\mathbb H}= {\mathbb C}\, {\mathbb F}$,
and can be solved as
 \begin{equation}   \label{MFA241q}
 \delta {R}_{\alpha}^{(0)}(t)  =
 \left[  e^{\,(t-t_{\rm i})\,{\mathbb C}\,{\mathbb F}}\right]_{\alpha}^{\;\beta}
  \delta{R}_{\beta}^{(0)} .
 \end{equation}
Here, as in the backward equations (\ref{MOC30q}),(\ref{MOC31q}), the dynamics
is governed by the product ${\mathbb C}\, {\mathbb F}$, so that the result
(\ref{MFA241q}) can alternatively be written as
  \begin{equation}   \label{MFA244q}
 \delta{R}_{\alpha}^{(0)}(t) =
 \left[  e^{\,(t-t_{\rm i})\,{\mathbb C}\,{\mathbb F}}\right]_{\alpha}^{\;\beta}
      {\rm Tr}\, \delta\tilde\DIh^{(0)} ( {\sf M}_{\beta} - {R}_{\beta}^{(0)} )
   = {\rm Tr}\, \delta\tilde\DIh^{(0)}
            [ {\sf M}_{\alpha}^{\rm H}(t,t_{\rm i}) - {R}_{\alpha}^{(0)} ] .
  \end{equation}
Within the variational treatment we recover for small deviations the equivalence
between Schr\"odinger and Heisenberg pictures, already exhibited in Eq.(\ref{MFA02q}).

If the matrix ${\mathbb F}$ is positive, one can change the basis
so as to diagonalize $i{\mathbb C}\,{\mathbb F}$ according to (\ref{MFA49q})
and use the new variables
$\delta{R}_{n{\rm b}}^{(0 )} = {\rm Tr}\,\delta\tilde\DIh^{(0 )}{\sf A}_n$
and
$\delta{R}_{p{\rm s}}^{(0 )} = {\rm Tr}\,\delta\tilde\DIh^{(0 )}{\sf Y}_p$,
which yields
\begin{eqnarray}   \label{MFA242q}
  && \delta{R}_{n{\rm b}}^{(0)}(t) =
          e^{\,-\,i\,{\Omega}_{n}\,(t-t_{\rm i})} \, \delta{R}_{n{\rm b}}^{(0)}\, ,
            \\      \nonumber
  && \delta{R}_{p{\rm s}}^{(0)}(t)  =    \delta{R}_{p{\rm s}}^{(0)} \,.
  \end{eqnarray}
These evolutions merely reflect the motion (\ref{MFA75q}) of the Heisenberg
operators $\{ {{\sf A}}^{\rm H},\,{{\sf Y}}^{\rm H}\}$.
Again the modes are decoupled, quasi-bosons (or oscillators) undergo
pure oscillations and quasi-scalars are static.

The above sinusoidal form of the dynamics entails Lyapunov stability.
This property is defined, according to \cite{Mac87}, as follows:
"The equilibrium point $x_{0}$ is said to be Lyapunov stable
if given any neighborhood $U$ of $x_{0}$, there is a sub-neighborhood
$V$ of $x_{0}$ such that if $x$ lies in $V$ then its orbit remains
in $U$ forever." In other words, the trajectories tend uniformly
to $\{{R}^{(0)}\}$ as their
initial point $\{{R}^{(0)}(t_{\rm i})\}$ tends to $\{{R}^{(0)}\}$.
Hence, the Lyapunov stability of the linearized motion is ensured if all eigenvalues
of ${\mathbb F}$ are positive, that is, if the approximate free energy
associated with $\tilde\DIh^{(0)}$ is a local minimum of $f\{{R}\}$.
Such a property is well known for fermions, both at zero \cite{TH60q} and
at finite temperature \cite{dClq,Mer63q}, in which case the minimization of
the Hartree-Fock (free) energy entails the reality of the RPA modes.
We find here, in a general variational context for any trial Lie group,
that the static stability of the approximate "state" $ {\DIh}^{(0)}$ 
implies the dynamic stability of motions  $\delta{R}^{(0)}(t)$ around it.

The matrix ${\mathbb F}$ may have vanishing eigenvalues. This occurs, for instance,
if a continuous invariance is broken by the approximation $\tilde\DIh^{(0)}$ for the 
equilibrium state; in this case $f\{{R}\}$ is minimum for a continuous set
of solutions $\{{R}^{(0)}\}$. Since ${\mathbb F}$ is then not invertible,
$i{\mathbb F}^{1/2}{\mathbb C}{\mathbb F}^{1/2}$ is no longer defined and
it is not ascertained that the matrix $i{\mathbb C}{\mathbb F}$ is diagonalizable
(some right eigenvectors may be missing). If $i{\mathbb C}{\mathbb F}$ 
is diagonalizable, all its vanishing eigenvalues
yield a constant contribution to the set $\delta{R}^{(0)}(t)$ so that Lyapunov
stability is still ensured. However, if it is not, contributions behaving as powers
of $t$ come out, so that the dynamics is unstable (for a more detailed discussion,
see Appendix B of \cite{BV89q}). Such a behavior
may be associated with Goldstone modes.
For instance, if the original problem is translationally invariant and if
this invariance is broken by a localized solution  ${\tilde{\DIh}}^{(0)}$,
a small shift in the initial conditions produces an instability,
characterized by a boost at some constant velocity.

\section{Classical structure of large amplitude motion} \label{sec10}

We have written in Sec.\,\ref{sec4.2} several equivalent dynamical equations
for the time-dependent expectation values $\{{R}^{(0)}(t)\}$ of the
operators $\{{\sf M}\}$. We now analyse the structure of these equations, 
which will turn out to have a classical form for any Lie group.

\subsection{Poisson structure} \label{sec10.1}

Let us consider $\{{R}\}$ as a set of classical dynamical variables, and let us 
show that the tensor
${\mathbb C}_{\alpha\beta}\{{R}\} \equiv {\Gamma}_{\alpha\beta}^{\;{\gamma}}\,{R}_{\gamma}$
can be regarded as the generator for this set of a Poisson structure 
issued from the Lie algebra (\ref{MFA00q}).
We first recall the definition of Poisson structures \cite{Wei83,Wei84}.
Consider some set of dynamical variables $\{{x}\}$ parametrizing points
on a manifold, and functions $f,g,h,...$ of these variables.
A Poisson structure is a mapping $\lpl {f}, {g} \lpr = {k}$
from a pair of functions $ {f}, {g} $ to a third function $ {k}$,
which obeys the following rules:

(i) bilinearity;

(ii) antisymmetry:  $\lpl {f}, {g} \lpr =  - \lpl {g}, {f} \lpr $;

(iii) Jacobi identity: $\lpl \lpl {f}, {g} \lpr, k\lpr
 + \lpl \lpl {k}, {f} \lpr, g\lpr + \lpl \lpl {g}, {k} \lpr, kf\lpr =0  $;

(iv) Leibniz derivation rule:
$\lpl {fg}, {k} \lpr =  f \lpl {g}, {k} \lpr +  \lpl {f}, {k} \lpr g $\,.

We consider here functions $g\{{R}\}$ of the variables $\{{R}\}$,
and define a Poisson structure through
 \begin{eqnarray}  \label{LP10q}
 \lpl {R}_{\alpha}, {R}_{\beta}  \lpr  =
  {\mathbb C}_{\alpha\beta}\{{R}\} \equiv {\Gamma}^{\gamma}_{\alpha\beta}{{R}_{\gamma}}
 \end{eqnarray}
and
  \begin{eqnarray}  \label{LP12q}
   \lpl g_{1}, g_{2} \lpr   =
     \frac  {  \partial {g_{1}} }
            { \partial {R}_{\alpha} }
   \,{\mathbb C}_{\alpha\beta}\{{R}\} \,
     \frac  {  \partial {g_{2}} }
            {  \partial {R}_{\beta} } \, .
   \end{eqnarray}

One can readily check that the above rules are satisfied, owing in particular to the 
properties of the structure constants ${\Gamma}^{\gamma}_{\alpha\beta}$, namely, 
antisymmetry and Jacobi identity. The Poisson structure (\ref{LP10q})-(\ref{LP12q}) 
is thus generated by the Lie algebra of the set $\{{\sf M}\}$.

The equations of motion (\ref{MFA13q}) for $\{{R}^{(0)}(t)\}$
can now be rewritten as
 \begin{eqnarray}   \label{LP18Aq}
 \frac{d{R}^{(0)}_{\alpha}(t)}{dt}=
 \lpl  {R}^{(0)}_{\alpha}(t),\;h\{{R}^{(0)}(t)\}  \lpr \, .
 \end{eqnarray}
These quantum variational equations are therefore identified with classical
equations involving the brackets (\ref{LP12q}) and
governed by a classical Hamiltonian $h\{{R}^{(0)}(t)\}$,
the {\it symbol of the quantum Hamiltonian} $H$.

The relation (\ref{MFA125q}) expresses the time-dependent expectation value
of the observable $\langle Q_j \rangle_t$ as a function of $\{{R}^{(0)}(t)\}$;
together with (\ref{LP18Aq}), it implies that $\langle Q_j \rangle_t$ evolves
according again to the classical dynamical equation
 \begin{eqnarray}   \label{LP14q}
 \frac{ d\langle Q_j \rangle_t  }{dt}=
 \lpl  q_{j}\{{R}^{(0)}(t)\},  \; h\{{R}^{(0)}(t)\}  \lpr \, ,
 \end{eqnarray}
which involves the symbols of $Q_{j}$ and $H$.
The variational approach, together with the introduction of symbols,
thus generate approximate dynamics of expectation
values  that have a classical structure, generated
by the Lie-Poisson bracket (\ref{LP12q}) and by a Hamiltonian.

The symbol $ h\{{R}^{(0)}(t)\}$ is obviously a constant of the motion.
Moreover, the von Neumann entropy $ S\{{R}^{(0)}(t)\}$ defined by (\ref{MFA8Dq})
is also a constant of the motion.
Indeed, using (\ref{MFA9A20Bq}) then (\ref{MFA10dq}), one finds
 \begin{eqnarray}   \label{LP20q}
 \lpl  {R}^{(0)}_{\beta}(t) ,  \; S\{{R}^{(0)}(t)\} \lpr =
   {\mathbb C}_{\beta\gamma}\{{R}^{(0)}\} \,
   \frac  { \partial  S\{{R}^{(0)}(t)\} }   { \partial {R}^{(0)}_{\gamma}(t) } =
 -\, {\mathbb C}_{\beta\gamma}\{{R}^{(0)}(t)\} \, {J}^{\gamma(0)}(t)  = 0  \, ,
 \end{eqnarray}
which implies that $ S\{{R}^{(0)}(t)\} $ remains constant during the evolution
(\ref{LP18Aq}) of $\{{R}^{(0)}(t)\}$.

Lie-Poisson structures for dynamical equations
issued from a variational principle have been recognized
in cases such as the Vlasov equation \cite{Wei83},
time-dependent Hartree-Fock equations \cite{BV89q},
time-dependent Hartree equations for bosons \cite{BF99q}
and for ${\phi}^{4}$ field theory \cite{CMq,MB98q}.

Proposals of non-linear extensions of quantum mechanics \cite{We89q}
have suggested a formulation in terms of a Poisson structure \cite{Jo93q}.
Here it is the restriction of the algebra of observables to the trial Lie algebra
which produces a Poisson structure within standard quantum mechanics.

\subsection{Canonical variables} \label{sec10.2}

We have seen in Secs.\,\ref{sec8.2} and \ref{sec8.4} that the diagonalization of
${\mathbb C}{\mathbb F}$ generates in the mapped space $\underline{\mathscr{H}}$
a linear transformation (\ref{MFA80q}),(\ref{MFA73q})
of the operators $\{{\underline{\sf M}}\}$,
which produces pairs of canonically conjugate operators
$\underline{\sf{X}}_{n}$, $\underline{\sf{P}}_{n}$
and scalars $\underline{\sf{Y}}_{p}$.
This corresponds in the original Hilbert space to a construction
of quasi-oscillator operators ${\sf{X}}_{n}$, ${\sf{P}}_{n}$
and quasi-scalars ${\sf{Y}}_{p}$ (Sec.\,\ref{sec8.5}).
Accordingly, the expectation values of their small deviations, defined by
 \begin{eqnarray} \nonumber
 && \delta{{{X}}_{n}(t)} \equiv  \sqrt{\hbar/2} \,
 [ \delta{R}_{n{\rm b}}^{(0 )}(t) + \delta{R}_{n{\rm b}}^{(0)*}(t) ]
   \equiv \sqrt{\hbar/2} \,
  {\rm Tr}\,\delta\tilde\DIh^{(0 )}(t)  \,
   ({\sf A}_{n} +{\sf A}_{n}^{\dagger} )\, ,
 \\ \nonumber
 && \delta{{P}}_{n}(t) \equiv  \sqrt{\hbar/2} \,
 [ \delta{R}_{n{\rm b}}^{(0 )}(t) - \delta{R}_{n{\rm b}}^{(0)*}(t) ] / i
       \equiv  \sqrt{\hbar/2}\,
  {\rm Tr}\,\delta\tilde\DIh^{(0 )}(t)  \,
   ({\sf A}_{n} -{\sf A}_{n}^{\dagger} ) / {i}\, ,
 \end{eqnarray}
evolve according to (\ref{MFA242q}) as conjugate variables of classical
harmonic operators with frequency ${\Omega}_n$, while the quantities
$\delta{{Y}}_{p} \equiv  \delta{R}_{p{\rm s}}^{(0 )}
\equiv {\rm Tr}\, \delta\tilde\DIh^{(0 )}(t){{Y}}_p $ remain constant
since  $\lpl Y_{p}, h \lpr=0 $ for any $h$.
A classical symplectic structure thus appears for linearized motions,
with ordinary Poisson brackets $\{ \delta{{X}}_{n}, \delta{{P}}_{n'}\}
    = \delta_{nn'} \,,
    \;
    \{ \delta{{X}}_{n}, \delta{{X}}_{n'}\} =
    \{ \delta{{P}}_{n}, \delta{{P}}_{n'}\} = 0 \, $
and Hamiltonian $ (1/2) \sum_{n}  {\Omega}_{n} \,
  ( {{{P}}_{n}}^{2} +  {{{X}}_{n}}^{2} ) $ \,.

The above canonical classical structure pertains to the dynamics of small
amplitude motions of $\{{R}^{(0)}(t)\}=\{{R}^{(0)}\} + \{\delta{R}^{(0)}(t)\}$
around equilibrium $\{{R}^{(0)}\}$. We have seen however (Sec.\,\ref{sec10.1}) that a more
elaborate Poisson structure occurs {\it for large amplitude motions}.
In this case, the bracket
$\lpl{R}_{\alpha}, {R}_{\beta}\lpr \equiv {\mathbb C}_{\alpha\beta}\{{R}\} $
depends on the dynamical variables, whereas for linearized motions we had
simply to diagonalize the constant matrix ${\mathbb C}_{\alpha\beta}\{{R}^{(0)}\}$
according to (\ref{MFA47q}). Moreover, the motion is no longer harmonic.

Nevertheless, one can re-express the classical dynamics of Sec.\,\ref{sec10.1} 
in terms of canonical variables by means of a non-linear change of variables
in the space $\{{R}\}$. Indeed, a "splitting theorem" \cite{Wei83} states
than an arbitrary Poisson structure can {\it locally} be split into
symplectic components and invariant components: There exist independent
(non-linear) functions ${\xi}_n,{\pi}_n,{\eta}_p$ of the coordinates $\{{R}\}$
such that their brackets (\ref{LP12q}) reduce to
 \begin{eqnarray} \label{CV14q}
 && \lpl {\xi}_{n}, {\pi}_{n'} \lpr = \delta_{nn'}\,,
  \;\;\;\;\;
  \lpl {\xi}_{n}, {\xi}_{n'} \lpr =
  \lpl {\pi}_{n}, {\pi}_{n'} \lpr = 0 \,,
   \\ \nonumber
  &&   \lpl {\xi}_{n}, {\eta}_{p} \lpr =
       \lpl {\pi}_{n}, {\eta}_{p} \lpr =
       \lpl {\eta}_{p}, {\eta}_{p'} \lpr = 0\,.
 \end{eqnarray}
In general such a transformation does not exist globally but only locally
in some neighborhood around each point of the trajectory of the point
$\{{R}^{(0)}(t)\}$ in the space $\{{R}\}$. The dynamical variables $\xi_n$
and $\pi_n$ are canonically conjugate in the elementary sense
(their brackets $\lpl\,,\,\lpr$ reduce to ordinary Poisson brackets).
These variables are dynamically coupled and their motion (\ref{LP14q})
is governed by Hamilton's equations, while the variables ${\eta}_p$
(the Casimir invariants) are structurally conserved in the flow.

The construction of the set $\{{\xi},{\pi},{\eta} \}$ is not simple and
does not result from the mere diagonalization of
${\mathbb C}_{\alpha\beta}\{{R}\}$ at each point, contrary to the
linearized dynamics.
In fact, the reduction of the Lie-Poisson structure to the form (\ref{CV14q})
has been achieved only in special cases; relevant to the present work are
the Vlasov equation for which a rigorous proof has been given \cite{Wei83}, the time-dependent
Hartree-Fock theory at zero \cite{MB98q,Ker76,Mar76,Bla78,Yam87} and finite \cite{BV89q} temperatures.

\subsection{Stability of equilibrium and of non-linearized motions}  \label{sec10.3}

In Sec.\,\ref{sec9.3} we have shown that small amplitude motions around thermodynamic
equilibrium $\{{R}^{(0)}\}$ governed by the linearized equations 
for $\{{R}^{(0)}(t)\}$ (with $H=K$) are Lyapunov stable if the trial
free-energy function $f\{{R}\}$ is minimum at $\{{R}^{(0)}\}$.
Let us show that this stability property also holds for motions around equilibrium
governed by the non-linearized equations of motion (\ref{MFA13q}).

To this aim, we rely on the classical form (\ref{LP18Aq}) of this equation
written in the Poisson formalism. We note moreover that, owing to the property
(\ref{LP20q}) of the entropy function $S\{{R}^{(0)}(t)\}$, the addition of
$-{\beta}^{-1}S\{{R}^{(0)}(t)\}$ to the classical Hamiltonian
$h\{{R}^{(0)}(t)\}=k\{{R}^{(0)}(t)\}$ does not affect the dynamics,
so that the free-energy function also generates the large amplitude
trajectories of $\{{R}^{(0)}(t)\}$:
 \begin{equation}   \label{MFA150q}
 \frac { d{R}^{(0)}_{\alpha}(t) } {dt} =
 \lpl  {R}^{(0)}_{\alpha}(t),\;f\{{R}^{(0)}(t)\}  \lpr
   \equiv
  {\mathbb C}_{\alpha\beta} \{ {R}^{(0)}(t) \} \,
   \frac  {  \partial {f}  \{ {R}^{(0)}(t) \}   }
          {  \partial {R}^{(0)}_{\beta}(t)   } \,.
 \end{equation}

The time-dependence of $f\{{R}^{(0)}(t)\}$ is given, according to (\ref{LP14q}),
by $df\{{R}^{(0)}(t)\}/dt=\lpl f\{{R}^{(0)}(t)\},\;f\{{R}^{(0)}(t)\}\lpr=0$.
Hence $f\{{R}^{(0)}(t)\}$ remains constant along any trajectory.
On the other hand, if $\{{R}^{(0)}\}$ is a stable static equilibrium point,
$f\{{R}\}$ has an isolated minimum equal to $f\{{R}^{(0)}\}=F$.
These two conditions ensure Lyapunov stability
according to the Lagrange-Dirichlet theorem (see for instance \cite{Wei84}):
the trajectory $\{{R}^{(0)}(t)\}$ uniformly tends to the point $\{{R}^{(0)}\}$
if the initial value $f\{{R}^{(0)}(t_{\rm i})\}$ tends to $F$.

Thus, for non-linearized as well as for linearized motions
of $\{{R}^{(0)}(t)\}$, the present variational approximations preserve
the following property of the exact dynamics near an equilibrium state:
{\it the static stability}, $f\{{R}\}$ minimum at $\{{R}\}=\{{R}^{(0)}\}$,
{\it entails the Lyapunov stability of the dynamics} in some neighborhood.

 \section  {R\'esum\'e of outcomes and directions for use} \label{sec11}

 We have dwelt at length on the derivation of the results of a
 variational approach based on the principles of Sec.\,\ref{sec2} and on the
 restriction of trial spaces to Lie groups. We summarize below some of
 the formal outcomes thus obtained; they are sufficient for practical
 applications to specific many-body problems.

\subsection {General features} \label{sec11.1}

 The initial step consists in selecting, among the whole set of
 operators in Hilbert space, a Lie algebra spanned by a set of
 operators $\{ {\sf M} \}$ labeled by the index ${\alpha}$. (We include in this Lie algebra 
 the unit operator $I$, denoted as $ {\sf M}_{0}$.) We gave above as
 seminal example a system of fermions for which the set $\{ {\sf M} \}$
 encompasses the single-particle operators
 $a^{\dagger}_{\mu}a_{\nu}$, the index ${\alpha}$ denoting the pair
 $({\nu},{\mu})$. The approach can be applied to many other systems.
 For fermions with pairing, the Lie algebra $\{ {\sf M} \}$ includes the
 operators  $a_{\mu}a_{\nu}$ and $a^{\dagger}_{\nu}a^{\dagger}_{\mu}$ 
(with ${\mu}>{\nu}$); here the index ${\alpha}$
 denotes not only the pair $({\nu},{\mu})$ but also distinguishes
 between the operators $a^{\dagger}_ {\mu}a_{\nu}$, $a_{\mu}a_{\nu}$
 and $a^{\dagger}_{\nu}a^{\dagger}_{\mu}$. For
 bosons with possible condensation, one can take as basis $\{ {\sf M} \}$
 for the Lie algebra the operators
 ${I}$, $a_{\mu}$, $a^{\dagger}_{\mu}$, 
$a^{\dagger}_{\mu}a_{\nu}$, $a_{\mu}a_{\nu}$ (${\mu} \ge {\nu}$)
 and $a^{\dagger}_{\nu}a^{\dagger}_{\mu}$ (${\mu} \ge {\nu}$);
 coherent states arise from the inclusion of the operators
$a_{\mu}$ and $a^{\dagger}_{\mu}$. If such a bosonic system is
translationally invariant, the algebra can be reduced to
${I}$, $a_{0}$, $a^{\dagger}_{0}$, $a^{\dagger}_{k}a_{k}$, $a_{k}a_{-k}$,
$a^{\dagger}_{-k}a^{\dagger}_{k}$ (where $k$ denotes the momentum of
single-particle states). The formalism also applies to other
sub-algebras, or to other systems such as mixtures of fermions and
bosons, spin systems or quantum fields as in \cite{CMq}, the only
restriction being the Lie algebraic structure of the set $\{ {\sf M} \}$. 
In all such cases the algebra $\{ {\sf M} \}$ is characterized by the
structure constants $\Gamma_{\alpha\beta}^{\gamma} $ entering the
commutation relations
\begin{equation} \label{MFA01bq}
 [ {\sf M}_{\alpha}, {\sf M}_{\beta}]
   = {i\,\hbar\,} \Gamma_{\alpha\beta}^{\gamma} \, {\sf M}_{\gamma}.
\end{equation}

As we wish to derive static or dynamic properties of some observables $Q_j$ 
of interest, we have relied on the variational evaluation of a generating 
functional (Secs.\,\ref{sec2.1} and \ref{sec2.3}), so as to deal simultaneously and consistently 
with all such properties. This has entailed the introduction of two sets of trial 
objects, namely, trial "generating operators" $\AIh$ that depend on the observables 
$Q_j$ and their associated (time-dependent) sources, and trial density operators $\DIh$. Both 
$\AIh$ and $\DIh$ are elements of the Lie group generated by the chosen Lie algebra 
$\{ {\sf M} \}$, that is, exponentials of elements of this algebra. We focused on expectation values 
and pair correlations of the observables $Q_j$; they are found by expansion of 
the generating functional in powers of the sources. We summarize below the results thus obtained.  

We parametrize the trial operators $\DIh$, which behave as non-normalized density
operators, by their normalization $ Z = {\rm Tr}\,\DIh $ and
by the numbers $ {R}_{\alpha} =  {\rm Tr}\,{\sf
M}_{\alpha}\,\tilde\DIh $ associated with the operators
${\sf M}_{\alpha}$ $(\alpha \ne 0)$, where $\tilde\DIh$
denotes the normalized operator $Z^{-1}\DIh$. 
The exponential form of the operator $\DIh$ is suited to investigate
systems at non-zero temperature; ground state problems are dealt 
with by taking a zero-temperature limit. We may thus approximately answer
questions about an equilibrium (unnormalized) state
$D=\exp(-{\beta}K)$; for a system in canonical equilibrium, $K$ is
the Hamiltonian. Other choices of $K$ allow us to deal with
non-equilibrium problems, $D=\exp(-{\beta}K)$ being then the initial state. 

An essential tool consists in the representation of the observables
$Q_j$, of the operator $K$ entering $D=\exp(-{\beta}K)$ and of the
Hamiltonian $H$ by their symbols $q_j\{{R}\}$, $k\{{R}\}$ and $h\{{R}\}$ 
within the Lie group (Sec.\,\ref{sec3.2}). Defined for any operator $Q$ and an 
arbitrary element $\tilde\DIh$ of the Lie group as
\begin{equation}  \label{resu11q}
  q\{{R}\} \equiv  {\rm Tr}\,Q\,\tilde\DIh \,,
\end{equation}
a symbol is a function of the scalar variables ${R}_{\alpha}$ ($\alpha \ne
0$) that parametrize $\tilde\DIh$. The practical implementation of
the approach is based on the possibility of evaluating explicitly
such symbols $q\{{R}\}$. (In the fermionic and bosonic examples,
this is feasible owing to Wick's theorem.) One also needs to express
in terms of the variables ${R}_{\alpha}$ the entropy function $
S\{{R}\} \equiv - {\rm Tr}\,\tilde\DIh \ln \tilde\DIh $.

\subsection {Static quantities} \label{sec11.2}

The generalized free energy
$F \equiv -\, {\beta}^{-1} \,\ln{\rm Tr}\,\exp(\,-\,{\beta}\,K)$ 
and the thermodynamic quantities issued from it are
obtained (Sec.\,\ref{sec4.1}) by looking for the minimum of the trial
free-energy function $f\{{R}\}  \equiv  k\{ {R} \} -T\,S\{ {R} \}$
(where $k\{{R}\}$ is the symbol of $K$); the values of 
${R}_{\alpha} = {R}_{\alpha}^{(0)}$ at this minimum are given by 
$\frac {\partial f\{ {R} \}}  {\partial{R}_{\alpha}} = 0$. 
The resulting, consistent, approximations are then, for the free energy:
\begin{eqnarray}  \label{sta10q}
F \equiv  -\,  {\beta}^{-1} \ln {\rm Tr}\,e^{\,-\,{\beta}{K}}
 \simeq   f\{ {R}^{(0)} \}
 = -\, {\beta}^{-1} \ln {\rm Tr}\,
 e^ { \left[ \,-\,{\beta} \,
 \frac { \partial k\{ {R}^{(0)} \} }  { \partial{R}_{\alpha}^{(0)}} 
\,  \sf M_{\alpha} \right]  }  \, ,
\end{eqnarray}
for the entropy:
\begin{eqnarray}  \label{stat12q}
 S \simeq
   S\{ {R}^{(0)} \}
    = - \, \frac { \partial f\{ {R}^{(0)} \} }  { \partial{T}  } \, ,
  \end{eqnarray}
for the energy (when $K$ is the Hamiltonian):
  \begin{equation}    \label{stat14q}
  \langle K \rangle  \simeq  k \{ {R}^{(0)} \} \, ,
 \end{equation}
and for the expectation values of the observables $Q_j$ in the state 
$\tilde{D}\propto \exp{(-{\beta}K)}$ (Sec.\,\ref{sec4.2}):
 \begin{equation}    \label{stat16q}
  \langle Q_j \rangle  \equiv
  \frac { {\rm Tr}\,Q^{\rm S}_j\, e^{\,-\,{\beta}K} }  { {\rm Tr}\,
e^{\,-\,{\beta}K} }
   \simeq q_{j}\{{R}^{(0)}\}    \,.
 \end{equation}
Thermodynamic coefficients, such as the specific heat
[Eq.(\ref{qbCTq})], are found from the second derivatives
\begin{equation} \label{stat18q}
 {\mathbb F}^{\alpha\beta} =
 \frac {\partial^{\,2} f\{{R}^{(0)}\}} {\partial{R}^{(0)}_{\alpha}\,
   \partial{R}^{(0)}_{\beta}}
 = {\mathbb K}^{\alpha\beta} - T\, {\mathbb S}^{\alpha\beta}
 \equiv  \frac {\partial^{\,2} k\{{R}^{(0)}\}} {\partial{R}^{(0)}_{\alpha}\,
   \partial{R}^{(0)}_{\beta}}
    -  T \,
 \frac {\partial^{\,2} S\{{R}^{(0)}\}} {\partial{R}^{(0)}_{\alpha}\,
  \partial{R}^{(0)}_{\beta}}\, ,
 \end{equation}
which appear as a matrix in the space of indices $\alpha$.

While the above expressions appear as mere extensions to arbitrary Lie groups 
of standard mean-field results, the approach has also provided
variational expressions for the correlations in the state $\tilde{D}
\propto \exp{(-{\beta}K)}$ (Secs.\,\ref{sec5.4} and \ref{sec5.5}). For correlations between the 
elements of the Lie algebra, we have obtained $(\alpha,\beta \ne 0)$
 \begin{equation}  \label{stat20q}
 \langle {\sf M}_{\alpha}{\sf M}_{\beta} \rangle
  - \langle {\sf M}_{\alpha} \rangle \,
    \langle {\sf M}_{\beta}  \rangle
   \simeq    {\mathbb B}_{\alpha\beta} \,  ,
 \;\;\;\;  
 {\mathbb B} =
      \frac      { i\,\hbar\,{\mathbb C}\, {\mathbb F} }
                 { \Id -\exp( {-i\,\hbar}\,\beta\,{\mathbb C}\,{\mathbb F} )  }
         \,{\mathbb F}^{-1} \, ,
 \end{equation}
where the matrix ${\mathbb C}$ is defined by $ {\mathbb C}_{\alpha\beta} =
   \Gamma^{\gamma}_{\alpha\beta} \,{R}^{(0)}_{\gamma} $. 
More generally, the correlations between two observables $Q_j$ and
$Q_k$ have been found as
\begin{equation}  \label{stat22q}
 \langle {Q}_{j}{Q}_{k} \rangle
  - \langle {Q}_{j} \rangle \,\langle {Q}_{k}  \rangle
   \simeq
     \frac { \partial q_k\{{R}^{(0)}\} } { \partial{R}_{\alpha}^{(0)} }
          \, {\mathbb B}_{\alpha\beta}  \,
       \frac { \partial q_k\{{R}^{(0)}\} } { \partial{R}_{\alpha}^{(0)} }, \,
 \end{equation}
in terms of the matrix ${\mathbb B}$ and of the symbols of the 
operators $Q_j$ and $Q_k$.
Fluctuations $\Delta {Q}_j$ result for $j = k$. In the classical
or high-temperature limit, ${\mathbb B}$ reduces to 
${({\beta}{\mathbb F})}^{-1}$.

 \subsection{Time-dependent quantities} \label{sec11.3}

 Here $\tilde{D}\propto \exp{(-{\beta}K)}$ is the exact
 initial state at the time $t_{\rm i}$, and the subsequent evolution
 is governed by the Hamiltonian $H$. The above results hold as
 variational approximations at the time $t_{\rm i}$. For later
 times, the optimization of the expectation value
 ${\langle {Q}_{j} \rangle}_t$ yields (Sec.\,\ref{sec4.2})
\begin{equation} \label{td10q}
{\langle {Q}_{j} \rangle}_t \equiv \frac {  {\rm Tr}\,
e^{\,-\,{\beta}K}\,e^{\,i\,H\,t/\hbar}\,Q^{\rm
S}_j\,e^{\,-\,i\,H\,t/\hbar}  }
 { {\rm Tr}\,e^{\,-\,{\beta}K} }
   \simeq  q_{j}\{{R}^{(0)}(t)\} \, ,
  \end{equation}
where the time-dependent expectation values ${R}^{(0)}_{\alpha}(t)$
of the operators ${\sf M}_{\alpha}$ are given by the equations
\begin{equation}    \label{td12q}
\frac {d{R}^{(0)}_{\alpha}(t)} {dt}=
 {\mathbb L}_{\alpha}^{\;\gamma}\{{R}^{(0)}(t)\}
\, {{R}^{(0)}_{\gamma}(t)} \,
\end{equation}
with the initial conditions ${R}^{(0)}_{\alpha}(t_{\rm
i})={R}^{(0)}_{\alpha}$. The effective Liouvillian ${\mathbb L}$ is
expressed self-consistently in terms of the symbol $h\{{R}\}$ of the
Hamiltonian by
\begin{equation}   \label{td14q}
{\mathbb L}_{\alpha}^{\;\gamma}\{{R}\} =
\Gamma_{\alpha\beta}^{\;\gamma}\,
 \frac { \partial{h\{{R}\}} } { \partial{R}_{\beta} } \,.
\end{equation}
(In the presence of non-vanishing structure constants
$\Gamma_{\alpha\beta}^{\;0}$, ${R}^{(0)}_{0}(t)$ should be replaced by 1 
in the term $\gamma=0$ of (\ref{td12q}).)

We have shown (Sec.\,\ref{sec10}) that the dynamical equations (\ref{td12q})
and the resulting ones for ${\langle {Q}_{j} \rangle}_t$ have a
classical structure, where $ {\mathbb C}_{\alpha\beta} \{{R}\} =
   \Gamma^{\gamma}_{\alpha\beta} \,{R}_{\gamma} $ appears as the
Lie-Poisson tensor associated with the classical coordinates
   $\{{R}\}$ and where $h\{{R}\}$ behaves as a classical Hamiltonian.

Here again, non standard variational results have been obtained 
(Sec.\,\ref{sec5.3}) for
two-time correlation functions. Their expression
involves an approximation (Sec.\,\ref{sec5.1}) for the observables 
in the Heisenberg picture defined by
\begin{equation}\label{td16q}
Q_j^{\rm H}(t',t) \equiv  e^{ \,i\,H(t'-t)/\hbar } \,Q_j^{\rm S}\,
 e^{ \,-\,i\,H(t'-t)/\hbar } \, ,
\end{equation}
where $t'$ is the usual final running time and $t$ is a reference
time at which $Q_j^{\rm H}(t+0,t)$ reduces to the Schr\"odinger
observable $Q_j^{\rm S}$. This approximation $Q_j^{\rm H}(t',t) \simeq 
{\sf Q}_j^{\rm H}(t',t)\equiv {\cal Q}_j^{{\rm H}\alpha}(t',t)  {\sf M}_{\alpha}$ is
characterized by the differential equations
 \begin{equation} \label{td18q}
\frac {d {\cal Q}^{{\rm H}\alpha}_j(t',t)}{dt}= -\,{\cal Q}^{{\rm
H}\beta}_j(t',t) \,
 \left(  {\mathbb L} + {\mathbb C}\,{\mathbb H} \right)_{\beta}^{\;\alpha}
\;\;\;\;(\alpha,\beta \ne 0)\, ,
\end{equation}
in terms of the reference time $t$ which runs backward from $t'$ to
the initial time $t_{\rm i}$. The boundary condition is given by
  \begin{equation}   \label{td20q}
 {\cal Q}^{{\rm H}\alpha}_j(t',t'-0) =
     \frac { \partial q_j\{{R}^{(0)}(t')\} } { \partial{R}_{\alpha}^{(0)}(t') }
   \, .
\end{equation}
In Eq.(\ref{td18q}), the three matrices ${\mathbb L}$ defined by (\ref{td14q}), 
$ {\mathbb C}_{\alpha\beta} \{{R}\} = \Gamma^{\gamma}_{\alpha\beta} \,{R}_{\gamma}$
and ${\mathbb H}$, which denotes the matrix of second
derivatives of the symbol $h\{{R}\}$ with respect to the
variables ${R}_{\alpha}$, are evaluated at the point ${R}={{R}^{(0)}(t)}$.

The optimized expression for the two-time correlation functions
$C_{jk}(t',t'')$ then reads (Secs.\,\ref{sec5.3} and \ref{sec5.4})
\begin{eqnarray}
 &&   C_{jk}(t',t'')  \equiv   \frac
  {  {\rm Tr}\, e^{\,-\,{\beta}\,K}\, Q_j^{\rm H}(t',t_{\rm i}) \,
  Q_k^{\rm H}(t'',t_{\rm i})  }  { {\rm Tr}\, e^{\,-\,{\beta}\,K}  }
   - \langle Q_j \rangle_{t'}  \langle Q_k\rangle_{t''}  \nonumber
 \\   \label{td22q}
  && \simeq   {\cal Q}_{j}^{{\rm H}\alpha}(t',t_{\rm i})\,
 {\mathbb B}_{\alpha\beta} \, {\cal Q}_{k}^{{\rm H}\beta}(t'',t_{\rm i})
   \;\;\;\;(t'>t'') \,,
\end{eqnarray}
which involves both the approximate correlations (\ref{stat20q}) at the initial
time and the approximate Heisenberg operators given by Eqs. (\ref{td18q}) and (\ref{td20q}).   

\subsection{Properties of the results and consequences} \label{sec11.4}

Special cases (Sec.\,\ref{sec6}) include time-dependent fluctuations, obtained
from $C_{jj}(t+0,t)$, and linear responses

\begin{eqnarray}
   \chi_{jk}(t',t'')  \!\!\!\!\!\!\!\! &&=
 ({1}/{i\hbar}) \theta(t'-t'') [C_{jk}(t',t'') - C_{kj}(t'',t') ]   \nonumber \\
                     &&\simeq      \label{td24q}  \theta(t'-t'') \,
{\cal Q}_j^{{\rm H}\alpha}(t',t)\, {\mathbb
C}_{\alpha\beta}\{{R}^{(0)}(t)\}\, {\cal Q}_k^{{\rm
H}\beta}(t'',t)\, ,
\end{eqnarray}
where $t$ is arbitrary in the interval $t_{\rm i} \le t \le t''$. If
the initial state $D \propto e^{-{\beta}K}$ is in equilibrium, with
a dynamics generated by $H$ equal to $K$ (or to $K$ plus a constant
of motion), the approximate expectation values
$\langle{Q_j}\rangle_t$ remain constant. The equations (\ref{td20q})
are solved as
 \begin{equation}   \label{td26q}
   {\cal Q}^{{\rm H}\alpha}_j(t',t_{\rm i}) =
   {\cal Q}_j^{\beta} \{{R}^{(0)}\}   \,
    \left[  e^{\,{\mathbb C}\,{\mathbb F}\,
    (t'-t_{\rm i})}\right]^{\,\alpha}_{\beta}\, .
 \end{equation}
Hence, two-time correlation functions are variationally given in this case by
 \begin{equation}   \label{td28q}
    C_{jk}(t',t'')  \simeq
   {\cal Q}_j^{\alpha} \{{R}^{(0)}\}   \,
    \left[  e^{\, {\mathbb C}\,{\mathbb F}\,(t'-t'')} \,
             \frac { i\,\hbar\,{\mathbb C}\, {\mathbb F}}
                 {\Id -\exp({-i\,\hbar}\,\beta\,{\mathbb C}\,{\mathbb F})}
         \,{\mathbb F}^{-1}          \right]_{\alpha\beta} \,
   {\cal Q}_k^{\beta} \{{R}^{(0)}\}  \, ,
   \end{equation}
an expression which, as it should, depends only on the time difference 
$t-t'$. Derived within a unified framework, these results satisfy
consistency properties and conservation laws. For instance, the
approximation preserves the relation between the static stability of
a thermodynamic equilibrium state ($f\{{R}\}$ minimum) and the dynamical 
stability of the motions (\ref {td12q}) around it.

Other special cases are also considered in Sec.\,\ref{sec6}, including classical, 
high-temperature and zero-temperature limits, as well as Kubo correlations. 
Results for small deviations are presented in Sec.\,\ref{sec9}.

In Secs.\,\ref{sec7} and \ref{sec8.3}-\ref{sec8.6}, we have exhibited, for any Lie group, an
attractive interpretation for the static correlations
(\ref{stat20q}) and for the two-time functions (\ref{td26q})-(\ref{td28q}). 
To this aim, the Lie algebra $\{{\sf M}\}$ has been
mapped into a simpler Lie algebra $\{\underline{\sf M}\}$, the
operators ${\underline{\sf M}}_{\alpha}$ of which are linear
combinations of creation and annihilation bosonic operators
$\underline{\sf{A}}_{n}$ and $\underline{\sf{A}}_{n} ^{\dagger}$ (or
equivalently of position and momentum operators
$\underline{\sf{X}}_{n}$ and $\underline{\sf{P}}_{n} $ of harmonic
oscillators), and of scalar random variables
$\underline{\sf{Y}}_{p}$. Then, the dynamics (\ref{td26q}) is mapped
onto a dynamics governed by an effective Hamiltonian
  \begin{eqnarray} \label{td30q}
  {\underline{{F}}}
  =   \sum_{n}  \hbar\,{\Omega}_{n} \, \frac{1}{2}   \left(
     {\underline{\sf{A}}_{n}^{\dagger}} \,   {\underline {\sf{A}}_{n}}
  +  {\underline {\sf{A}}_{n}} \,  {\underline{\sf{A}}_{n}^{\dagger}}  \right)
   +  \sum_{p}\frac {1}{2\beta} \, \underline{\sf{Y}}_{p}^{2}
   \\ \nonumber
   =    \sum_{n}  {\Omega}_{n} \,
  ( {\underline{\sf {P}}_{n}}^{2} +  {\underline{\sf {X}}_{n}}^{2} )
   + \frac {1}{2\beta}  \sum_{p} \, {\underline{\sf{Y}}_{p}}^{2} \,,
 \end{eqnarray}
which describes uncoupled modes. The above correlations (\ref{stat20q})
and (\ref{td28q}) take for an arbitrary Lie group the form of expectation values of pairs of
quasi-boson operators and of scalars in a canonical equilibrium state with
Hamiltonian ${\underline{{F}}}$.

The explicit expression
 \begin{eqnarray}  \label{td32q}
 C_{jk}(t',t'') &= & \sum_{n} \left[{\cal Q}^{n}_{j{\rm b}}  \,
  \frac {e^{\,-\,i\,{\Omega}_{n}(t'-t'')}} {1- e^{\,-\,\beta\,\hbar\,\Omega_n}}\,
                       {( {\cal Q}^{n}_{k{\rm b}} )}^{*}      \right.
  \\ \nonumber
 && + \left.   {( {\cal Q}^{n}_{j{\rm b}} )}^{*}   \,
  \frac { e^{\,i\,{\Omega}_{n}(t'-t'')} } { e^{\,\beta\,\hbar\,\Omega_n} - 1} \,
    {\cal Q}^{n}_{k{\rm b}}  \right]
 +  \sum_{p}  {\cal Q}^{p}_{j{\rm s}} \,{\cal Q}^{p}_{k{\rm s}} \, ,
  \;\;\;\;(t'>t'')
\end{eqnarray}
derived in Sec.\,\ref{sec8} relies on the diagonalization (\ref{MFA49q}) of
the matrix $i\,{\mathbb C}\,{\mathbb F}$, which enters the formalism 
at several places. The coefficients ${\cal
Q}^{n}_{j{\rm b}}$ and ${\cal Q}_{j{\rm s}}^{p}$ appear as
coordinates, in the new basis of the Lie algebra, of the observable
$Q_j$, their explicit form being given by Eqs.(\ref{MFA58aq}) in
terms of the eigenvalues $\pm{\Omega}_n$ and eigenvectors of the
matrix $i\,{\mathbb C}\,{\mathbb F}$ (which satisfy the equations
(\ref{MFA45Aq}), (\ref{MFA48q}) and (\ref{MFA47q})). Special cases of
(\ref{td32q}) are the static correlations written in the diagonalized 
form (\ref{MFA50q}) and the linear responses, which satisfy the Kramers-Kronig 
dispersion relations and the Kubo fluctuation-dissipation relation.

\section  {Conclusion} \label{sec12}

 We have presented above a variational approach to the determination
 of various physical quantities pertaining to many-body systems. Although our
 scope has been formal, the generality and flexibility of the treatment appear
 well suited to many specific problems. The results, listed in
 Sec.\,\ref{sec11}, have been obtained by merging several ingredients:

 (i)The evaluation of a {\it generating functional} (Sec.\,\ref{sec2.1}) has
 allowed the simultaneous optimization of different quantities such
 as the (static or time-dependent) expectation values and
 correlations of the observables of interest: they are obtained by expanding
 in powers of the sources the functional
 $\psi\{\xi\} \equiv \ln {\rm Tr}{\boldmath A}(t_{\rm i})\, D $.
 Deriving the correlation functions as second-order contributions in the
 sources of this generating functional provides for them non-trivial
 approximations, even for a simple restricted trial space. As an
 example, for a system of interacting fermions, the use of
 independent-particle trial objects leads to standard mean-field
 theories for expectation values but to expressions of the form
 (\ref{td22q}) for correlation functions. It is the dependence of the trial
 objects on the sources which leads to such elaborate results. Moreover,
 approximating all quantities in a unique framework preserves
 consistency properties.

(ii)The {\it variational principle} for the optimization of the
generating functional is built by means of a general method
(Sec.\,\ref{sec2.3}). The object to be optimized is characterized by some
simple equations, regarded as {\it constraints} on the ingredients, namely,
on the initial state $D \propto \exp(-{\beta}K)$ and on the time-dependent
observables of interest $Q_j^{\rm H}(t',t)$. Lagrange multipliers
are then associated with these constraints.

(iii) For dynamical problems, the Heisenberg observables $Q_j^{\rm
H}(t',t)$ enter the generating functional, together with the sources
$ \xi_j(t)$, through the "generating operator"
\begin{equation}      \label{conc10q}
 {\boldmath A}(t)
\equiv  T e^{\textstyle \,i\int^\infty_{t} dt' \sum_j \xi_j(t') \,
 {\boldmath Q}_j^{\rm H}(t',t)}.
\end{equation}
The operator $A(t)$ is characterized by the differential equation
(\ref{A3q}) expressing $\frac {dA(t)}{dt}$, which plays the role of a
constraint in the variational principle. This Eq.(\ref{A3q}) has been
derived as a consequence of the {\it backward Heisenberg equation}
\begin{equation}     \label{conc12q}
\frac {dQ_j^{\rm H}(t',t)} {dt} = - \frac {1} {i\,\hbar} \,
[Q_{j}^{\rm H}(t',t), H] \,.
\end{equation}
We recall that, while the standard (forward) Heisenberg equation
describes the variation of the Heisenberg  observable $Q_{j}^{\rm
H}(t',t)$ as function of the running time $t'$, (\ref{conc12q}) is a
differential equation with respect to a reference time $t$ which
runs backward from $t'$ to the initial time $t_{\rm i}$. We have
explained in Sec.\,\ref{sec2.2} why this backward dynamics is the suitable
one.

(iv) As regards the initial state $D=e^{-{\beta}K}$, we have
characterized it by the {\it Bloch equation}
\begin {equation} \label{conc14q}
\frac {{\rm d}{\DIh}(\tau)}{{\rm d}\tau } + K\,{\DIh}(\tau)=0\,,
\end{equation}
where $\tau$ runs from $0$ to $\beta$. We thus deal with finite
temperature. Ground state problems are treated in the limit ${\beta}
\to 0$; in fact, this procedure turns out to be more convenient than
the direct implementation of ground states.

(v) The variational equations associated with the above constraints
on $A(t)$ and $D(\tau)$ have been worked out by taking a {\it Lie
group} as trial space for $A(t)$, $D(\tau)$ and for their associated
Lagrange multipliers (Sec.\,\ref{sec3}), thus replacing operators in Hilbert 
space by functions, their symbols. The approximate states that occur in
mean-field theories belong to such Lie groups, for instance static
and dynamic states in Hartree-Fock approximations for fermions,
Hartree-Fock-Bogoliubov states for fermions with pairing, or coherent
states for bosons. However, the trial operators $\DIh$ used here  
depend on the sources; they are not overall approximations for the exact 
state $D$, but serve only to optimize the generating 
functional (and hence all quantities of interest).

We have recalled in Sec.\,\ref{sec11} the outcomes of the above approach. Some
well-known approximate results have been recovered in special
cases, for instance, static and dynamic mean-field or RPA treatments
which now appear within a general unified framework based on the use
of a Lie group as trial space. Moreover, the new variational results
(\ref{stat20q}), (\ref{stat22q}), (\ref{td22q}) and (\ref{td28q}) have been derived 
for correlations and fluctuations. In particular, we have shown for $H=K$ that
quasi-bosons (or quasi-oscillators) come out {\it for any Lie group}, and
not only in the usual context of zero-temperature RPA for fermions. Let
us add a few comments.

In order to determine the generating functional $\psi\{\xi\} \equiv
\ln {\rm Tr}{\boldmath A}(t_{\rm i})\, D $, we have characterized
${\boldmath A}(t_{\rm i})$ and $D=e^{-{\beta}K}$ by introducing the
functions $A(t)$ and $D(\tau)=e^{-{\tau}K}$ determined from the
simple differential equations (\ref{A3q}) and (\ref{conc14q}). To
build a tractable variational principle, we have been led to
introduce the trial time-dependent quantities $\DIh(\tau)$ and
$\AIh(t)$ whereas we need only their boundary values $\DIh(\beta)$
and $\AIh(t_{\rm i})$. Moreover, the number of variables is {\it doubled} by
the introduction of the Lagrange multipliers $\AIh(\tau)$,
associated with the equation for $D(\tau)$, and $\DIh(t)$,
associated with the equation for $A(t)$. These quantities are not
coupled by the exact equations of motion. However, in a restricted
trial space, the stationarity conditions entail a {\it coupling}
between them which allows a better optimization of the generating
functional $\psi\{\xi\}$.

In spite of their resemblance with density operators and
exponentials of observables, the trial objects $\DIh(\tau)$,
$\AIh(t)$, $\AIh(\tau)$, $\DIh(t)$ are {\it only computational tools} for
the evaluation of expectation values and correlation functions. In
particular the trial quantity $\DIh(t)$, which looks formally like a
density operator in the Schr\"odinger picture, adapts itself to the
question asked, namely to the value of $\psi\{\xi\}$. It thus
depends on the sources (and is not even hermitian). It cannot be
interpreted as an approximate state of the system, and in fact there
exists no approximate density operator in the original Hilbert space
that would produce both the optimized expectation values
(\ref{td10q}) and correlations (\ref{stat20q}), (\ref{stat22q}), (\ref{td22q}).

Since all the results have been derived from the same variational
principle, they naturally satisfy some {\it consistency properties}
fulfilled by exact quantities. For instance, if an observable
$Q_j$ belongs to the Lie algebra and commutes with $H$, exact
conservation laws express the constancy in time of the expectation
value ${\langle Q_j \rangle}_t $ and of the fluctuation
${\Delta}Q_j(t)$. These two properties are ensured by the
variational approximations (Sec.\,\ref{sec6.3}). In contrast, a naive evaluation of 
fluctuations based on the approximation 
${\rm Tr}\, {{Q}_j}^{2}\,\tilde{D}(t) \simeq {\rm Tr}\,
{{Q}_j}^{2}\,\tilde\DIh^{(0)}(t)$ where $\tilde\DIh^{(0)}(t)$ evolves according
to (\ref{qb23q}), would provide an unphysical time dependence: The
Schr\"odinger picture for $\tilde\DIh^{(0)}(t)$ is variationally suited to
the evaluation of expectation values, but not of correlation
functions or fluctuations, which involve here the approximate Heisenberg
operators ${\sf Q}_j^{\rm H}(t,t')$. The variational approximation
(\ref{td28q}) fulfils another consistency property, as discussed in Sec.\,\ref{sec6.1},
namely that any correlation function $C_{jk}(t',t'')$ depends only
on the time difference $t'-t''$ when $H=K$. 
We have also stressed
that the approach unifies static and dynamical properties,
generalizing theorems well known for fermion systems \cite{TH60q,dClq,Mer63q} : 
When the thermodynamic equilibrium is stable, i.e., when the trial
free energy $f\{{R}\}$ is minimum at $\{{R}\} = \{{R}^{(0)}\}$, the
matrix ${\mathbb F}$ of second derivatives is positive, and this
implies that the eigenvalues of $i{\mathbb C}{\mathbb F}$ are real.
The latter property ensures dynamical stability: a small deviation of
$\{{R}\}$ around $\{{R}^{(0)}\}$ is never amplified (Sec.\,\ref{sec9.3}).

The approximate equations (\ref {stat16q}) and (\ref {td10q}) determine 
the expectation values ${\langle Q_j \rangle}_t$ at the initial time $t_{\rm i}$ 
and for arbitrary $t$, express these quantities in terms of the
expectation values $ {R}^{(0)}_{\alpha}(t) =  {\rm Tr}\,{\sf
M}_{\alpha}\,\tilde\DIh^{(0)}(t)$, which are the
symbols of the elements ${\sf M}_{\alpha}$ of the Lie algebra. Thus,
at first order in the sources, these elements ${\sf M}_{\alpha}$ are
replaced by the c-numbers ${R}^{(0)}_{\alpha}(t)$ as if the algebra
were replaced by a commutative one. Accordingly, the approximate dynamical 
equations then have a classical Lie-Poisson structure (Sec.\,\ref{sec10}). At 
second order in the sources,
we have shown (Sec.\,\ref{sec7}) that in the initial state the fluctuations
and correlations ${\mathbb B}$ of the operators ${\sf M}_{\alpha}$
have the same form as if the Lie algebra were replaced by a mapped
algebra in which the commutators ${[{\sf M}_{\alpha},\,{\sf M}_{\beta}]}$ 
are replaced by c-numbers (\ref{MOC10q}),
which are their expectation values (\ref{MFA14q}). We have also seen that
this amounts to replace the deviations (\ref{MFA81q}) by linear combinations
of operators of non-interacting quasi-bosons (or of harmonic
oscillators) together with quasi-scalars. This property extends for
$H=K$ to two-time correlations. We thus acknowledge {\it two successive
modifications of the Lie algebra}: For first-order quantities, the
operators ${\sf M}_{\alpha}$ are approximated by scalars, whereas for
 second-order quantities it is their commutators which are approximated by scalars.

Although formal and technical, the present approach is systematic,
flexible, and consistent owing to the optimization of the generating
functional. It allows a variational evaluation of correlation
functions, which lie beyond the realm of standard mean-field
theories. It generalizes some known approaches to arbitrary Lie
groups, setting them within a unified framework. It may be 
applied to various questions of statistical
physics and field theory, static or dynamic, at zero or non-zero
temperature, at equilibrium or off equilibrium.

\section*{Acknowledgments} During the long elaboration of this work, Marcel 
V\'en\'eroni constantly manifested his deep gratitude to Marie-Th\'er\`ese 
Commault for her efficient and friendly assistance in setting the manuscript 
into shape. Denis Lacroix from the IPN at Orsay has kindly revised the 
references.

\appendix

\section{Derivation of the correlation matrices ${\mathbb B}$ and 
${\mathbb B}^{\rm K}$}
\label{Appendix A}

\subsection{Ordinary correlations}

As indicated in Sec.\,\ref{sec5.4}, the determination of the correlation
matrix ${\mathbb B}$ involves the solution at first order of the
coupled equations (\ref{GS200q}),(\ref{GS201q}) for
$\DIh(\tau)$,\,$\AIh(\tau)$ in the range $0 \le \tau \le \beta$,
with the boundary conditions ${\DIh}^{(1)}_{\alpha}(0)=0$ and
${\AIh}^{(1)}_{\alpha}(\beta)={\sf M}_{\alpha}$. The zeroth order
operators $\DIh^{(0)}(\tau)=\exp[-{\tau}\,{\sf K}^{(0)}]$,
$\AIh^{(0)}(\tau)=\exp[({\tau}-{\beta}){\sf K}^{(0)}]$ and
$\DIh^{(0)}=\exp[-{\beta}\,{\sf K}^{(0)}]$, where we denote for
shorthand ${\sf K}\{{R}^{(0)}\} \equiv {\sf K}^{(0)}$, have been
derived in Sec.\,\ref{sec4.1}. At first order, we have parametrized the
combination $\DIh(\tau)$\,$\AIh(\tau)$ by $\{{R}^{(1)}(\tau)\} \sim
\{{R}^{\DIh\AIh}(\tau)\} - \{{R}^{(0)}\}$, that is,
\begin{eqnarray}    \label{App10q}
  {R}^{(1)}_{\alpha\beta}(\tau)     \sim
   {\rm Tr} [ \DIh^{(0)}(\tau)\,\AIh^{(1)}_{\alpha}(\tau)
  + \DIh^{(1)}_{\alpha}(\tau)\,\AIh^{(0)}(\tau) ]
   ( {\sf M}_{\beta} - {R}^{(0)}_{\beta} )
   /  {\rm Tr}\, {\DIh}^{(0)} \,.
 \end{eqnarray}
The matrix element ${\mathbb B}_{\alpha\beta}$ was then identified with
\begin{eqnarray}   \label{MFC22q}
   && {\mathbb B}_{\alpha\beta}    =
 {R}^{(1)}_{_{\alpha}\beta}(\tau=\beta)
   \\ \nonumber
   && \;\;\;\;\;\;={\rm Tr}\, {\sf M}_{\alpha}
 ( {\sf M}_{\beta} - {R}^{(0)}_{\beta} ) {\tilde\DIh}^{(0)}
  + \frac {  {\rm Tr}\, {\DIh}^{(1)}_{\alpha}(\beta)
  ( {\sf M}_{\beta} - {R}^{(0)}_{\beta} )  }
          {  {\rm Tr}\,{\DIh}^{(0)}  } \,.
 \end{eqnarray}

A preliminary task is to find the $\tau$-dependence of 
${R}^{(1)}_{\alpha\beta}(\tau)$. This is done through the equation
  \begin{eqnarray} \label{App14q}
      \frac {d} {d{\tau}} [ {\DIh(\tau)}{\AIh(\tau)} ] =
      \left[ {\DIh(\tau)}{\AIh(\tau)}, \,{\sf K}\{{R}^{\DIh\AIh}\}   \right] \, ,
  \end{eqnarray}
a consequence of (\ref{GS200q}),(\ref{GS201q}) equivalent to
 \begin{eqnarray} \label{App16q}
   \frac {d{R}^{\DIh\AIh}_{\beta}(\tau)}{d{\tau}} = -\,i\,\hbar \,
 {\Gamma}^{\,\delta}_{\beta\gamma} \, {\cal K}^{\gamma}\{{R}^{\DIh\AIh}\}\,
{R}^{\DIh\AIh}_{\delta}(\tau)\,.
 \end{eqnarray}
This equation, since the boundary conditions
${\DIh}^{(1)}_{\alpha}(\tau =0)=0$
and ${\AIh}^{(1)}_{\alpha}(\tau =\beta)={\sf M}_{\alpha}$
are imposed at different times $\tau$,
cannot fully determine the quantity ${R}^{(1)}_{\alpha\beta}(\tau)$
but it will provide its dependence on $\tau$.
Expanding (\ref{App16q}) up to first order involves the kernel
\begin{eqnarray}  \label{App12q}
  {\sf K}\{ {R}^{\AIh\DIh}(\tau) \}
     \approx  {\sf K}^{0} +  {R}^{(1)}_{\alpha\beta}(\tau)\,
  {\mathbb K}^{\beta\delta}\,{\sf M}_{\delta} \, ,
  \end{eqnarray}
 which depends self-consistently
on $\DIh^{(1)}_{\alpha}(\tau)$ and $\AIh^{(1)}_{\alpha}(\tau)$
through $\{{R}^{(1)}(\tau)\}$.
Using (\ref{App12q}), then (\ref{MFA10dq}),(\ref{MFA10fq}) and (\ref{MFA20q}),
one obtains
 \begin{eqnarray} \label{App18q}
  \frac {d{R}^{(1)}_{\alpha\beta}(\tau)}{d{\tau}} = -\,i\,\hbar \,
  {\mathbb C}_{\beta\gamma}\,{\mathbb F}^{\gamma\delta} \,{R}^{(1)}_{\delta}(\tau) \, ,
 \end{eqnarray}
which is readily solved, in terms of the still unknown quantities
${R}^{(1)}_{\alpha\beta}(\tau=\beta)={\mathbb B}_{\alpha\beta}$, as
 \begin{eqnarray} \label{App20q}
 {R}^{(1)}_{\alpha\beta}(\tau) =
  \left[ e^{\, i\,\hbar\,{\mathbb C}\,{\mathbb F}\,(\beta-\tau) } \right]_{\beta}^{\;\gamma}
 \,{R}^{(1)}_{{\alpha\gamma}}({\beta})
  =  {\mathbb B}_{\alpha\gamma}  \left[
 e^{\, i\,\hbar\,{\mathbb F}\,{\mathbb C}\,(\tau-\beta) } \right]_{\;\beta}^{\gamma} \, .
 \end{eqnarray}

We now turn to the determination of ${\DIh}^{(1)}_{\alpha}(\tau)$
which enters (\ref{MFC22q}) for $\tau=\beta$.
(We shall not need ${\AIh}^{(1)}_{\alpha}(\tau)$ for $0 \le {\tau} < \beta$.)
The first order contribution to (\ref{GS200q}) takes the form
  \begin{eqnarray} \label{App22q}
   \frac {d} {d{\tau}} [ {\DIh}^{(1)}_{\alpha}(\tau)\,{\AIh}^{(0)}(\tau) ] =
      \left[ {\DIh}^{(1)}_{\alpha}(\tau)\,{\AIh}^{(0)}(\tau) , \,{\sf K}^{(0)} \right]
  - {R}^{(1)}_{\alpha\beta}(\tau)\,{\mathbb K}^{\beta\gamma}\,
    {\sf M}_{\gamma}\,{\DIh}^{(0)}\,.
  \end{eqnarray}
Using the boundary condition $ {\DIh}^{(1)}_{\alpha}(0)=0$,
we can solve (\ref{App22q}) as
\begin{eqnarray} \label{App24q}
   {\DIh}^{(1)}_{\alpha}(\tau)\,{\AIh}^{(0)}(\tau)  =   - \int_{0}^{\tau} d{\tau'}
   {R}^{(1)}_{\alpha\beta}(\tau')\, {\mathbb K}^{\beta\gamma}\,
   e^{ {\sf K}^{(0)} ( {\tau'}-{\tau} ) } \, {\sf M}_{\gamma}\,
    e^{ {\sf K}^{(0)}( {\tau}-{\tau'} ) } \, {\DIh}^{(0)}\,.
  \end{eqnarray}
The integrand involves the transform of the element ${\sf M}_{\gamma}$
of the Lie algebra by the element
$e^{ {\sf K}^{(0)}( {\tau'}-{\tau} ) }$ of the group,
with $ {\sf K}^{(0)} = {\sf M}_{\beta} {\cal K}^{(0)\beta}$.
This transform is also an element of the algebra given by
the automorphism (\ref{MFA10gq})
for $\DIh={\DIh}^{(0)}$ and $\lambda=(\tau-{\tau}')/{\beta}$ as
  \begin{eqnarray} \label{App28q}
   e^{{\sf K}^{(0)}({\tau'}-{\tau})}\,{\sf M}_{\gamma}\,
   e^{{\sf K}^{(0)}({\tau}-{\tau'})}
     = \left[ e^ {  \,i\,\hbar \, {\beta}^{-1} \,
   {\mathbb C}\,{\mathbb S} \,({\tau}'-{\tau}) } \right]^{\,\delta}_{\gamma}
  \,{\sf M}_{\delta}\, .
  \end{eqnarray}
 Inserting (\ref{App20q}) and (\ref{App28q}) into (\ref{App24q}) leads to
  \begin{eqnarray} \label{App30q}
   &&{\DIh}^{(1)}_{\alpha}(\tau)\,{\AIh}^{(0)}(\tau) =
   \\ \nonumber
   && - \int_{0}^{\tau} d{\tau'} \, {\mathbb B}_{\alpha\beta}  \left[
   e^{ i\,\hbar \,{\mathbb F}\,{\mathbb C}\, ({\tau}'-{\beta}) }
   \left(  {\mathbb F} + {\beta}^{-1} \,{\mathbb S}    \right)
    e^{  i\,\hbar\,{\beta}^{-1} \,{\mathbb C}\,{\mathbb S} \,({\tau}'-{\tau})  }
   \right]^{\beta\gamma }
    {\sf M}_{\gamma}\,{\DIh}^{(0)}\,.
   \end{eqnarray}

 We note that the bracket in Eq.(\ref{App30q}) is the derivative with respect to ${\tau'}$ of
  \begin{eqnarray}  \nonumber
    \frac
 { e^{\,i\,\hbar\,{\mathbb F}\,{\mathbb C}\,(\tau'-\beta) } - {\mathbb I} }
 { i\,\hbar\,{\mathbb F}\,{\mathbb C} }
    \,{\mathbb F}\,  e^{ \,i\,\hbar\,{\beta}^{-1}\,{\mathbb C}\,{\mathbb S}\,(\tau'-\tau) }
    +    {\mathbb S} \,   \frac
 {  e^{  \, i\,\hbar\,{\beta}^{-1} \,{\mathbb C}\,{\mathbb S}\, ({\tau}'-{\tau})  }
    - {\mathbb I} }
 { i\,\hbar\,{\mathbb C}\,{\mathbb S} } \, ,
   \end{eqnarray}
 where $(e^{x}-1)/x$ is defined by continuity as $1$ for $x=0$.
Thus, by integration of (\ref{App30q}), we obtain for $\tau = \beta$:
 \begin{eqnarray}   \label{App34q}
 {\DIh}^{(1)}_{\alpha}(\beta) =  {\mathbb B}_{\alpha\beta}     \left[
   \frac
 { e^{\,-\,i\,\hbar\,{\beta}\,{\mathbb F}\,{\mathbb C} } - {\mathbb I} }
 { i\,\hbar\,{\mathbb F}\,{\mathbb C} }
   \,{\mathbb F}\,  e^{ \,-\,i\,\hbar\,{\mathbb C}\,{\mathbb S} }
  +  {\mathbb S} \, \frac
   { e^{  \,-\,i\,\hbar\,{\mathbb C}\,{\mathbb S}  } - {\mathbb I} }
   { i\,\hbar\,{\mathbb C}\,{\mathbb S} }
    \right]^{\beta\gamma}   {\sf M}_{\gamma} \,{\DIh}^{(0)} \,.
  \end{eqnarray}

We have thus found the expression of ${\DIh}^{(1)}_{\alpha}(\beta)$ to be inserted
into (\ref{MFC22q}) so as to obtain ${\mathbb B}_{\alpha\beta}$.
The resulting two terms in (\ref{MFC22q}) involve a correlation in the state
${\tilde\DIh}^{(0)} \propto \exp[-{\beta}{\sf K}^{(0)}]$
between two operators of the basis $\{{\sf M}\}$ of the Lie algebra,
${\sf M}_{\alpha}$ and ${\sf M}_{\beta}$,
or ${\sf M}_{\beta}$ and ${\sf M}_{\gamma}$, respectively.
Such correlations have been evaluated in Sec.\,\ref{sec3.3}
(within replacement of $\DIh$ by ${\DIh}^{(0)}$)
and are provided by Eq.(\ref{MFA10hq}).
Inserting both terms in (\ref{MFC22q}), we find
\begin{eqnarray}   \label{App110q}
  {\mathbb B} = && 
   \frac  { i\,\hbar\,{\mathbb C}\,{\mathbb S} }
          { {\mathbb I} - e^{\,i\,\hbar\,{\mathbb C}\,{\mathbb S} } }
                        \, {\mathbb S}^{-1}
    \\  \nonumber
                 &&  + \, {\mathbb B}
                  \left(      {\mathbb F}\,
    \frac { e^{\,-\,i\,\hbar\,{\beta}\,{\mathbb C}\,{\mathbb F} } - {\mathbb I} }
               { i\,\hbar\,{\mathbb C}\,{\mathbb F} }
           \,  e^{ \,-\,i\,\hbar\,{\mathbb C}\,{\mathbb S} }
    +  \frac  { e^{  \,-\,i\,\hbar\,{\mathbb S}\,{\mathbb C}  } - {\mathbb I} }
                 { i\,\hbar\,{\mathbb S}\,{\mathbb C} }
        \, {\mathbb S} \,  \right)
    {\mathbb S}^{-1}  \frac  { i\,\hbar\,{\mathbb S}\,{\mathbb C} }
    { e^{\,- \,i\,\hbar\,{\mathbb C}\,{\mathbb S} } - {\mathbb I}   } \, ,
    \end{eqnarray}
   and hence
  \begin{eqnarray} \label{App120q}
   {\mathbb B} =
   \frac  { i\,\hbar\,{\mathbb C}\, {\mathbb F} }
          {  \Id - e^{ \,-i\,{\hbar}\,{\beta}\,{\mathbb C}\,{\mathbb F} }  }
         \,{\mathbb F}^{-1} \, .
  \end{eqnarray}

\subsection{Kubo correlations}

In order to find a variational approximation for the Kubo correlation
 \begin{eqnarray}   \label{AppA21}
     {\rm Tr} \, \frac{1}{\beta}   \int_{{0}}^{\beta}\,d{\tau} \,
       e^{ \,   {\tau}\,{K} } \, {\sf M}_{\alpha}\,
       e^{ \,-\,{\tau}\,{K} }  \,{\sf M}_{\beta}  \, \tilde{D}
  -  {\rm Tr} \,{\sf M}_{\alpha}  \,\tilde{D} \,
     {\rm Tr} \,{\sf M}_{\beta}  \, \tilde{D} \, ,
  \end{eqnarray}
we first replace $A(t)$ by $I$ in the generating functional
$\ln {\rm Tr}{A(t_{\rm i})}\, D$ and introduce, 
instead of the sources $\xi_j(t)$ entering the exponent of $A$,
small sources $\delta{J}^{\alpha}$ entering terms 
$-{\beta}^{-1}\delta{J}^{\alpha}{\sf M}_{\alpha}$
added to $K$ in the exponent of $D=\exp(-{\beta}K)$.
We now deal with a single exponential operator instead of a product of two,
and Kubo correlations are the second-order terms
in $\{\delta{J}\}$ in the expansion of $\ln{\rm Tr}\,D$.

The variational approximation ${\mathbb B}^{\rm K}$
for Kubo correlations is then obtained through
the formalism of Sec.\,\ref{sec4.1} within replacement of the image ${\sf K}$
by ${\sf K} -{\beta}^{-1}\delta{J}^{\alpha}{\sf M}_{\alpha} $.
This yields
$\ln{\rm Tr}\exp (-{\beta}K+\delta{J}^{\alpha}{\sf M}_{\alpha}) \simeq
-{\beta}f\{{R}\} + \delta{J}^{\alpha}{R}_{\alpha}$
where $\{{R}\}=\{{R}^{(0)}+\delta{R}\}$ is given by the stationarity condition
${\beta}\,{\partial}f\{{R}^{(0)}+\delta{R}\} / \partial{{R}_{\alpha}} =
{\beta}\,{\mathbb F}^{\alpha\beta}\delta{R}_{\beta}=\delta{J}^{\alpha}$.
As in the evaluation of the thermodynamic coefficients,
the second-order terms in $\{\delta{J}\}$ are finally found as
\begin{eqnarray}   \label{AppA22}
 {\mathbb B}^{\rm K}_{\alpha\beta}
  = \frac{1}{\beta} ({\mathbb F}^{-1})_{\alpha\beta} = 
[({\beta}\,{\mathbb K}-{\mathbb S})^{-1}]_{\alpha\beta} \,.
\end{eqnarray}

Here as in the case of ${\mathbb B}$, this approximation for (\ref{AppA21})
is obtained by replacing $-{\mathbb S}$ by
${\beta}\,{\mathbb F}={\beta}\,{\mathbb K}-{\mathbb S}$ in 
the naive expression (\ref{MFA010q}) written for ${\tilde{\DIh}}=
{\tilde{\DIh}}^{(0)}$.

\section  {The variational results and the effective state} 
\label{Appendix B}

We prove in this Appendix that the variational approximations
${\rm Tr}\,{\sf M}_{\alpha} \tilde{D} \simeq {\langle {\sf M}_{\alpha} \rangle}_{\rm app}
= {R}^{(0)}_{\alpha} $
and ${\rm Tr}\,{\sf M}_{\alpha}{\sf M}_{\beta}\tilde{D}
     \simeq {\langle {\sf M}_{\alpha}{\sf M}_{\beta} \rangle}_{\rm app}
   = {\mathbb B}_{\alpha\beta} + {R}^{(0)}_{\alpha} {R}^{(0)}_{\beta} $
found in Secs.\,\ref{sec4} and \ref{sec5} are reproduced in the mapped Hilbert space
$\underline{\mathscr{H}}$ as exact
expectation values of ${\underline{\sf M}}_{\alpha}$ and ${\underline{\sf M}}_{\alpha}{\underline{\sf M}}_{\beta}$
over the effective state $\tilde{\underline{D}}$ defined by Eq.(\ref{MOC12q}).

Let us introduce the generating function
 \begin{eqnarray} \label{MOC14q}
\phi\{ {\lambda} \}   \equiv   \ln      \frac
 {  \underline {\rm Tr} \, \exp (  \,-\,{\beta}\,\underline{F} + {\lambda}^{\alpha}\, {\underline{\sf M}}_{\alpha}  )  }
 {  \underline {\rm Tr} \, \exp(  \,-\,{\beta}\,\underline{F} )  }
\end{eqnarray}
which will produce the expectation values
$ \langle {\underline{\sf M}}_{\alpha} \rangle_{\rm map} $
and the Kubo correlations
${\langle {\underline{\sf M}}_{\alpha}\,
  {\underline{\sf M}}_{\beta} \rangle}_{\rm map}^{\rm K}$ in the state $\tilde{\underline{D}}$
by derivation with respect to the sources $ {\lambda}^{\alpha}$.
We can rewrite it as
\begin{eqnarray} \label{MOC15q}
\phi\{ {\lambda} \}   \equiv        \ln       \frac
 {    \underline {\rm Tr} \, \exp \left[   \, -\,\frac{1}{2}\,{\beta} \,
      {\underline{\sf M}}_{\alpha}^{'}
        \, {\mathbb F}^{\alpha\gamma} \,
      {\underline{\sf M}}_{\gamma}^{'}
      + {\lambda}^{\alpha}\, {R}^{(0)}_{\alpha}
      +\frac{1}{2}\,{\beta}^{-1} \, {\lambda}^{\alpha} {\left( {\mathbb F}^{-1} \right)}_{\alpha\gamma}
  \, {\lambda}^{\gamma} \right]   }
      {  \underline {\rm Tr} \, \exp(  \,-\,{\beta}\,\underline{F} )  }
 \end{eqnarray}
where the operators $ {\underline{\sf M}}_{\alpha}^{'}$ in the mapped space $\underline {\mathscr{H}}$
are defined through the shift (we drop $ {\underline{\sf M}}_{0}$):
\begin{eqnarray} \label{MOC16q}
 {\underline{\sf M}}_{\alpha}^{'}  = {\underline{\sf M}}_{\alpha}
  - {\beta}^{-1}{\left( {\mathbb F}^{-1} \right)}_{\alpha\gamma} \, {\lambda}^{\gamma} - {R}^{(0)}_{\alpha} \, .
\end{eqnarray}
These operators obey the same commutation relations as (\ref{MOC10q}), so that the replacement of
$\{ {\underline{\sf M}} \}$ by $\{{\underline{\sf M}}^{'}\}$ does not modify the trace.
Hence, we find
\begin{eqnarray} \label{MOC17q}
  \phi\{ {\lambda} \}    =   {\lambda}^{\alpha}\, {R}^{(0)}_{\alpha}
      +\frac{1}{2\,{\beta}} \, {\lambda}^{\alpha} {\left( {\mathbb F}^{-1} \right)}_{\alpha\gamma}  {\lambda}^{\gamma}  \,.   \end{eqnarray}

We now expand
$\underline {\rm Tr}\,\exp[ \,-\,{\beta}\,\underline{F} + {\lambda}^{\alpha} \,{\underline{\sf M}}_{\alpha} ]$
in powers of the sources $ {\lambda}^{\alpha} $:
\begin{eqnarray} \label{MOC18q}
&&\underline {\rm Tr}\,e^{ \,-\,{\beta}\,\underline{F} + {\lambda}^{\alpha} \,{\underline{\sf M}}_{\alpha} }  \approx
  \\  \nonumber
  &&    \underline {\rm Tr}\, e^{ \,-\,{\beta}\,\underline{F} }
    +  {\lambda}^{\alpha} \, \underline {\rm Tr}
      \left( e^{ \,-\,{\beta}\,\underline{F} }\,{\underline{\sf M}}_{\alpha} \right)
    + \frac{1}{2\,{\beta}}\, {\lambda}^{\alpha}\, {\lambda}^{\gamma} \,
      \int_{{0}}^{\beta}\,d{\tau} \,
     \underline {\rm Tr} \left( e^{ \,-\,({\beta}-{\tau})\,\underline{F} }\,\,{\underline{\sf M}}_{\alpha}
      \,e^{ \,-\,{\tau}\,\underline{F} }\,\,{\underline{\sf M}}_{\gamma}  \right)\, .
\end {eqnarray}
Inserting in (\ref{MOC14q}) and identifying with (\ref{MOC17q}), we recover at first order
\begin{eqnarray}   \label{MOC19q}
   \langle {\underline{\sf M}}_{\alpha} \rangle_{\rm map}  = {R}^{(0)}_{\alpha} \, .
  \end {eqnarray}
 At second order, we obtain in the space $\underline {\mathscr{H}}$ the Kubo
correlations of the operators $\{ {\underline{\sf M}}\}$:
  \begin{eqnarray}   \label{MOC20qq}
  &&{ \langle  ( {\underline{\sf M}}_{\alpha} -  {R}^{(0)}_{\alpha} )
  ( {\underline{\sf M}}_{\beta} - {R}^{(0)}_{\beta} ) \rangle }_{\rm map}^{\rm K} =
  \\ \nonumber
  &&\left \langle \frac{1}{\beta} \,  \int_{{0}}^{\beta}\,d{\tau} \,
     e^{ \,{\tau}\underline{F} }
   \,( {\underline{\sf M}}_{\alpha} - {R}^{(0)}_{\alpha} )
        \,   e^{ \,-\,{\tau}\underline{F} }  \,
  ( {\underline{\sf M}}_{\gamma}-{R}^{(0)}_{\gamma} ) \right \rangle_{\rm map}
  = \frac {1}{\beta} \, \left( {\mathbb F}^{-1} \right)_{\alpha\gamma}  \,.
\end{eqnarray}

In order to derive therefrom the ordinary correlations we proceed as in 
Sec.\,\ref{sec3.3}.
We note that
 \begin{eqnarray}   \label{MOC21q}
       \frac{d}{d{\tau}}  \,  e^{ \,{\tau}\underline{F} }  \,
    {\underline{\sf M}}_{\alpha} \, e^{ \,-\,{\tau}\underline{F} }
  =  e^{\,{\tau}\underline{F}} \,[\underline{F} ,\,{\underline{\sf M}}_{\alpha}] \,
      e^{ \,-\,{\tau}\underline{F} }
  = -\,i\,\hbar\, ({\mathbb C}\,{\mathbb F})_{\alpha}^{\;\gamma}
   \, e^{ \,{\tau}\underline{F} }
  \, ( {\underline{\sf M}}_{\gamma} -  {R}^{(0)}_{\gamma} )
    \,e^{\,- \,{\tau}\underline{F} } \,,
 \end{eqnarray}
where we used the expression (\ref{MOC12q}) of $\underline{F} $ and the commutation relations (\ref{MOC10q}).
Integration over ${\tau}$ then yields
 \begin{eqnarray}   \label{MOC22q}
   e^{ \,{\tau}\underline{F} }
  \, ( {\underline{\sf M}}_{\alpha} - {R}^{(0)}_{\alpha} ) \,
    e^{ \,-\,{\tau}\underline{F} }
  = \left(  e^{  \,-\,i\,\hbar\,{\tau} \,{\mathbb C}\,{\mathbb F} }
    \right) _{\alpha}^{\;\beta}
  ( {\underline{\sf M}}_{\beta} - {R}^{(0)}_{\beta} )   \, ,
 \end{eqnarray}
  and hence, through a new integration as in (\ref{MFA012q}),
 \begin{eqnarray}   \label{MOC23q}
   &&{ \langle  ( {\underline{\sf M}}_{\alpha} -  {R}^{(0)}_{\alpha} )
   ( {\underline{\sf M}}_{\beta} - {R}^{(0)}_{\beta} ) \rangle }_{\rm map}^{\rm K} =
   \\   \nonumber
   && \left(  \frac {\Id-\exp [\,-\,i\,\hbar\,{\beta}\,{\mathbb C}\,{\mathbb F}]}
      {{i\,\hbar}\,{\beta}\,{\mathbb C}\,{\mathbb F}}  \right)_{\alpha}^{\;\beta}
     \,\langle  ( {\underline{\sf M}}_{\beta} -  {R}^{(0)}_{\beta} )
   ( {\underline{\sf M}}_{\gamma} - {R}^{(0)}_{\gamma} ) \rangle _{\rm map} \, .
 \end{eqnarray}
 By combining (\ref{MOC20q})
and (\ref{MOC23q}) we find that the ordinary correlations of the operators
$\{ {\underline{\sf M}}\}$ are given by
 \begin{eqnarray}   \label{MOC24qq}
  \langle  ( {\underline{\sf M}}_{\alpha} -  {R}^{(0)}_{\alpha} )
         ( {\underline{\sf M}}_{\beta} - {R}^{(0)}_{\beta} ) \rangle _{\rm map}
 =   \left(   \frac { i\,\hbar\,{\mathbb C}\,{\mathbb F}}
                 {  \Id -\exp[ {-i\,\hbar}\,\beta\,{\mathbb C}\,{\mathbb F} ]  }
 \,{\mathbb F}^{-1}    \right)_{\alpha\beta}
  =  {\mathbb B}_{\alpha\beta}  \, ,
 \end{eqnarray}
 so that we recover here the matrix ${\mathbb B}$ of Eq.(\ref{MFA19Aq}), now  
derived as an exact correlation in the mapped space $\underline{\mathscr{H}}$.


\end{document}